\documentclass[iop,apj,appendixfloats]{emulateapj}
\usepackage{graphicx}
\usepackage{natbib}
\usepackage{amssymb}
\usepackage{amsmath}
\usepackage{float}
\usepackage{gensymb}
\usepackage{afterpage}

\usepackage{chngcntr}

\def\beq{\begin{equation}}
\def\eeq{\end{equation}}

\def\Msol{\ensuremath{M_{\odot}}}

\def\abt{$\sim$}

\def\shen{Shen EOS}
\def\ls{LSEOS}

\def\background{$\bar\nu_e + \nu_x$ background}
\def\backgrounds{$\bar\nu_e + \nu_x$ backgrounds}

\def\cer{Cherenkov}

\def\lmax{$L^n_{\nu_e,\mathrm{max}}$}
\def\tmax{$t_{\mathrm{max}}$}
\def\trise{$t_{\mathrm{rise},\nobreak 1/2}$}
\def\tfall{$t_{\mathrm{fall},\nobreak 1/2}$}

\def\ar40{$^{40}$Ar}
\def\nue{$\nu_e$}
\def\anue{$\bar\nu_e$}
\def\nux{$\nu_x$}
\def\nuxpart{$\nu_{\mu,\nobreak \tau}$}
\def\nuxanti{$\bar\nu_{\mu,\nobreak \tau}$}

\begin{document}

\title{Detecting the Supernova Breakout Burst in 
Terrestrial Neutrino Detectors}

\author{Joshua Wallace\footnote{joshuajw@astro.princeton.edu}, Adam Burrows}
%\email{joshuajw@astro.princeton.edu}
\affil{Department of Astrophysical Sciences, Princeton
  University, Princeton, NJ 08544, USA}

\and

\author{Joshua C. Dolence}
\affil{Computational Physics Group (CCS-2), MS-K784, Los
  Alamos National Laboratory, P.O. Box 1663, Los Alamos, NM 87545,
  USA}

\begin{abstract}
We calculate the distance-dependent performance of a few representative terrestrial 
neutrino detectors in detecting and measuring the properties of the $\nu_e$
breakout 
burst light curve in a Galactic core-collapse supernova. The breakout 
burst is a signature phenomenon of core collapse and offers a probe
into 
the stellar core through collapse and bounce. We examine cases of 
no neutrino oscillations and oscillations due to normal and inverted 
neutrino-mass hierarchies. For the normal hierarchy, other neutrino 
flavors emitted by the supernova overwhelm the $\nu_e$ signal, making a 
detection of the breakout burst difficult. For the inverted hierarchy (IH), 
some detectors at some distances should be able to see the $\nu_e$ breakout 
burst peak and measure its properties. For the IH, the 
maximum luminosity of the breakout burst can be measured at 10 kpc to 
accuracies of \abt 30\% for Hyper-Kamiokande (Hyper-K) and \abt 60\%
for the Deep Underground Neutrino Experiment (DUNE). Super-Kamiokande
(Super-K) and Jiangmen Underground Neutrino Observatory (JUNO) lack 
the mass needed to make an accurate measurement. 
For the IH, 
the time of the maximum luminosity of the breakout burst can be
measured 
in Hyper-K to an accuracy of \abt 3 ms at 7 kpc, in DUNE \abt 2 ms at 4 kpc, 
and JUNO and Super-K can measure the time of maximum luminosity to 
an accuracy of \abt 2 ms at 1 kpc. 
Detector backgrounds in IceCube render a measurement of the $\nu_e$
breakout burst unlikely.
For the inverted hierarchy, a 
measurement of the maximum luminosity of the breakout burst could 
be used to differentiate between nuclear equations of state. 

\end{abstract}


\section{Introduction}
The core-collapse supernova (CCSN) explosion mechanism is a
long-standing unsolved problem in astrophysics. Extant CCSN models
universally indicate that neutrino emission is a key aspect
of a CCSN, with $\sim$99\% of a CCSN's energy carried away by
neutrinos. Such a neutrino signature was confirmed in broad outline 
by the detection of 
neutrinos from SN 1987A (\citealp{bionta1987,hirata1987}).  
Since the events that lead to a CCSN 
are entirely contained 
in the obscured core of the exploding star, the measurement of neutrinos 
(which are able to stream from the core of a star) is vital for testing CCSN theory.  

CCSNe occur as successive stages of nuclear burning build a
degenerate core to the Chandrasekhar limit. At this point, the 
core collapses in on
itself.  During collapse, electron capture causes the core to become neutron-rich.
  The implosion of the core proceeds until the material reaches
densities near nuclear, at which point the implosion
rebounds and produces an outward-propagating shock wave. 
  As the shock propagates, it causes nuclei to dissociate.
Electron capture on the now-free protons produces a large
number of \nue's in the region behind the shock. Initially, the optical
depth seen by these neutrinos prevents their escape from the star, 
but as the shock crosses the \nue-neutrinospheres, the
\nue's produced by the shock (as well as \nue's produced previously 
that have diffused to the neutrinosphere) create a very
luminous (${\sim} 3.5 \times 10^{53}$ erg s$^{-1}$) spike (``breakout
burst'') in the
\nue\ emission, which lasts for ${\sim}10$ ms.  After
this breakout spike,
neutrinos of all types radiate from the proto-neutron
star for ${\sim}10$ s or more  
(\citealt{burrowslattimer1986}).  Neutrino oscillations are likely to convert
the \nue's of the breakout burst partially or entirely to other
neutrino flavors (\citealt{mirizzietal2015}).

Prior to the \nue\ breakout burst, 
there is a smaller \nue\ luminosity peak
due to \nue's produced by the neutronization of the core during
 collapse. As the density and temperature of the core increase, the
 opacity increases and eventually these neutrinos are trapped,
 producing a peak and subsequent decrease in luminosity
 (``pre-breakout neutronization peak'').  The peak
 luminosity reached is $\sim$5$\times 10^{52}$ erg s$^{-1}$. 

The breakout burst of a CCSN is a signature
phenomenon that must exist for current CCSN theories to be 
valid. Hence, unambiguous detection of the breakout burst 
in the next Galactic 
CCSN is vital to validating theory, and a measurement of 
the properties of the breakout burst
would be important for testing and discriminating between CCSN models. Since 
Galactic CCSNe occur at a rate of $\sim$3 per century 
(\citealt{adamsetal2013}),  to take advantage of the next
Galactic CCSN, we must constantly be ready to
take data.  The Supernova Early Warning System (SNEWS) provides such
constant vigilance
 (\citealp{antoniolietal2004,scholberg2008}).

The very property of neutrinos that allows them to stream through the
stellar mantle also makes them very difficult to detect. Only 19
neutrinos were detected in the Kamiokande II and IMB detectors from SN
1987A (\citealp{bionta1987,hirata1987}). This was sufficient to confirm
general details of CCSN theory but lacked sufficient discriminating
power to 
truly differentiate between models, as well as lacked detail to see
the \nue\ breakout burst.  The current generation of neutrino
detectors promises much larger integral signals for a Galactic CCSN,
but will they be adequately sensitive to detect and characterize the
inaugural breakout burst?
This work seeks to answer this question in the context of
current and likely future neutrino detectors.

The CCSN
models used for our analysis are introduced in
Section~\ref{sec:model}, and the expected evolution of
the detected breakout burst signal is discussed in
Section~\ref{sec:signalevolution}.
The various classes of 
neutrino detectors are
discussed in Section~\ref{sec:detection}. The method of our analysis is
outlined in Section~\ref{sec:method} and the results
of our analysis are explored in Section~\ref{sec:discussion} (which
includes results from the no-oscillation case, as well as
those due to the normal and inverse neutrino-mass hierarchies).
We then conclude in 
Section~\ref{sec:conclusion}.

\section{The Supernova Models}
\label{sec:model}

The models used in this study were produced using F{\sc{ornax}} (J. Dolence 
\& A. Burrows 2015, in preparation).
F{\sc{ornax}} is a code written in generally covariant form for 
multidimensional,
self-gravitating, radiation hydrodynamics that is second-order accurate in
space and time and was designed from scratch with the CCSN 
problem in mind. The code solves the equations of compressible
hydrodynamics with an arbitrary equation of state (EOS), coupled to the
multigroup two-moment equations for neutrino transport. The hydrodynamics
in F{\sc{ornax}} is based on a directionally unsplit Godunov-type finite-volume
method. Fluxes at cell faces are computed with the fast and accurate HLLC
approximate Riemann solver, with left and right states given by limited
parabolic reconstructions of the underlying volume-averaged states. The
multigroup two-moment equations for neutrino transport are formulated in
the comoving frame and include all terms to $\mathcal{O}$($v/c$). 
The moment hierarchy
is closed with the ``M1" model. F{\sc{ornax}} adopts a Godunov-type approach for
treating the transport-related divergence terms that requires that we
solve a generalized Riemann problem at each face. All of the these
transport-related terms are treated explicitly in time. After core bounce,
the fastest hydrodynamic signal speeds in the CCSN 
problem are within a factor of a few of the speed of light, so explicit
time integration is not only simpler and generally more accurate but it is
also faster than globally coupled time implicit transport solves that are
typically employed in radiation hydrodynamics. The source and sink terms
that transfer energy between the radiation and the gas are operator-split
and treated implicitly. These terms are purely local to each cell and do
not introduce any global coupling.

 Progenitors with mass of 12, 15, 20, and
25 \Msol\ from \cite{woosleyheger2007} 
were simulated in 1D with the Lattimer--Swesty high-density
nuclear EOS with $K=220$ MeV (\ls;
\citealt{lattimerswesty1991}), while a progenitor with mass 15
\Msol\ was simulated using the Shen nuclear EOS (\shen;
\citealp{shenetaljul1998,shenetalnov1998}).  This set of models is
chosen to give a good span of potential progenitors for an actual
CCSN. 
 Since in this study we focus on the breakout burst, the full ${\sim}10$ s of
core neutrino cooling is not important for our study and is, thus, not
discussed further.

Figure~\ref{fig:lumallt} shows the early (unoscillated) neutrino light curve for the
neutrino species in the models: \nue, \anue, and \nux\ (which represents
$\nu_\mu$, $\nu_\tau$, and their antiparticles collectively), with
$L_{\nu_i}$ representing the $\nu_i$ energy luminosity. In this work,
we use \nuxpart\ to refer to $\nu_\mu$ and $\nu_\tau$ and use \nuxanti\
to refer to $\bar\nu_\mu$ and $\bar\nu_\tau$.
Additionally, $L_{\nu_x}$ refers to the luminosity due to all of
the \nux's together, while  $L_{\nu_{\mu,\nobreak \tau}}$ and
$L_{\bar\nu_{\mu,\nobreak \tau}}$ refer to the luminosity due to just
one of their constituent species.
The initial peak in \nue\ luminosity
from the preshock neutronization of the core is evident, followed
by the much steeper rise and
larger peak in \nue\ luminosity due to the breakout burst. After the 
breakout burst, the \anue\ and
\nux\ luminosities rise to nearly constant values.  The \nue\ luminosity
then falls to a comparably constant value.  In studying the
breakout burst, \anue\ and \nux\ serve as a background 
to the \nue\ signal we want to detect.

Figure~\ref{fig:nuelumt} shows the unoscillated-\nue\ number luminosity as a function of
time during breakout for the various models. The light curves are
centered around their maximum values. The \shen\ model has the largest
peak luminosity,
while the \ls\ models are grouped together at a slightly lower peak
luminosity. 
Results from \cite{sullivanetal2015} suggest that the higher peak 
\nue\ luminosity and smaller light curve width associated with the \shen\ 
are due to its smaller electron-capture rate (relative to that 
using the \ls), particularly on infall.  The different 
electron-capture rates are due to different free proton 
abundances and nucleon chemical potentials, the latter of which 
also affect the stimulated absorption correction of electron capture.
This tight grouping of peak luminosities for all progenitors, 
if in fact real,
can be used as a standard candle to determine the distance to a
supernova (SN).  \cite{kachelriessetal2005} suggest that an SN at 10 kpc
could have its distance determined to a precision of \abt 5\%, if these
theoretical predictions play out.

The goal of our analysis here is to examine, for a given
detector, for a given SN distance, how well the various properties of
this breakout burst can be constrained based on expected neutrino
signals, as well as determine whether our different CCSN models can
be differentiated based on a measurement of the breakout burst.  To
quantify various properties of the breakout burst, we define the
following physical parameters: 
the maximum number 
luminosity of breakout burst
($L^n_{\nu_e,\mathrm{max}}$),
the time of maximum luminosity ($t_{\mathrm{max}}$), the width of the
breakout burst peak ($w$, calculated using the FWHM), 
the rise time ($t_{\mathrm{rise},1/2}$, calculated
using the width left of peak at half-maximum), and the fall time
($t_{\mathrm{fall},1/2}$, calculated using
the width right of peak at half-maximum).  For the preshock
neutronization peak we define the following physical parameters: the
maximum number luminosity of the preshock peak
($L^n_{\nu_e,\mathrm{max,pre}}$) and the time of this maximum luminosity 
($t_{\mathrm{max,pre}}$).
Table~\ref{tab:numpeakfit_realparm} shows the values the physical
parameters of the breakout burst take in our  number luminosity light
curves, and Table~\ref{tab:smallbumpphysicalparms} shows the values
the physical parameters of the preshock neutronization peak take in
the same curves.
The fact that  $t_{\mathrm{max}}=0$ for all models is to be
expected, since the zero point in time for all the models is defined
to be the time of maximum luminosity.
\begin{deluxetable}{lccccc}
%\begin{deluxetable*}{lccccc}
\tablewidth{\linewidth}
\tablecolumns{7}
\tablecaption{Physical Parameters of the Breakout Peak Fit to the Number Luminosity, L$^n_{\nu_e}$\label{tab:numpeakfit_realparm}}
\tablehead{
\colhead{Model} & \multicolumn{1}{c}{$L^n_{\nu_e,\mathrm{max}}$} & \colhead{$t_{\mathrm{max}}$}
& \colhead{$w$} & \colhead{$t_{\mathrm{rise},1/2}$}
& \colhead{$t_{\mathrm{fall},1/2}$}
%& \colhead{$L^n_{\nu_e,\mathrm{max,pre}}$}
\\ [0.35ex] %& \colhead{$t_{\mathrm{max,pre}}$}\\ [0.35ex]
\colhead{(\Msol)} &  \colhead{$(10^{58} $ s$^{-1})$} &  \colhead{(ms)} &  \colhead{(ms)}
&  \colhead{(ms)}
&  \colhead{(ms)} }%& \colhead{$(10^{58} $ s$^{-1})$} & \colhead{(ms)}}
\startdata
12 & 1.97 & 0.00 & 8.85 & 2.23 & 6.62 \\%& 0.552 & -6.51\\
15 & 2.01 & 0.00 & 9.29 & 2.24 & 7.06 \\%& 0.577 & -6.48\\
15 S & 2.23 & 0.00 & 8.71 & 1.70 & 7.01 \\%& 0.629 & -5.73\\
20 & 2.03 & 0.00 & 10.2 & 2.22 & 7.96 \\%& 0.597 & -6.43\\
25 & 2.02 & 0.00 & 10.4 & 2.28 & 8.10 %& 0.604 & -6.48
\enddata
%\end{deluxetable*}
\end{deluxetable}
\begin{deluxetable}{lcc}
\tablewidth{0pc}
\tablecolumns{5}
\tablecaption{Physical Parameters of the Preshock Neutronization Peak Fit to the Number Luminosity, L$^n_{\nu_e}$\label{tab:smallbumpphysicalparms}}
\tablehead{
\colhead{Model} & \multicolumn{1}{c}{$L^n_{\nu_e,\mathrm{max,pre}}$} & \colhead{$t_{\mathrm{max,pre}}$} \\
  \colhead{(\Msol)} & \colhead{(10$^{58}$ s$^{-1}$)} &\colhead{(ms)} }
\startdata
12 & 0.552 & -6.51\\
15 &  0.577 & -6.48\\
15 S & 0.629 & -5.73\\
20 & 0.597 & -6.43\\
25 & 0.604 & -6.48
\enddata
\end{deluxetable}

We construct an analytic
model for the main breakout peak of our numerical light curves similar to 
equation (10) of \cite{burrowsmazurek1983}. This analytic model will
later be used to fit simulated observations of the \nue\ breakout
burst light curve to measure various physical parameters of the
breakout burst.  The function we use to 
fit the main peak is
\begin{equation}
\label{eq:analytic}
L(t) = \frac{A}{((t-t_{c})/[\textrm{ms}])^\alpha} \exp\left[ -\left(\frac{b}{t-t_{c}}\right)^\beta\right] + L_{\textrm{base}},
\end{equation}
where $A$ is a scaling parameter with units of (for energy luminosity) erg
s$^{-1}$ or (for number luminosity) s$^{-1}$, $b$ is a
width parameter, $t$ is the time since the maximum 
\nue\ number luminosity, $t_{c}$
is a number used to center the fit appropriately,
$\alpha$ and
$\beta$ are exponents similar to those used in
\cite{burrowsmazurek1983}, 
and $L_{\textrm{base}}$ (in units of erg s$^{-1}$ for energy luminosity or s$^{-1}$ for
number luminosity) is used to set the floor of the fit.  Since the
intent of Equation~(\ref{eq:analytic}) is to provide a fit to our models
for subsequent analysis, we do not spend much time in this work 
examining its physical significance. We refer readers to
\cite{burrowsmazurek1983} for a motivation of its functional form.  
We do note that the floor set by $L_{\textrm{base}}$ can be thought of as 
the level 
to which the \nue\ luminosity decays as the luminosity source
transitions from the electron capture that dominates the breakout
burst to the accretion, deleptonization, and cooling phases of the proto-neutron star.

\begin{deluxetable*}{lcccccc}
%\begin{deluxetable}{lcccccc}
\tablewidth{\linewidth}
\tablecolumns{7}
\tablecaption{Breakout Peak Fit to the Energy Luminosity, L$_{\nu_e}$\label{tab:peakfit}}
\tablehead{
\colhead{Model} & \multicolumn{1}{c}{$A$} & \colhead{$b$}
& \colhead{$t_c$} & \colhead{$\alpha$}
& \colhead{$\beta$} & \colhead{$L_{\textrm{base}}$} \\ 
\colhead{(\Msol)} &  \colhead{$(10^{57} $ erg s$^{-1})$} &
\colhead{(ms)} &  \colhead{(ms)}
&  \colhead{} &  \colhead{}
&  \colhead{$(10^{53} $ erg s$^{-1})$}  }
\startdata
12 & 0.620 & 12.2 & -4.69 & 3.06 & 1.09 & 0.360\\
15 & 1.52 & 20.6 & -4.33 & 3.20 & 0.849 & 0.418\\
15 S & 1.00e14 & 6.65e8 & -2.49 & 6.26 & 0.182 & 0.345\\
20 & 1.00e14 & 2.33e7 & -3.47 & 6.96 & 0.220 & 0.501\\
25 & 2.33e5 & 7090 & -3.69 & 4.91 & 0.349 & 0.511
\enddata
\end{deluxetable*}
%\end{deluxetable}
%
Figure~\ref{fig:fit}
displays the fits of
Equation~(\ref{eq:analytic}) to our numerical models.  The fits 
match the numerical models quite well.    The
analytic fitting parameters to the energy luminosity of  all the
numerical models are given in 
Table~\ref{tab:peakfit}, and the  fitting parameters to the number 
luminosity of  all the
numerical models are given in 
Table~\ref{tab:numpeakfit}.  The \shen\ model is represented by ``S'' in the
``Model'' column; those models not marked with an ``S'' use
the \ls. This is a convention we take throughout this work. 
Two of the parameters ($A$ and $b$) show
variation over many orders of magnitude across the fits of the
different models.  Because of this, we set a maximum value for $A$,
which equates to (for energy luminosity) $10^{71}$ erg s$^{-1}$ and
(for number luminosity) $10^{71}$ s$^{-1}$.  The choice of this
particular maximum value is arbitrary.  It is based on our experience
with the behavior of the fits with unbounded parameters.  Since the fits are so
good, and since we care about the physical parameters that 
are derived from the fits rather than the fit parameters themselves, 
we see no harm in doing this.  Additionally, we 
force $\alpha$, $\beta$,
$L_{\textrm{base}}$, and $A$ to be positive.  We wish to emphasize that
although the parameters of Table~\ref{tab:peakfit} vary to a 
large degree between models, the important consideration in our analysis is
not the parameters themselves, but the characteristics of the fit they
provide (i.e. the physical parameters introduced previously and
shown in Table~\ref{tab:numpeakfit_realparm}), so the
fidelity of the fits in representing the numerical models is far more
important than the values the fit parameters
take. A reason for the large variation in the 
best-fit values of
$A$ and $b$ is the very large, positive covariance between these two
parameters.  Since the specific values of these parameters are not
important, we do not focus on the covariance
data in this work. 

\begin{deluxetable}{lcccccc}
\tablewidth{\linewidth}
\tablecolumns{7}
\tablecaption{Breakout Peak Fit to the Number Luminosity, L$^n_{\nu_e}$\label{tab:numpeakfit}}
\tablehead{
\colhead{Model} & \multicolumn{1}{c}{$A$} & \colhead{$b$}
& \colhead{$t_c$} & \colhead{$\alpha$}
& \colhead{$\beta$} & \colhead{$L^n_{base}$} \\ [0.35ex]
%& \colhead{$\beta$} & \colhead{$L^n_{base}$} \\ [0.1ex]
\colhead{(\Msol)} &  \colhead{$(10^{61} $ s$^{-1})$} &  \colhead{(ms)} &  \colhead{(ms)}
&  \colhead{} &  \colhead{}
&  \colhead{$(10^{57} $ s$^{-1})$}  }
\startdata
12 & 3.29 & 16.3 & -4.30 & 2.94 & 0.892 & 2.88\\
15 & 39.7 & 60.1 & -3.90 & 3.39 & 0.621 & 3.35\\
15 S & 1.00e10 & 2.71e7 & -2.38 & 4.95 & 0.198 & 2.76\\
20 & 1.00e10 & 8.44e5 & -3.31 & 5.68 & 0.251 & 3.92\\
25 & 5.18e9 & 4.89e5 & -3.41 & 5.67 & 0.260 & 4.06
\enddata
\end{deluxetable}

Additionally, we fit the preshock neutronization peak with a
modified lognormal curve,
\begin{equation}
\label{eq:smallpeak}
L(t) =
C \exp\left(\frac{-(\ln(t' -\theta)-\mu)^2}
{2\sigma^2}\right)(t'-\theta)^{-1},
\end{equation}
where $t'=-t/$[ms], $C$ is a scaling factor, and all the other
parameters have their
usual meaning in relation to the lognormal distribution: $\sigma$ is
the standard deviation, $\theta$ is the location parameter, and $\mu$
is the median of the distribution.  The best-fit values to the
preshock  energy
luminosity for the various models are shown in
Table~\ref{tab:smallbumpfit}, and  Table~\ref{tab:numsmallbumpfit} shows the
best-fit values for the preshock 
number luminosity. Figure~\ref{fig:fit}
displays the fits of
Equation~(\ref{eq:smallpeak}) to our numerical models. 
The physical parameters we define for the 
preshock neutronization peak are (as introduced previously) the
maximum number luminosity of the preshock peak
($L^n_{\nu_e,\mathrm{max,pre}}$) and the time of this maximum luminosity 
($t_{\mathrm{max,pre}}$).  The model values of these parameters are
shown in Table~\ref{tab:smallbumpphysicalparms}. It is the fits
themselves (Equations~\ref{eq:analytic} and~\ref{eq:smallpeak}), with
the appropriate fitting parameters, that we use as our baseline models for our
analysis over the time ranges fitted by them, not the numerical data 
from F{\sc{ornax}}.  The numerical data are used in the time ranges
not fitted by Equations~\ref{eq:analytic} and~\ref{eq:smallpeak}
(i.e., the time ranges before to the pre-breakout neutronization peak
and after the breakout burst).

\begin{deluxetable}{lcccc}
\tablewidth{0pc}
\tablecolumns{5}
\tablecaption{Preshock Neutronization Peak Fit to the Energy Luminosity, L$_{\nu_e}$\label{tab:smallbumpfit}}
\tablehead{
\colhead{Model} & \multicolumn{1}{c}{$C$} & \colhead{$\sigma$}
& \colhead{$\theta$} & \colhead{$\mu$} \\
  \colhead{(\Msol)} & \colhead{(10$^{53}$ erg s$^{-1}$)} &\colhead{} &\colhead{} &\colhead{}  }
\startdata
12 & 3.93 & 0.822 & 1.96 & 2.08\\
15 & 4.28 & 0.809 & 1.79 & 2.12\\
15 S & 4.78 & 0.790 & 0.925 & 2.08\\
20 & 5.17 & 0.708 & 0.653 & 2.20\\
25 & 5.41 & 0.682 & 0.404 & 2.22
\enddata
\end{deluxetable}

\begin{deluxetable}{lcccc}
\tablewidth{0pc}
\tablecolumns{5}
\tablecaption{Preshock Neutronization Peak Fit to the Number Luminosity, L$^n_{\nu_e}$\label{tab:numsmallbumpfit}}
\tablehead{
\colhead{Model} & \multicolumn{1}{c}{$C$} & \colhead{$\sigma$}
& \colhead{$\theta$} & \colhead{$\mu$} \\
  \colhead{(\Msol)} & \colhead{(10$^{58}$ s$^{-1}$)} &\colhead{} &\colhead{} &\colhead{}  }
\startdata
12 & 3.74 & 0.834 & 1.72 & 2.26\\
15 & 3.91 & 0.857 & 1.79 & 2.28\\
15 S & 4.32 & 0.817 & 0.809 & 2.26\\
20 & 4.63 & 0.754 & 0.596 & 2.33\\
25 & 4.40 & 0.807 & 1.22 & 2.31
\enddata
\end{deluxetable}

The energy spectrum of all neutrino 
types varies during the shock breakout.  Figure~\ref{fig:energyspectrum} 
shows the \nue\ energy spectrum as a
function of time through the breakout burst for a specific model, 
the 15 \Msol\ \ls\ model, derived using F{\sc{ornax}}.
We define the average neutrino energy $\overline{E}_{\nu_i}$ as
\beq
\label{eq:averageenergy}
\overline{E}_{\nu_i}(t) \equiv \frac{\int L_{\nu_i}(E_{\nu_i},t)~dE_\nu}{\int L_{\nu_i}(E_{\nu_i},t)/E_{\nu_i}~dE_{\nu_i}}.
\eeq
  Figure~\ref{fig:avgenergy} shows the average neutrino energy as
a function of time for all the models (with the zero point in time set
to be the time of maximum number luminosity). 
All the models show the same
general behavior. 
$\overline{E}_{\nu_e}$ increases from the onset
of the breakout burst, peaks near \tmax, and then
decays slightly.  However, during the breakout, $\overline{E}_{\nu_e}$ 
does not change radically and is similar from model to
model.
 There is a slight trend for the average \nue\ energy during
breakout to be slightly higher for the lower-mass progenitors.  In
addition, the \shen\ results in a slightly higher mean \nue\ energy than
the \ls.  In all cases, the average energy peaks $\lesssim$1 ms later than
the maximum number luminosity.

% LocalWords:  ornax lccccccc 35ex lcccccc 00e14 65e8 33e7 33e5 00e10 71e7 44e5
% LocalWords:  18e9 89e5 lcccc 0pc

\section{Signal Evolution}
\label{sec:signalevolution}
In a neutrino detector, for a given neutrino detection channel, the
expected number of detected neutrinos $N_{\textrm{det}}$ as a function of time is given by
\beq
\label{eq:Nint}
N_{\textrm{det}}(t) =  \frac{N_t}{4\pi D^2} \int \frac{dL_\nu(E_\nu,t)}{dE_\nu}\frac{1}{E_\nu}\sigma(E_\nu)\epsilon(E_e)dE_\nu,
\eeq
where $N_t$ is the number of target particles inside the
detector, $dL_\nu/dE_\nu$
is the energy luminosity spectrum of neutrinos, $E_\nu$ is the
neutrino energy, $E_e$ is the energy of the final-state electron, 
$D$ is the distance between the detector and the SN, 
$\sigma$ is the interaction cross section, and $\epsilon$ is
the efficiency of detection.

In a detector, there is not a one-to-one mapping between the energy
of an interacting neutrino and the detected energy of the products.
To understand exactly what
a signal will look like in a detector, we have to understand how the
spectrum of the detectable products relates to the spectrum of the
incident neutrinos.  The number of events produced in a specific
channel with product $x$ with observed kinetic energy $E_x$
is
\begin{equation}
%\frac{dN_y}{dE_y} = \frac{N_t}{4\pi D^2} 
%L_\nu(E_\nu) \frac{d\sigma (E_y\prime,E_\nu)}{dE_y\prime}
\frac{dN_x}{dE_x dE_\nu dt} = \frac{N_t}{4\pi D^2} 
\frac{dL_\nu(t,E_\nu)}{dE_\nu}\frac{1}{E_\nu} \frac{d\sigma
(E_{x}',E_\nu)}{dE_{x}'}, %\delta(t-t'),
\end{equation}
where $E_\nu$ is the energy of the interacting neutrino, $E_x'$ is
the energy of the product (usually an electron or positron),
and $t$ is the detector time (with light-travel time $D/c$ subtracted).  

Neutrino oscillations that occur as a result of the
Mikheyev--Smirnov--Wolfenstein (MSW) effect 
in the envelope
of a star will likely alter the measured \nue\ signal from the
breakout burst.  There are two main realms of flavor conversions to be
considered in the \nue\ breakout burst, connected with the two
potential neutrino-mass hierarchies: the normal hierarchy (NH), 
where the $\nu_3$ mass
eigenstate is of larger mass than either of $\nu_1$ or $\nu_2$; and
the inverted hierarchy (IH), where the $\nu_3$ mass
eigenstate is of smaller mass than either of $\nu_1$ or $\nu_2$. In
the case of the NH, the observed luminosity of a neutrino species
$L^{obs}_{\nu_i}$ is (\citealt{mirizzietal2015})
\begin{align}
L^{\textrm{obs}}_{\nu_e} &= L_{\nu_{\mu,\tau}}, \\
L^{\textrm{obs}}_{\bar \nu_e} &= \cos^2{\theta_{12}}L_{\bar \nu_e}
+ \sin^2{\theta_{12}}L_{\bar \nu_{\mu,\tau}},
\end{align}
where $\theta_{12}$ is the 1, 2 mixing angle.  We use
$\sin^2{\theta_{12}} = 0.308$ in this work (\citealt{oliveetal2014}).
In the case of the IH, the observed luminosity of a neutrino species
is (\citealt{mirizzietal2015})
\begin{align}
L^{\textrm{obs}}_{\nu_e} &= \sin^2{\theta_{12}}L_{\nu_e}
+ \cos^2{\theta_{12}}L_{\nu_{\mu,\tau}}, \\
L^{\textrm{obs}}_{\bar \nu_e} &= L_{\bar \nu_{\mu,\tau}}. \label{eq:anue_ih}
\end{align}

% LocalWords:  det

\section{Terrestrial Neutrino Detectors}
\label{sec:detection}
 There are four main classes
of detectors relevant for detecting the \nue\ 
breakout burst: \ar40\ detectors, water-\cer\
detectors, long-string detectors, and 
scintillation detectors.    
Each of these detector classes has its
own strengths and weaknesses with regard to the detection of
the breakout burst.

Table~\ref{tab:detectors} lists the detectors we highlight in our
study, as well as some other detectors of interest for detection of
the breakout burst.  
It consists of both detectors currently running and detectors
that are expected to come online in the coming years.  The list is not
exhaustive, but is rather a representative ``short list'' of detectors we
think will provide the best opportunity to examine the breakout
burst.  We now discuss specifics of each class of detector.

\begin{deluxetable*}{lccc}
%\begin{deluxetable}{cccc}
\tablecolumns{4}
\tablecaption{Selected Neutrino Detectors\label{tab:detectors}}

\tablehead{\colhead{Detector} & \colhead{Detection Medium} & \colhead{Mass} & \colhead{Status} \\ 
\colhead{} & \colhead{} & \colhead{(ktonne)} & \colhead{} } 

\startdata
DUNE &  \ar40 & 40 & planning\tablenotemark{a}\\
Hyper-K & water & 560 (fiducial) & proposed\tablenotemark{b}\\
Super-K & water & 22.5 (fiducial)\tablenotemark{c} & running \\
IceCube & water ice & $\sim$900\tablenotemark{d} & running\\
JUNO  & scintillator & 20\tablenotemark{e} & construction\\
\hline\\ [-1.25ex]

ICARUS & \ar40 & .476 (active)\tablenotemark{f} & being refurbished\\
KamLAND & scintillator & 1\tablenotemark{g}  & running  \\ 
LVD & scintillator & $\sim$1\tablenotemark{h} & running \\
NO$\nu$A\tablenotemark{i} & scintillator & 14\tablenotemark{j} & running
\enddata

\tablenotetext{a}{Mass and status from \cite{deoliviera2015};}
\tablenotetext{b}{Mass and status from \cite{abeetal2011};}
\tablenotetext{c}{Mass from \cite{ikedaetal2007};}
\tablenotetext{d}{Based on the energy-dependent effective detection volume given in \cite{abbasietal2011};}
\tablenotetext{e}{Mass from \cite{li2014};}
\tablenotetext{f}{Mass from \cite{rubbiaetal2011};}
\tablenotetext{g}{Mass from \cite{eguchietal2003};}
\tablenotetext{h}{Mass from http://www.bo.infn.it/lvd/;}
\tablenotetext{i}{NO$\nu$A is located at the surface;}
\tablenotetext{j}{Mass from \cite{pattersonetal2012}.}
\end{deluxetable*}
%\end{deluxetable}

\subsection{\ar40\ Detectors}
Of all the detector types we consider, 
\ar40\ detectors have the highest sensitivity to \nue's (the primary
neutrino emission of the breakout burst, ignoring oscillations). The 
\nue's interact
with \ar40\ nuclei via charged-current (CC) 
capture; it is the large cross section of
this interaction  that gives \ar40\ 
detectors such great \nue\ sensitivity.  The electron created in this
process deposits its kinetic energy 
along an ionization trail through the detection medium.  The \ar40\
detectors we mention  in this paper are all time-projection chambers
(TPCs).  In a TPC, a voltage is applied
across the detection medium, causing the particles ionized by the
product electrons to drift
toward a wire mesh that (when combined with the timing of the
formation of the ionization trail) gives spatial information about the
interaction inside the detector.  The \anue's also undergo CC
absorption with the \ar40\ nuclei.  The cross section for this
interaction, for the neutrino energies in an SN, is 2--3
orders of magnitude smaller than for \nue\ CC absorption
in the energy range relevant for this study,
and so this interaction serves as a small background to the \nue\
signal in the case of no oscillations.  Neutrinos of all flavors can also be
detected  via elastic scattering off the electrons.
Electrons produced from electron scattering are 
indistinguishable from the electrons produced by CC
absorption of \nue's on the \ar40\ nuclei, but electron scattering has a
much smaller cross section (factor of ${\sim}100$ for 10 MeV neutrinos) 
than the CC interaction channel on \ar40.
 Because of the dominance of
the cross sections through which \nue's are detected and the dominance
of the \nue\ flux relative to the \anue\ and \nux\ flux through the breakout
burst, we assume the ability of these detectors to separate or ignore
the background \anue\ and \nux\ signals 
and consider only the \nue\ luminosity in our analysis for the case of
no neutrino oscillations.  The mixing of the \nuxpart\ and \nue\ flux
in the case of neutrino oscillations complicates this and we cannot
assume, for either the NH or IH, that the \nuxpart\ backgrounds are negligible. 
Also, \ar40\ detectors can measure
the energies of the electrons produced in the neutrino interactions
inside them.  
In principle, gamma rays from nuclear de-excitation
could be detected, allowing the tagging of \nue\ CC
absorption events and their separation from electron-scattering
events, as well as detection of neutral-current (NC) scatterings off of
nuclei.  The detectability of these gamma rays is still under study
(A. Rubbia 2015, private communication),
and we do not assume the ability to detect them.

The largest \ar40\ detector operated so far is ICARUS (Imaging 
Cosmic And Rare Underground Signals), formerly located in the Gran Sasso 
underground laboratory in Italy.  The detector had an active mass of
476 tonnes (\citealt{rubbiaetal2011}).  The detector is currently in
the process of being refurbished for later installation in the USA 
at Fermilab.  
In our analysis, we assume a detection efficiency
of 100\% across all product energies for the \ar40\ detectors. 
We note that the interaction cross sections set their own
threshold for neutrino detection, which is incorporated with our
cross-section 
calculations.  We refer interested readers to the Appendix for
a further discussion of these cross sections.  
We calculate that a detector of ICARUS's size 
will detect $\sim$1 \nue\ in a 10 ms period over the
breakout burst at a distance of 10 kpc in the case of no neutrino
oscillations (even less in either NH- or IH-case oscillations). 
Because of the expected small
signal from ICARUS, we do not use it in our analysis.

There are plans for a 40 ktonne (fiducial) \ar40\ detector to be
constructed at the Sanford Lab in the
Homestake Mine in South Dakota as a part of
part of the Deep Underground Neutrino Experiment    
 (DUNE; \citealt{deoliviera2015}).  We calculate that an \ar40\ detector of
this size will detect $\sim$120 \nue's in a 10 ms period over the
breakout burst at a distance of 10 kpc in the case of no neutrino
oscillations.  For the same situation as just outlined, the number of
detected (original) \nue's in the NH case will be \abt 2 and in the IH
case will be \abt 40. The timing resolution of DUNE depends on its 
photon detection capabilities. We assume a DUNE that will have 
timing much better than the time bin width used in our analysis (1 ms).

\subsection{Water-\cer\ detectors}
Water-\cer\ detectors are large tanks of purified water
primarily sensitive to \anue's through inverse $\beta$ decay (IBD) on
protons (hydrogen nuclei): $\bar\nu_e + p \rightarrow e^+ + n$. 
The positron produced by IBD emits \cer\ light, which is detected by
a photomultiplier tube (PMT) array placed around the detection
volume. Neutrinos of all flavors can be detected through elastic
scattering on electrons, $\nu_i + e^- \rightarrow \nu_i + e^-$.  
The final-state electrons are detected through their
\cer\ emission.  \nue\ and \anue\ can also undergo CC absorption 
on the oxygen
nuclei  and are detected through the electrons/positrons formed in
these interactions, as well as through photons emitted via nuclear
de-excitation.  
Additionally, neutrinos of all types may undergo NC
interactions with the oxygen nuclei, which (if the interaction puts
the nucleus in an excited state) can be detected by the photon emitted
upon de-excitation of the nucleus.

In standard water-\cer\ detectors, positrons from
IBDs and electrons from 
electron scatterings can be statistically distinguished by the
forward-peaked directionality of the electron-scattering products and the
nearly isotropic products of IBD.
 The IBD
detection channel has the largest cross section and so 
dominates when there is a \anue\ flux.  Figure~\ref{fig:lumallt} shows
that the \anue\ flux is not significant until after the
\nue\ peak of the breakout burst.
Thus, previous to this, almost all detections can be attributed
to \nue's (in the no-oscillation case).
After this the \anue's (and, less so, the \nux's) 
must be accounted
for.  It is also possible to 
employ gadolinium (Gd) in water-\cer\ detectors 
to tag the final-state neutrons and allow the IBD and
electron elastic scattering signals to be separated
(\citealt{vagins2012}; \citealt{lahabeacom2014}).  The large neutron
capture cross section of Gd allows neutrons formed in IBD events to be
quickly ($\sim$20 $\mu$s) captured, emitting three to four gamma rays with a total
energy of 8 MeV (\citealt{beacomvagins2004}).  
The coincidence of a neutrino
detection and a gamma ray from neutron capture on Gd 
allows the neutrino detection to be associated with an IBD.  We
assume the presence of Gd in our analysis.  In practice, using Gd to
tag IBD events will not be perfect, although the fraction of neutrons
that are captured onto Gd is very high even for modest additions of Gd
to water (\citealt{beacomvagins2004}).  The signal remaining from
untagged IBDs can be statistically subtracted from the remaining signal.
Because of this, and because of the ability to statistically distinguish
\nue's and \anue's based on direction, in our analysis we assume that
IBDs can be separated from the rest of the signal.  In the case of no
oscillations, this leaves the \nue\ flux dominant.
\cite{lahabeacom2014} also state that \nux\ 
information from scintillation detectors 
will allow those events to be statistically subtracted,
leaving only the \nue\ events.  Based on this, we assume the
ability to subtract or ignore all \backgrounds\footnote{In this work,
the \backgrounds\ to which we refer are due to \anue\ and \nux\
emission from the CCSN itself, not the detector backgrounds or any
ambient neutrino background (for instance, solar neutrinos or the
diffuse SN neutrino background).} in the no-oscillation
case.  In the case of neutrino oscillations, the detectability of the 
background \nux\ flux
will increase and cannot be so easily ignored.

Super-Kamiokande (Super-K) is a 22.5 ktonne (fiducial) water-\cer\ 
detector located
in the Kamioka Mine in Japan (\citealp{fukudaetal2003,ikedaetal2007}).
Super-K
 IV, for solar
 neutrino analysis, reports a 99\% triggering efficiency for 4.0--4.5 
MeV and a 100\% triggering efficiency above 4.5 MeV
(\citealt{sekiya2013}).  The exact threshold and detection volume that
can be used for neutrino detection in a CCSN depend on the background
rate and the signal rate, the latter of which depends on distance to
the SN.  Thus, there is some ambiguity regarding which detector
threshold would be most appropriate to use.
In our analysis, we assume a Heaviside step function with 
step at 4.0 MeV as our detection efficiency function for Super-K, for
all SN distances.  Since electron scattering
has a spread of electron energies that can be produced for a given
neutrino energy, neutrinos with energies well above the detector
threshold may result in product electrons that fall below the
threshold.  For 10 MeV \nue's (comparable to the average neutrino
energy through the breakout burst), we find that \abt 40\% of electron
scatterings produce electrons with energies below our chosen 4 MeV
threshold.
This represents a significant reduction in the detected \nue\ flux
relative to what otherwise could be measured that could possibly be measured. 
Any improvements that can be made to the detector threshold for
a CCSN have significant potential to 
improve the results presented throughout this work.
  We calculate 
that Super-K  will detect $\sim$7 \nue's in a 10 ms period over the
breakout burst at a distance of 10 kpc in the case of no neutrino
oscillations.  For the same situation as just outlined, the number of
detected (original) \nue's in the NH case will be \abt 0 and in the IH
case will be \abt 1.

Super-K has recently been approved to have Gd added, following
years of study as to the  feasibility and impact of adding Gd to 
Super-K (\citealp{beacomvagins2004,watanabe2009,vagins2012,mori2013}).  
\cite{beacomvagins2004} suggest that with 0.2\% (by mass) 
Gd added to Super-K,
${\sim}90\%$ of the IBD events could be tagged. The remaining
IBD events (as well as the \anue\ absorption events on $^{16}$O) 
can then be statistically subtracted from the remaining
signal.  

Hyper-Kamiokande (Hyper-K) is a proposed 560 ktonne (fiducial) water-\cer\
detector planned to be located at the Kamioka Mine in Japan
(\citealt{abeetal2011}). The detector threshold for Hyper-K depends on
the PMT coverage fraction, which is not yet
finalized.  If this coverage fraction is smaller than that of Super-K,
then the detector threshold for Hyper-K will be greater than for
Super-K.  Because the final detector design is not yet finalized, and
for simplicity, for Hyper-K we assume the same 4 MeV 
detector threshold that
we assume for Super-K.  We calculate 
that a water-\cer\ detector of Hyper-K's size  will detect 
$\sim$160 \nue's in a 10 ms period over the
breakout burst at a distance of 10 kpc in the case of no neutrino
oscillations.  For the same situation as just outlined, the number of
detected (original) \nue's in the NH case will be \abt 30 and in the IH
case will be \abt 70.

Both Super-K and Hyper-K could provide an estimate of the direction of the neutrino 
flux.  The scattering of \nue's  off of electrons results in electron 
propagation that is forward peaked relative 
to the incident neutrino's motion.  The direction of
the final-state electrons can be measured using information from the
electrons' \cer\ light cones.  Because the electrons produced in these
scatterings are not perfectly forward peaked, and because of the
subsequent straggling of the electrons as they scatter within the
detector, the precision of such a direction measurement is
limited.  \cite{andosato2002} calculate that Super-K can measure the
location of a CCSN at 10 kpc to within a circle of
$\sim$9\degree\ radius, using \nue's measured over the whole neutrino
event (and not just the breakout burst).  \cite{tomasetal2003}
calculate that Super-K, with Gd added, could measure an SN
position to an accuracy of 3.2\degree--3.6\degree,
depending on the neutron tagging efficiency.  They also calculate that
a megatonne water detector with Gd and 90\% tagging efficiency would measure
the direction to an accuracy of 0.6\degree, and  \cite{abeetal2011} state
that Hyper-K would be able to measure the direction for a CCSN at 10 kpc
to an accuracy of $\sim$2\degree.

\subsection{Long-string Detectors}

IceCube is a long-string detector embedded in the Antarctic ice at the
South Pole (\citealp{achterbergetal2006,abbasi2010}). It
is optimized for the detection of neutrinos with TeV energies, much
higher than the $\mathcal{O}(10)$~MeV energies expected for the
breakout burst neutrinos. 
However, the
neutrinos from a CCSN will create a correlated rise in the measured
detector background across all the individual PMTs of IceCube
(\citealp{pryoretal1988,halzenetal1996}).  
%Estimates for the 
%volume probed for CCSN neutrino
%detection by the entire
%ensemble of detectors in IceCube vary from $\sim$2.6 Mtonne
%in \cite{halzenrodrigues2010} (scaling their estimate of
%$\sim$2.4 Mtonne to that of the completed IceCube) to $\sim$3.5 Mtonne
%in \cite{abbasietal2011}.  We use a detection mass of 3.5 Mtonne  
%in our analysis.
Each individual PMT effectively monitors 
an energy-dependent volume of ice surrounding it, with the 
size of the effective volume depending linearly on the 
energy of the interaction products (\citealt{abbasietal2011}).  
Using a cross-section-weighted average $\nu_e$ energy of 13 MeV 
for the 15 \Msol\ CCSN model with the LSEOS and the spread of interaction 
product energies, IceCube corresponds to a \abt 900 ktonne 
CCSN neutrino detector.
Although real-time SN monitoring in IceCube bins data 
in 2 ms time bins, data on the individual photon detections 
are saved in a 90 s window around a putative SN 
event (\citealt{aartsenetal2013}), allowing for arbitrary binning.
IceCube
currently lacks the ability to measure the energy of the neutrinos it
would detect from a CCSN, so it would be able to measure a light curve, but
not a spectrum for the breakout burst.  For our analysis, we assume no
detector threshold for IceCube.  We calculate that 
IceCube  will detect 
$\sim$1600 \nue's in a 10 ms period over the
breakout burst at a distance of 10 kpc in the case of no neutrino
oscillations.  For the same situation as just outlined, the number of
detected (original) \nue's in the NH case will be \abt 300 and in the IH
case will be \abt 700.  We also note that because
IceCube does not reproduce neutrinos from CCSNe on an event-by-event
basis, it cannot provide any pointing information by itself, as can
Super-K.  However, its large breakout yield may be useful
in a triangulation calculation using multiple detectors.

%IceCube also lacks the ability to discriminate between neutrino
%species.  However, \nue's dominate through the breakout burst peak,
%and in the no-oscillation case the peak in \nue\ luminosity is still
%visible against the \backgrounds.  Because of this, because of the
%potential of using information from other detectors to subtract 
%the \backgrounds\ (especially the \anue\ IBD background), and for
%consistency with the other detectors discussed here, we assume the
%ability to subtract the \backgrounds\ in the no-oscillation case.  As
%with the other detectors, neutrino oscillations complicate things
%and the \backgrounds\ cannot be so easily subtracted in the IH and NH
%cases.
IceCube also lacks the ability to discriminate between neutrino species.  The 
    breakout burst light curve is dominated by \nue's, but the higher-energy 
    \anue's and \nux's are favored by the energy-dependent effective detection 
    volume and may swamp the \nue\ signal.  Additionally, although the detector 
    background rate is stable, random fluctuations around the average rate (540 Hz 
    per PMT with no dead time and 286 Hz per PMT 
    with a 250 $\mu$s dead time; \citealt{abbasietal2011}) can also swamp the 
    \nue\ signal.  We relegate a quantitative discussion of the detectability of 
    the \nue\ signal against the backgrounds to Section~\ref{sec:discussion}.

\subsection{Scintillation Detectors}
Scintillation detectors are tanks of hydrocarbon 
scintillators.  They are very similar to water-\cer\
detectors in that 
they employ a proton-rich medium for neutrino detection 
and, as such, are most sensitive  to
\anue's. The final-state electrons and positrons that result from
electron scattering and IBD are detected via their scintillation light
using PMT's. 
 Scintillation detectors have a much lower energy detection threshold  
(${\sim}0.2$~MeV, \citealt{lahaetal2014}) than
 water-\cer\ detectors.  In our analysis, 
for the detector efficiency we assume a Heaviside step function 
with step at 0.2 MeV. In addition to detecting
neutrinos through IBD and electron-scattering reactions, \nue\
and \anue\ absorption on the carbon nuclei produce detectable
products, and the scattering
of all neutrino types on the carbon nuclei can 
in principle  be detected via photon
emission from de-excitation, much as for oxygen in water-\cer\
detectors.  Scintillation detectors can also make a measurement of the
neutrino spectrum by measuring the energies of the final-state
products of the neutrino interactions.

Scintillation detectors have the advantage that electron-scattering 
and IBD events are distinguishable: 99\% of the neutrons formed in IBD
events will quickly (${\sim}0.2$ ms) 
combine with a proton, producing a
2.2 MeV gamma ray (\citealt{abeetal2008}), which can be detected.  
A coincidence in time and space of an electron/positron signal with a
neutron capture gamma ray allows the identification of that signal as
a positron.
 In addition, experiments have shown that
scintillation detectors are able to differentiate electrons
and positrons through pulse shape discrimination 
(\citealp{kinoetal2000,francoetal2011}), which
allows for further differentiation between electron-scattering
and \anue\ absorption events.  Pulse shape discrimination 
has been demonstrated in active scintillation detectors
(\citealp{abeetal2014,bellinietal2014}), and we anticipate its
continued use in future scintillation detectors.  Because of these
things, in our analysis we assume the ability to tag all IBDs.  For
the no-oscillation case, this corresponds to the \anue\ flux.
Scintillation detectors can detect
\nux\ through $\nu_x + p \rightarrow \nu_x + p$
(\citealp{oberaueretal2005,lahabeacom2014}), and  \nux\ can also, in
principle, be measured via $\nu_x + {^{12}}\textrm{C} \rightarrow  
\nu_x + {^{12}}\textrm{C*}$ (\citealt{ryazhskaya1992}), so we assume in
our analysis that \nux\ can be differentiated from other types.  Thus,
we only care about the \nue\ flux in our analysis for the
no-oscillation case.  Again, complications due to neutrino
oscillations do not permit so straightforward a subtraction of the
\backgrounds\ in the case of oscillations.

Although the exact ratio of carbon to hydrogen varies in 
the scintillators employed in detectors, 
it does not depart too much from a
C$_{n}$H$_{2n}$ stochiometry, which is the chemical form 
assumed in our analysis.

There are currently two scintillation detectors with detection mass
$\sim$1 ktonne: the Kamioka Liquid Scintillator Antineutrino Detector
(KamLAND) in the Kamioka Mine in Japan
(\citealt{eguchietal2003}) and the Large Volume Detector (LVD) in 
the Gran Sasso underground laboratory in 
Italy (\citealt{agliettaetal1992}). There are also several smaller detectors.
We calculate 
that a scintillation detector with fiducial mass 1 ktonne will detect 
$\sim$0-1 \nue's in a 10 ms period over the
breakout burst at a distance of 10 kpc in the case of no neutrino
oscillations (even less in the case of neutrino oscillations). 
Because of this small signal,
we do not consider scintillation detectors of this size (and smaller) further.
There is a 14 ktonne
scintillation detector in operation, the NO$\nu$A far
detector in Ash River, Minnesota
(\citealt{pattersonetal2012}), which is located at the surface.  
Because of the
high backgrounds in this detector, we do not consider it.

The Jiangmen Underground Neutrino Observatory (JUNO), currently under
construction,\footnote{http://english.ihep.cas.cn/rs/fs/juno0815/PPjuno/201501/
t20150112\_135044.html} is a 20-ktonne
scintillation detector located in Jiangmen, China (\citealt{li2014}).  We calculate 
that a scintillation detector of JUNO's size will detect 
$\sim$10 \nue's in a 10 ms period over the
breakout burst at a distance of 10 kpc in the case of no neutrino
oscillations.  For the same situation as just outlined, the number of
detected (original) \nue's in the NH case will be \abt 2 and in the IH
case will be \abt 5.  We take the JUNO mass of 20
ktonne as the representative mass for scintillation detectors in our
analysis. 

In summary, \ar40\ detectors have the highest sensitivity to \nue's. 
Scintillation detectors  have the best intrinsic 
particle identification abilities.
Functional and material
considerations make water-\cer\ detectors (including long-string
detectors) less expensive to build with a large detection volume.
\ar40, scintillation, and water-\cer\ detectors are 
all able to measure the energies expected for the
final-state products in a CCSN, while long-string detectors 
are currently unable to measure the energies in that range.  
Table~\ref{tab:numdetections} summarizes, in the case of no neutrino
oscillations, 
our calculations of how many \nue's each of our representative
detectors would be able to detect in a 10 ms period during the
breakout burst at a selection of
distances. Table~\ref{tab:numdetections_NH} shows the same for the NH
case, and  Table~\ref{tab:numdetections_IH} shows the same for the IH case.
Table~\ref{tab:numdetectionsprebreakout}  shows
the same as Table~\ref{tab:numdetections}, in the case
of no neutrino oscillations, for the
pre-shock neutronization peak.  
Table~\ref{tab:numdetectionsprebreakout_NH} shows the same for the NH
case, and  Table~\ref{tab:numdetectionsprebreakout_IH} shows the 
same for the IH case.
The numbers for the pre-shock
neutronization peak are significantly smaller than those of the
breakout burst, because of the lower \nue\ number flux and lower average
\nue\ energy in the pre-shock neutronization peak relative to the
breakout burst peak (Figure~\ref{fig:avgenergy}).  These tables do not
take any \backgrounds\ into account.  In general, the no-oscillation
case causes the largest number of \nue\ detections, while the NH
case causes the smallest number of \nue\ detections, for all
detectors. For the water-\cer, scintillation, and long-string
detectors, this is owing to the smaller electron-scattering cross
section for \nux's as opposed to \nue's (a factor of \abt 6 smaller
for \nuxpart\ and \abt 7 smaller for \nuxanti).
For \ar40\ detectors, this is due to the dominance of the \nue\ CC
absorption channel in the measured signal.
\begin{deluxetable}{cccccc}
\tablewidth{0pc}
\tablecolumns{6}
\tablecaption{Approximate Number of $\nu_e$'s Detected in 10 ms
Interval during Breakout in the No-oscillation Case\label{tab:numdetections}}
\tablehead{
\colhead{Distance} & \colhead{DUNE} & \colhead{Super-K}
& \colhead{Hyper-K} & \colhead{IceCube}
& \colhead{JUNO} \\ \colhead{(kpc)} & \colhead{} & 
\colhead{} & \colhead{} & \colhead{} & \colhead{} }
\startdata
%1 & 12000 & 1000 & 25000 & 160000 & 1100\\
%4 & 740 & 62 & 1600 & 9800 & 68\\
%7 & 240 & 20 & 510 & 3200 & 22\\
%10 & 120 & 10 & 250 & 1600 & 11\\
%20 & 30 & 3 & 62 & 390 & 3
%1 & 12000 & 660 & 16000 & 160000 & 1100\\
%4 & 740 & 41 & 1000 & 10000 & 68\\
%7 & 240 & 13 & 330 & 3300 & 22\\
%10 & 120 & 7 & 160 & 1600 & 11\\
%20 & 30 & 2 & 41 & 400 & 3
1 & 12000 & 660 & 16000 & 44000 & 1100\\
4 & 740 & 41 & 1000 & 2700 & 68\\
7 & 240 & 13 & 330 & 890 & 22\\
10 & 120 & 7 & 160 & 440 & 11\\
20 & 30 & 2 & 41 & 110 & 3
\enddata
\end{deluxetable}

\begin{deluxetable}{cccccc}
\tablewidth{0pc}
\tablecolumns{6}
\tablecaption{Approximate Number of $\nu_e$'s Detected in 10 ms Interval during Breakout in the NH Case\label{tab:numdetections_NH}}
\tablehead{
\colhead{Distance} & \colhead{DUNE} & \colhead{Super-K}
& \colhead{Hyper-K} & \colhead{IceCube}
& \colhead{JUNO} \\ \colhead{(kpc)} & \colhead{} & 
\colhead{} & \colhead{} & \colhead{} & \colhead{} }
\startdata
%1 & 230 & 180 & 4500 & 28000 & 220\\
%4 & 14 & 11 & 280 & 1700 & 14\\
%7 & 5 & 4 & 91 & 570 & 5\\
%10 & 2 & 2 & 45 & 280 & 2\\
%20 & 1 & 0 & 11 & 70 & 1
%1 & 230 & 110 & 2800 & 28000 & 220\\
%4 & 14 & 7 & 170 & 1700 & 14\\
%7 & 5 & 2 & 56 & 570 & 5\\
%10 & 2 & 1 & 28 & 280 & 2\\
%20 & 1 & 0 & 7 & 70 & 1
1 & 230 & 110 & 2800 & 5800 & 220\\
4 & 14 & 7 & 170 & 360 & 14\\
7 & 5 & 2 & 56 & 120 & 5\\
10 & 2 & 1 & 28 & 58 & 2\\
20 & 1 & 0 & 7 & 14 & 1
\enddata
\end{deluxetable}

\begin{deluxetable}{cccccc}
\tablewidth{0pc}
\tablecolumns{6}
\tablecaption{Approximate Number of $\nu_e$'s Detected in 10 ms
Interval during Breakout in the IH Case\label{tab:numdetections_IH}}
\tablehead{
\colhead{Distance} & \colhead{DUNE} & \colhead{Super-K}
& \colhead{Hyper-K} & \colhead{IceCube}
& \colhead{JUNO} \\ \colhead{(kpc)} & \colhead{} & 
\colhead{} & \colhead{} & \colhead{} & \colhead{} }
\startdata
%1 & 3800 & 440 & 11000 & 69000 & 500\\
%4 & 240 & 27 & 680 & 4300 & 31\\
%7 & 78 & 9 & 220 & 1400 & 10\\
%10 & 38 & 4 & 110 & 690 & 5\\
%20 & 10 & 1 & 27 & 170 & 1
%1 & 3800 & 280 & 6900 & 69000 & 490\\
%4 & 240 & 17 & 430 & 4300 & 31\\
%7 & 78 & 6 & 140 & 1400 & 10\\
%10 & 38 & 3 & 69 & 690 & 5\\
%20 & 10 & 1 & 17 & 170 & 1
1 & 3800 & 280 & 6900 & 18000 & 490\\
4 & 240 & 17 & 430 & 1100 & 31\\
7 & 78 & 6 & 140 & 360 & 10\\
10 & 38 & 3 & 69 & 180 & 5\\
20 & 10 & 1 & 17 & 44 & 1
\enddata
\end{deluxetable}

\begin{deluxetable}{cccccc}
\tablewidth{0pc}
\tablecolumns{6}
\tablecaption{Approximate Number of $\nu_e$'s Detected in 10 ms Interval during Preshock Neutronization Peak in the No-oscillation Case\label{tab:numdetectionsprebreakout}}
\tablehead{
\colhead{Distance} & \colhead{DUNE} & \colhead{Super-K}
& \colhead{Hyper-K} & \colhead{IceCube}
& \colhead{JUNO} \\ \colhead{(kpc)} & \colhead{} & 
\colhead{} & \colhead{} & \colhead{} & \colhead{} }
\startdata
%1  & 880 & 180 & 4500 & 29000 & 170\\
%4  & 55 & 11 & 280 & 1800 & 11\\
%7  & 18 & 4 & 91 & 590 & 4\\
%10  & 9 & 2 & 45 & 290 & 2\\
%20  & 2 & 0 & 11 & 72 & 0
%1 & 890 & 80 & 2000 & 30000 & 170\\
%4 & 56 & 5 & 120 & 1900 & 11\\
%7 & 18 & 2 & 41 & 600 & 3\\
%10 & 9 & 1 & 20 & 300 & 2\\
%20 & 2 & 0 & 5 & 74 & 0
1 & 890 & 80 & 2000 & 4600 & 170\\
4 & 56 & 5 & 120 & 290 & 11\\
7 & 18 & 2 & 41 & 93 & 3\\
10 & 9 & 1 & 20 & 46 & 2\\
20 & 2 & 0 & 5 & 11 & 0
\enddata
\end{deluxetable}

\begin{deluxetable}{cccccc}
\tablewidth{0pc}
\tablecolumns{6}
\tablecaption{Approximate Number of $\nu_e$'s Detected in 10 ms
Interval during Pre-breakout Neutronization Peak in the NH
Case\label{tab:numdetectionsprebreakout_NH}}
\tablehead{
\colhead{Distance} & \colhead{DUNE} & \colhead{Super-K}
& \colhead{Hyper-K} & \colhead{IceCube}
& \colhead{JUNO} \\ \colhead{(kpc)} & \colhead{} & 
\colhead{} & \colhead{} & \colhead{} & \colhead{} }
\startdata
%1 & 45 & 30 & 750 & 4800 & 29\\
%4 & 3 & 2 & 47 & 300 & 2\\
%7 & 1 & 1 & 15 & 99 & 1\\
%10 & 0 & 0 & 8 & 48 & 0\\
%20 & 0 & 0 & 2 & 12 & 0
%1 & 45 & 12 & 290 & 4800 & 28\\
%4 & 3 & 1 & 18 & 300 & 2\\
%7 & 1 & 0 & 6 & 99 & 1\\
%10 & 0 & 0 & 3 & 48 & 0\\
%20 & 0 & 0 & 1 & 12 & 0
1 & 45 & 12 & 290 & 700 & 28\\
4 & 3 & 1 & 18 & 44 & 2\\
7 & 1 & 0 & 6 & 14 & 1\\
10 & 0 & 0 & 3 & 7 & 0\\
20 & 0 & 0 & 1 & 2 & 0
\enddata
\end{deluxetable}

\begin{deluxetable}{cccccc}
\tablewidth{0pc}
\tablecolumns{6}
\tablecaption{Approximate Number of $\nu_e$'s Detected in 10 ms
Interval during Pre-breakout Neutronization Peak in the IH
Case\label{tab:numdetectionsprebreakout_IH}}
\tablehead{
\colhead{Distance} & \colhead{DUNE} & \colhead{Super-K}
& \colhead{Hyper-K} & \colhead{IceCube}
& \colhead{JUNO} \\ \colhead{(kpc)} & \colhead{} & 
\colhead{} & \colhead{} & \colhead{} & \colhead{} }
\startdata
%1 & 310 & 78 & 1900 & 12000 & 74\\
%4 & 19 & 5 & 120 & 780 & 5\\
%7 & 6 & 2 & 40 & 250 & 2\\
%10 & 3 & 1 & 19 & 120 & 1\\
%20 & 1 & 0 & 5 & 31 & 0
%1 & 310 & 33 & 820 & 12000 & 71\\
%4 & 19 & 2 & 51 & 780 & 4\\
%7 & 6 & 1 & 17 & 250 & 2\\
%10 & 3 & 0 & 8 & 120 & 1\\
%20 & 1 & 0 & 2 & 31 & 0
1 & 310 & 33 & 820 & 1900 & 71\\
4 & 19 & 2 & 51 & 120 & 4\\
7 & 6 & 1 & 17 & 39 & 2\\
10 & 3 & 0 & 8 & 19 & 1\\
20 & 1 & 0 & 2 & 5 & 0
\enddata
\end{deluxetable}

% LocalWords:  lccc 25ex TPCs lahabeacom2014 Antineutrino Jiangmen t20150112
% LocalWords:  cccccc 0pc

\section{Method}
\label{sec:method}
Our goal in this paper is to determine, for highlighted detectors, 
for each SN
model, and over a range of distances, how well one can expect to measure 
the shape and features of the \nue\ breakout light curve, as well as
examine how well different CCSN models can be discriminated using this
light curve. To do this, we use
Equation~(\ref{eq:Nint}) to calculate the total expected number of
neutrino interactions in a given detector for a given distance over a
time range that includes 
the breakout burst peak.  A cumulative distribution function of
the arrival times of the detected neutrinos is also calculated.  We
then perform a Monte Carlo sampling of observations in the given
detector, using the cumulative distribution function of detection
times and the calculated total expected number of neutrino
interactions to create simulated realizations of individual
observations.  
The cross sections used in our analysis are detailed in
the Appendix.
The time range we use in the Monte Carlo sampling is
larger than that used for our subsequent analysis, so that the total
number of neutrinos actually used in the analysis is allowed to
fluctuate randomly, instead of being set by our calculated total expected
number of neutrino interactions.   A collection of $5\times10^4$ 
sample observations are thus assembled for each detector and SN
distance.  The simulated data
are binned in time (in time bins of 1 ms width), 
and standard deviations for each time bin are
calculated using the values across the $5\times10^4$ different
simulated observations.
These standard deviations are then applied to each of the individual simulated
observations for the purposes of fitting our analytic equations to the
simulated data.  For simplicity, we assume symmetric errors in each
time bin of the light curve.  Since the distribution of the number 
of detections in
each time bin is expected to be Poissonian (an expectation that we
verified in our simulated data), the distribution is asymmetric.  
However, in the larger detectors
and at smaller distances a sufficient number of neutrinos will be
detected in each time bin for the Poissonian distribution to approach a
symmetric Gaussian, so our assumption holds in these cases.  For
smaller detectors and larger distances, with fewer detections expected
in each time bin, our assumption is not valid, but we hold to it both
for simplicity and also because the sparse data expected with smaller
detectors and at larger distances will themselves create large errors in the
distribution of fits to our simulated observations, larger than those
that could be corrected by providing asymmetric errors in each time bin.

As a specific example of a calculation, 
Figure~\ref{fig:hyperkhistogram} shows the results of one of the $5\times10^4$
realizations of a CCSN detection in Hyper-K for each of the distances 4, 7, and
10 kpc for the 15 \Msol\ \ls\ progenitor model.

After the simulated observations are assembled, 
Equation~(\ref{eq:analytic}) is
then fit to the resultant number luminosity histograms (with error bars)
using the \texttt{curve\_fit()} function
of SciPy.  
\texttt{curve\_fit()} implements
the Levenberg--Marquardt algorithm to fit data to a function with arbitrary
parameters.  The function we fit is derived from 
Equation~(\ref{eq:Nint}), as follows.  First, the energy luminosity spectrum is
converted to number luminosity spectrum, $dL^{n}_{\nu}/dE_\nu$,
\beq
\frac{dL^{n}_{\nu}(E_\nu,t)}{dE_\nu} = \frac{dL_\nu(E_\nu,t)}{dE_\nu}\frac{1}{E_\nu}.
\eeq
The number luminosity spectrum is then normalized,
\beq
\label{eq:numlumnorm}
\frac{dL^{n \prime}_{\nu}}{dE_{\nu}} = \frac{dL^{n}_{\nu}/dE_{\nu}}{\int dL^{n}_{\nu}/dE_{\nu}~~dE_{\nu}}.
\eeq
Since
\beq
L^n_\nu(t) = \int \frac{dL^n_\nu(E_\nu,t)}{dE_\nu} dE_\nu,
\eeq
we can rearrange Equation~(\ref{eq:numlumnorm}) to get
\beq
\label{eq:substitution}
\frac{dL^{n}_{\nu}(E_\nu,t)}{dE_\nu} = L^n_\nu(E_\nu,t)\frac{dL^{n \prime}_{\nu}}{dE_{\nu}}.
\eeq
 Equation~(\ref{eq:substitution}) can be substituted into 
Equation~(\ref{eq:Nint}) to obtain
\begin{equation}
\label{eq:blackbox}
N_{\textrm{det}}(t) =  \frac{N_{t}}{4\pi D^2}
L^n_\nu(t,\boldsymbol{p})
\int 
\frac{dL^{n \prime}_{\nu}}{dE_{\nu}}\sigma(E_\nu)\epsilon(E_e)dE_\nu,
\end{equation}
where $L^n_\nu$ is given by
Equation~(\ref{eq:analytic}), the superscript $n$
specifies the use of number luminosity instead of energy luminosity, 
$\boldsymbol{p}$ represents the parameter values being used in 
Equation~(\ref{eq:analytic}) or Equation~(\ref{eq:smallpeak}), $dN'/dE_\nu$ is the
normalized number spectrum, and the other symbols have the same
meanings as in Equation~(\ref{eq:Nint}). To use
Equation~\ref{eq:blackbox}, the distance to the SN must be
known.  If the SN is visible, an independent measurement of $D$
can be made.  If the SN is obscured, the distance will likely
have to be estimated from the neutrino signal.  In our analysis, we
assume knowledge of the distance. In this analysis, we have the
advantage of knowing the energy spectrum from our models, and that is
the spectrum that is used in Equation~\ref{eq:blackbox} for our
analysis.  An actual detection might entail the measurement of the
energy spectrum of the neutrinos and will likely have an additional
function and parameterization that will be used to fit the spectrum.
We do not perform such a full analysis in this work, but in principle
it would be straightforward. Figures~\ref{fig:energyspectrum}
and \ref{fig:avgenergy} show that the energy distribution and average
energy do not vary appreciably over the duration of the breakout
burst, so even something as simple as assuming a constant spectrum
through the breakout burst would be reasonable.  
Therefore, measurements of the spectrum could be integrated over
the time of the breakout burst to provide higher statistics in
measuring the energy spectrum than measuring a time-dependent
energy spectrum.  
For a given detector (which has
multiple detection channels), Equation~(\ref{eq:blackbox}) can be applied
to all the interaction channels and the results summed together.
The equation we give to
the \texttt{curve\_fit()} function to fit the simulated data is
Equation~(\ref{eq:blackbox}), while the parameters that are being used
in the fitting algorithm are those of the intrinsic number
luminosity. 
The Levenberg--Marquardt algorithm requires an initial guess for the
parameters, for which we provide the values from
Table~\ref{tab:numpeakfit}.

After the best-fit fitting parameters are calculated, the physical
parameters are derived from the fit.  
We emphasize again that it is not the fitting parameters but rather the
physical parameters that are important to
our analysis.  For the main breakout burst peak, the physical parameters
calculated are the
 maximum number luminosity of breakout burst
($L^n_{\nu_e,\mathrm{max}}$),
the time of maximum luminosity ($t_{\mathrm{max}}$), the width of the
peak ($w$),
the rise time ($t_{\mathrm{rise},1/2}$), and the fall time
($t_{\mathrm{fall},1/2}$).  
%For the pre-shock
%neutronization peak, the physical parameters calculated are the
%maximum number luminosity of the pre-shock peak
%($L^n_{\nu_e,\mathrm{max,pre}}$) and the time of this maximum luminosity 
%($t_{\mathrm{max,pre}}$).

\section{Results}
\label{sec:discussion}

We first discuss the results obtained without neutrino oscillations
taken into account, followed by the results expected based on the
neutrino oscillation scenarios due to the NH and IH. For
the purpose of this analysis, we take one model (15 \Msol, \ls) as an
example. 
Throughout this section (and this work), we use the 95\% uncertainties
as the basis for our discussion.

\subsection{Results without Neutrino Oscillations}

We first consider the case of no neutrino oscillations.  While this is
not likely to be the case, it provides a good baseline for
quantifying the capabilities of neutrino detectors in measuring the
properties of the \nue\ breakout burst.  This
is for two reasons. The first is that the no-oscillation case 
represents the case with
the largest detectable \nue\ flux, since the \nuxpart's to which the \nue's
oscillate, either partially (in the IH) or entirely (in the NH), 
have systematically smaller interaction cross sections
than do \nue's in the detectors of our analysis. The 
second reason is that 
any oscillations of \nuxpart's to \nue's or \nuxanti's to \anue's open 
interaction cross sections to these species that are larger than those 
they otherwise could access, and so the \background\ levels increase.
 Thus, the no-oscillation case represents the
maximum performance level of the detectors in our analysis in terms of
maximizing the \nue\ signal and minimizing \backgrounds.

IceCube's 540 Hz average background rate per PMT, when 
    multiplied by the total number of PMTs (5160), gives a total 
    background rate of 2786 ms$^{-1}$.  Assuming Poissonian noise, the fluctuations
    on this rate are $\sqrt{2786} = 53$ ms$^{-1}$.  As can be seen from Table~\ref{tab:numdetections}, the 
    expected $\nu_e$ detection rate through the breakout burst peak for a CCSN at 
    10 kpc in the case of no oscillations is 44 ms$^{-1}$, smaller than the expected fluctuations in the detector 
    background rate.  The \nue\ count rate is only lower for both
    of the two oscillation scenarios.  Smaller CCSN distances will provide a higher count rate,
    but even a CCSN at 7 kpc will have a count rate of only 89 ms$^{-1}$, somewhat 
    larger but still comparable to the detector background fluctuations.  Even with 
    introducing a 250 $\mu$s dead time to lower the background rate to 
    286 Hz (which leads to a \abt 13\% dead time total; \citealt{abbasietal2011}), the 
    Poissonian fluctuations on the detector background rate are 37 ms$^{-1}$, as 
    compared with the reduced 38 ms$^{-1}$ \nue\ signal rate for CCSNe at 10 
    kpc and 77 ms$^{-1}$ for CCSNe at 7 kpc.  Even if a CCSN was sufficiently 
    close to distinguish a signal against the detector background fluctuations, there 
    is still the issue of extracting the \nue\ signal from the
    \backgrounds, 
which our calculations show begin dominating in the first few milliseconds 
    after the peak \nue\ luminosity.  In light of all this, it is doubtful that IceCube
    will be able to extract a meaningful signal of the breakout burst in a Galactic
    CCSN, and we do not consider IceCube further in our quantitative analysis of the
    performance of our highlighted neutrino detectors in measuring the properties of the
    breakout burst.

For the figures and tables in this section, 
we take one model as an example model (15 \Msol, \ls).

\subsubsection{Physical Parameter Probability Distribution Functions}

We show here the probability distribution functions (PDFs) of the
physical parameters derived from our simulated observations for each
detector we consider in our analysis without neutrino oscillations
taken into account.  Each detector is representative
of a given detector type.  In order of presentation in this section,
they are Super-K and
Hyper-K, representing water-\cer\ detectors; the 40 ktonne DUNE far
detector (hereafter referred to simply as ``DUNE''), 
for \ar40\ detectors; and JUNO, for scintillation detectors.

%Figure~\ref{fig:icecubephysicalparms_hist} shows the PDFs of the
%physical parameters derived from the fits to the simulated
%observations for IceCube in the no-oscillation case. 
%For comparison purposes, the horizontal
%scales of Figure~\ref{fig:icecubephysicalparms_hist} 
%are the same as for the PDF's of the
%corresponding quantities in the next several figures.  In the
%no-oscillation case, IceCube would be able
%to constrain these parameters quite well at both 4 and 10 kpc relative
%to the other detectors examined in our study, as will be shown. 

Figure~\ref{fig:superkphysicalparms_hist} shows the PDFs of the
physical parameters derived from the fits to the simulated
observations for Super-K in the no-oscillation case.  
It is important to
note is that, in the no-oscillation case, 
the distributions of \lmax\ and \tmax\ are relatively
symmetric about the model value (vertical green line), 
while the distributions of $w$, \trise, and \tfall\ are 
asymmetric, being skewed to higher values.  This causes the mode of the
distributions for these parameters to occur at a value
smaller than the model value.

Figure~\ref{fig:hyperkphysicalparms_hist} shows the PDFs of the
physical parameters derived from the fits to the simulated
observations for Hyper-K in the no-oscillation case.  
The widths of the distributions are less than 
those of Super-K, consistent with Hyper-K's mass being
greater than Super-K's.  The asymmetries in the
distributions of 
$w$, \trise, and \tfall\ that are seen in
Super-K are also exhibited in Hyper-K, though to a lesser degree.

Figure~\ref{fig:dunephysicalparms_hist} shows the PDF of the
physical parameters derived from the fits to the simulated
observations for DUNE in the no-oscillation case.  The widths of the
distributions are less than 
those of Super-K but comparable to those of Hyper-K, again consistent with the number of
detected neutrinos expected in DUNE relative to both of these
detectors.  The asymmetries of $w$, \trise, and \tfall\ seen in
Super-K are present in DUNE.  

Figure~\ref{fig:junophysicalparms_hist} shows the PDF of the
physical parameters derived from the fits to the simulated
observations for JUNO in the no-oscillation case.

For each parameter, a PDF is calculated and the 95\%
confidence values are calculated.  By repeating this process over a
set of distances for the detectors highlighted in this study, we can
determine, for a given detector and SN model, how well the various
features of the breakout burst light curve can be determined as a
function of distance.

\subsubsection{Detector Performance for Measuring \lmax}
Figure~\ref{fig:15maxvalfuncD} shows the 95\% uncertainty in
measuring \lmax\ (the maximum value of the breakout burst luminosity)
in the no-oscillation case as a function of distance for the detectors in our
analysis for the 15 \Msol\ model employing the \ls. 
Table~\ref{tab:maxvalpercenterrortable} shows, for each representative
detector and as a function of distance, in the no-oscillation case, 
the mode of the PDF obtained
in each case for \lmax, as well as the percent
errors associated with 95\% uncertainties.  For both
Figure~\ref{fig:15maxvalfuncD} and
Table~\ref{tab:maxvalpercenterrortable},  when the uncertainty values for a
  specific detector get either
  too large or too small relative to the model value, we no longer
  represent 
  the uncertainty at that distance and greater distances.  For
  Figure~\ref{fig:15maxvalfuncD} (and subsequent figures in this
  section), 
this simply means that the data are no
  longer plotted for these distances.  For
  Table~\ref{tab:maxvalpercenterrortable} (and subsequent tables in
  this section), the missing data are
  represented with ellipses.
The uncertainty values for 
Super-K and JUNO were cut off
  at 7 kpc if the previous criteria were not met at 7 kpc because of
  the small number of events for an SN beyond that
  distance.
The mode is obtained
by fitting a Gaussian curve to the peak of the PDF.
 For this and all the other parameters, there is a clear
hierarchy in the detectors' abilities to precisely measure the
parameters.  Detectors expected to detect a larger number of \nue's in
a breakout burst in the no-oscillation case 
(owing to their larger mass and/or use of a more
\nue-sensitive detection medium) can more accurately measure
the physical parameters for a given SN distance.  
Specifically, Super-K's
and JUNO's smaller detection volumes and cross sections make them least
likely to make an accurate measurement in the no-oscillation case, with the accuracy of  
Hyper-K and DUNE being greater, owing to their larger detection
volumes and (for DUNE) larger detection cross sections. 

%\begin{deluxetable*}{ccccc}
\begin{deluxetable}{cccccc}
\tablewidth{0pc}
\tablecolumns{6}
\tablecaption{Most Likely Value and Percent Error for Measuring
$L^n_{\nu_e,\mathrm{max}}$, 
  for the 15 \Msol\ Model Employing the \ls, Based on
  95\% Error Bounds, in the No-oscillation Case \label{tab:maxvalpercenterrortable}}
\tablehead{
\multicolumn{1}{c}{Distance} & \colhead{Super-K} &
\colhead{Hyper-K} & \colhead{DUNE} & \colhead{JUNO}\\
\colhead{(kpc)} & 
\colhead{($10^{58}$ s$^{-1}$)} & \colhead{($10^{58}$ s$^{-1}$)} 
& \colhead{($10^{58}$ s$^{-1}$)} & \colhead{($10^{58}$ s$^{-1}$)}}
\startdata
%1  & 2.0$^{+0.8\%}_{-0.99\%}$ & 2.0$^{+12\%}_{-9.8\%}$ & 2.0$^{+2.3\%}_{-2.3\%}$ & 2.0$^{+2.6\%}_{-2.6\%}$ & 2.0$^{+9.2\%}_{-7.7\%}$\\
%4  & 2.0$^{+2.9\%}_{-2.9\%}$ & 2.0$^{+57\%}_{-31\%}$ & 2.0$^{+9.6\%}_{-7.7\%}$ & 2.0$^{+11\%}_{-8.9\%}$ & 2.0$^{+40\%}_{-27\%}$\\
%7  & 2.0$^{+5.0\%}_{-4.6\%}$ & ... & 2.0$^{+18\%}_{-13\%}$ & 2.0$^{+20\%}_{-15\%}$ & 2.1$^{+101\%}_{-39\%}$\\
%10  & 2.0$^{+7.4\%}_{-6.4\%}$ & ... & 2.0$^{+25\%}_{-18\%}$ & 2.0$^{+29\%}_{-20\%}$ & ...\\
%13.3  & 2.0$^{+10\%}_{-8.4\%}$ & ... & 2.0$^{+33\%}_{-23\%}$ & 2.0$^{+43\%}_{-25\%}$ & ...\\
%16.7  & 2.0$^{+13\%}_{-10\%}$ & ... & 2.0$^{+45\%}_{-26\%}$ & 2.0$^{+64\%}_{-29\%}$ & ...\\
%20  & 2.0$^{+16\%}_{-12\%}$ & ... & 2.0$^{+57\%}_{-31\%}$ & 2.0$^{+98\%}_{-32\%}$ & ...\\
%23.3  & 2.0$^{+19\%}_{-14\%}$ & ... & 2.0$^{+79\%}_{-32\%}$ & ... & ...\\
%26.7  & 2.0$^{+22\%}_{-16\%}$ & ... & 2.0$^{+109\%}_{-36\%}$ & ... & ...\\
%30  & 2.0$^{+24\%}_{-18\%}$ & ... & ... & ... & ...
1  & 2.0$^{+12\%}_{-9.8\%}$ & 2.0$^{+2.3\%}_{-2.3\%}$ & 2.0$^{+2.6\%}_{-2.6\%}$ & 2.0$^{+9.2\%}_{-7.7\%}$\\
4  & 2.0$^{+57\%}_{-31\%}$ & 2.0$^{+9.6\%}_{-7.7\%}$ & 2.0$^{+11\%}_{-8.9\%}$ & 2.0$^{+40\%}_{-27\%}$\\
7  & ... & 2.0$^{+18\%}_{-13\%}$ & 2.0$^{+20\%}_{-15\%}$ & 2.1$^{+101\%}_{-39\%}$\\
10  & ... & 2.0$^{+25\%}_{-18\%}$ & 2.0$^{+29\%}_{-20\%}$ & ...\\
13.3  & ... & 2.0$^{+33\%}_{-23\%}$ & 2.0$^{+43\%}_{-25\%}$ & ...\\
16.7  & ... & 2.0$^{+45\%}_{-26\%}$ & 2.0$^{+64\%}_{-29\%}$ & ...\\
20  & ... & 2.0$^{+57\%}_{-31\%}$ & 2.0$^{+98\%}_{-32\%}$ & ...\\
23.3  & ... & 2.0$^{+79\%}_{-32\%}$ & ... & ...\\
26.7  & ... & 2.0$^{+109\%}_{-36\%}$ & ... & ...\\
30  & ... & ... & ... & ...
\enddata
%\end{deluxetable*}
\end{deluxetable}

Table~\ref{tab:numpeakfit_realparm} shows that, 
 in order to use a measurement of \lmax to 
differentiate between progenitor masses assuming the \ls, a
measurement accuracy of $\sim$1--2\% is needed.  Specifically, to
differentiate between a 12 \Msol\ and a 15 \Msol\ progenitor, an
accuracy of $\sim$2\% is needed.  For Super-K and JUNO, in the
 no-oscillation case, this level of
accuracy is not obtained for any of the distances examined in our
analysis.  DUNE and Hyper-K, for an SN at 1 kpc in 
the no-oscillation case, 
approach this accuracy but do not quite achieve it. 
More likely is the ability to
differentiate between the \ls\ and the \shen\ by measuring \lmax.
Table~\ref{tab:numpeakfit_realparm} shows that, for the
15 \Msol\ model, an accuracy of \abt 10\% is needed to differentiate
between the two EOSs.  Super-K is close to having this
 accuracy at 1~kpc, JUNO does
obtain this accuracy for SN distances of \abt 1 kpc and smaller
 in the no-oscillation case, and 
DUNE and Hyper-K for distances of \abt 4 kpc and smaller, 
all in the case of no oscillations.

%For IceCube, Hyper-K, and DUNE, the percentage value of the
%lower bound of the 95\% uncertainty in the no-oscillation case 
%does not change much as a function
%of model, but the upper bound increases modestly both with model progenitor
%mass and substituting the \shen\ for the \ls.  

\subsubsection{Detector Performance for Measuring \tmax}
Figure~\ref{fig:15maxlocfwhmfuncD} shows the  95\% uncertainty in
measuring \tmax\ (the time of the maximum luminosity of the breakout
burst) in the no-oscillation case as a function of 
distance for the detectors in our
analysis.  
Table~\ref{tab:maxlocpercenterrortable} shows, for each representative
detector and as a function of distance, the mode of the PDF obtained
in each case for \tmax, as well as the 
errors associated with 95\% uncertainty values, in the no-oscillation case.
Hyper-K, in the no-oscillation case,  can
determine \tmax\ to within \abt 1 ms of the model value out to a
distance of \abt 7 kpc.  
Table~\ref{tab:maxlocpercenterrortable} also shows that the value of
\tmax\ most likely to be measured (the mode of the PDF of \tmax\ in our
analysis) is displaced from the model \tmax\  through many of the 
SN distances under examination.  
However, this offset of the most likely measured value is
only a fraction of the error expected in a measurement of \tmax\ in
Hyper-K for reasonable SN distances ($\gtrsim$7 kpc) 
and so is less important.  
DUNE can measure \tmax\ to an 
accuracy of \abt 1 ms out to \abt 7 kpc in the no-oscillation case.  
Again, for a
given distance the measurement has a possibility of being 
slightly less accurate with
increasing model progenitor mass and more accurate for the \shen.  
JUNO and Super-K, in the no-oscillation case,  
cannot make a measurement within \abt 1 ms of the
model value for an SN at distances greater 
than \abt 2 kpc.  For all distances and
models, in the no-oscillation case, 
Hyper-K will be the most likely to accurately measure \tmax.

We have defined
\tmax\ in such a way that it is not useful for distinguishing between
progenitor models and EOSs, but an accurate measurement
of \tmax\ in multiple detectors could be useful in
triangulating the position of the SN.  

\begin{deluxetable}{ccccc}
\tablewidth{0pc}
\tablecolumns{6}
\tablecaption{Most Likely Value and Error for Measuring $t_{\mathrm{max}}$ 
 for the 15 \Msol\ Model Employing the \ls, Based on
  the 95\% Error Bounds, in the No-oscillation Case\label{tab:maxlocpercenterrortable}}
\tablehead{
\multicolumn{1}{c}{Distance} & \colhead{Super-K} &
\colhead{Hyper-K} & \colhead{DUNE} & \colhead{JUNO}  \\
\colhead{(kpc)} & \colhead{(ms)} & \colhead{(ms)} &
\colhead{(ms)} & \colhead{(ms)}}
\startdata
%1  & 0.08$^{+0.05}_{-0.06}$ & 0.0$^{+0.75}_{-0.59}$ & 0.08$^{+0.13}_{-0.14}$ & 0.06$^{+0.16}_{-0.17}$ & 0.0$^{+0.6}_{-0.44}$\\
%4  & 0.08$^{+0.17}_{-0.17}$ & 0.04$^{+2.3}_{-2.5}$ & 0.05$^{+0.56}_{-0.51}$ & 0.02$^{+0.7}_{-0.63}$ & 0.03$^{+1.9}_{-1.9}$\\
%7  & 0.08$^{+0.29}_{-0.29}$ & ... & 0.0$^{+1.0}_{-0.85}$ & -0.02$^{+1.2}_{-1.1}$ & 0.05$^{+3.2}_{-3.9}$\\
%10  & 0.08$^{+0.42}_{-0.41}$ & ... & 0.0$^{+1.3}_{-1.2}$ & 0.0$^{+1.5}_{-1.6}$ & ...\\
%13.3  & 0.0$^{+0.65}_{-0.47}$ & ... & -0.06$^{+1.7}_{-1.6}$ & 0.0$^{+1.9}_{-2.2}$ & ...\\
%16.7  & 0.02$^{+0.76}_{-0.62}$ & ... & -0.06$^{+2.0}_{-2.0}$ & 0.0$^{+2.3}_{-2.8}$ & ...\\
%20  & 0.02$^{+0.9}_{-0.76}$ & ... & -0.06$^{+2.4}_{-2.4}$ & ... & ...\\
%23.3  & -0.01$^{+1.0}_{-0.84}$ & ... & 0.02$^{+2.7}_{-2.9}$ & ... & ...\\
%26.7  & 0.0$^{+1.2}_{-0.99}$ & ... & 0.0$^{+3.1}_{-4.0}$ & ... & ...\\
%30  & 0.0$^{+1.3}_{-1.1}$ & ... & ... & ... & ...
1  & 0.0$^{+0.75}_{-0.59}$ & 0.08$^{+0.13}_{-0.14}$ & 0.06$^{+0.16}_{-0.17}$ & 0.0$^{+0.6}_{-0.44}$\\
4  & 0.04$^{+2.3}_{-2.5}$ & 0.05$^{+0.56}_{-0.51}$ & 0.02$^{+0.7}_{-0.63}$ & 0.03$^{+1.9}_{-1.9}$\\
7  & ... & 0.0$^{+1.0}_{-0.85}$ & -0.02$^{+1.2}_{-1.1}$ & 0.05$^{+3.2}_{-3.9}$\\
10  & ... & 0.0$^{+1.3}_{-1.2}$ & 0.0$^{+1.5}_{-1.6}$ & ...\\
13.3  & ... & -0.06$^{+1.7}_{-1.6}$ & 0.0$^{+1.9}_{-2.2}$ & ...\\
16.7  & ... & -0.06$^{+2.0}_{-2.0}$ & 0.0$^{+2.3}_{-2.8}$ & ...\\
20  & ... & -0.06$^{+2.4}_{-2.4}$ & ... & ...\\
23.3  & ... & 0.02$^{+2.7}_{-2.9}$ & ... & ...\\
26.7  & ... & 0.0$^{+3.1}_{-4.0}$ & ... & ...\\
30  & ... & ... & ... & ...
\enddata
\end{deluxetable}

\subsubsection{Detector Performance for Measuring $w$}
Figure~\ref{fig:15maxlocfwhmfuncD} shows the 95\% uncertainty in
measuring $w$ (the width of the breakout burst peak) in the
no-oscillation case as a function of distance for the detectors in our
analysis. 
Table~\ref{tab:fwhmpercenterrortable} shows, for each representative
detector and as a function of distance, the mode of the PDF obtained
in each case for $w$ as well as the 
errors associated with 95\% uncertainty values, in the no-oscillation case.
To use a measurement of $w$ to differentiate between the 12 \Msol\ and
15 \Msol\ models using the \ls,
Table~\ref{tab:numpeakfit_realparm}
 shows that $w$ needs to be measured to an accuracy  of $\sim$0.4 ms.
To differentiate between the \ls\ and the \shen\ for the
15 \Msol\ model an accuracy of $\sim$0.6 ms is needed.  However, for
$w$ there appears to be a degeneracy between progenitor mass and
EOS.  For instance, the 12 \Msol\ model with \ls\ has a
value of $w$ that is close to that of the 15 \Msol\ model with \shen,
much closer than any other two values for the models under
consideration.  Thus, by itself, 
a measurement of $w$ seems unable to
specify a particular progenitor mass and EOS, but rather
possible combinations of these two.

Super-K and JUNO are unable to make a determination of $w$ to an accuracy of
0.4 ms (the difference between the 15 \Msol\ and 12 \Msol\ models with
the \ls) for any distances under consideration here, 
in the no-oscillation case.  DUNE is close to
being able to measure this accuracy at 1 kpc, but it takes a
detector such as Hyper-K before such an accuracy can be achieved for an
SN at \abt 1 kpc in the no-oscillation case.
For differentiating
between the 15 \Msol\ model and the 20 or 25 \Msol\ models employing
the \ls, the accuracy needed is \abt 0.9 ms.  In the no-oscillation
case, Hyper-K and
 DUNE would obtain such an accuracy for SNe out to \abt 2 kpc, but JUNO/Super-K would be
unable to obtain this accuracy at any distance examined in this
work.  We thus conclude that a measurement of
$w$ sufficiently accurate to discriminate between SN progenitor
models is not likely to happen in the event of a galactic SN,
in the no-oscillation case, 
since the distances needed to obtain a sufficiently accurate
measurement only encompass a minority fraction of the Galaxy.

\begin{deluxetable}{ccccc}
\tablewidth{0pc}
\tablecolumns{6}
\tablecaption{Most Likely Value and Error for Measuring $w$ for the 15 \Msol\ Model Employing
  the \ls, Based on the 95\% Error Bounds, in the No-oscillation Case\label{tab:fwhmpercenterrortable}}
\tablehead{
\multicolumn{1}{c}{Distance} & \colhead{Super-K} &
\colhead{Hyper-K} & \colhead{DUNE} & \colhead{JUNO}   \\
\colhead{(kpc)} & \colhead{(ms)} & \colhead{(ms)} &
\colhead{(ms)} & \colhead{(ms)}}
\startdata
%1  & 9.4$^{+0.15}_{-0.13}$ & 9.5$^{+2.0}_{-1.8}$ & 9.5$^{+0.42}_{-0.38}$ & 9.5$^{+0.49}_{-0.46}$ & 9.5$^{+1.5}_{-1.4}$\\
%4  & 9.4$^{+0.53}_{-0.47}$ & 8.0$^{+11}_{-3.6}$ & 9.5$^{+1.6}_{-1.5}$ & 9.5$^{+2.0}_{-1.8}$ & 8.3$^{+8.5}_{-3.0}$\\
%7  & 9.5$^{+0.85}_{-0.79}$ & ... & 9.3$^{+3.1}_{-2.3}$ & 9.2$^{+4.0}_{-2.6}$ & ...\\
%10  & 9.5$^{+1.2}_{-1.1}$ & ... & 9.0$^{+4.9}_{-2.7}$ & 8.8$^{+6.4}_{-2.8}$ & ...\\
%13.3  & 9.4$^{+1.7}_{-1.5}$ & ... & 8.3$^{+7.3}_{-2.6}$ & 8.2$^{+9.0}_{-3.1}$ & ...\\
%16.7  & 9.4$^{+2.1}_{-1.9}$ & ... & 8.0$^{+9.4}_{-2.9}$ & 8.0$^{+11}_{-3.8}$ & ...\\
%20  & 9.4$^{+2.7}_{-2.2}$ & ... & ... & ... & ...\\
%23.3  & 9.3$^{+3.2}_{-2.5}$ & ... & ... & ... & ...\\
%26.7  & 9.2$^{+3.8}_{-2.6}$ & ... & ... & ... & ...\\
%30  & 9.1$^{+4.5}_{-2.7}$ & ... & ... & ... & ...
1  & 9.5$^{+2.0}_{-1.8}$ & 9.5$^{+0.42}_{-0.38}$ & 9.5$^{+0.49}_{-0.46}$ & 9.5$^{+1.5}_{-1.4}$\\
4  & 8.0$^{+11}_{-3.6}$ & 9.5$^{+1.6}_{-1.5}$ & 9.5$^{+2.0}_{-1.8}$ & 8.3$^{+8.5}_{-3.0}$\\
7  & ... & 9.3$^{+3.1}_{-2.3}$ & 9.2$^{+4.0}_{-2.6}$ & ...\\
10  & ... & 9.0$^{+4.9}_{-2.7}$ & 8.8$^{+6.4}_{-2.8}$ & ...\\
13.3  & ... & 8.3$^{+7.3}_{-2.6}$ & 8.2$^{+9.0}_{-3.1}$ & ...\\
16.7  & ... & 8.0$^{+9.4}_{-2.9}$ & 8.0$^{+11}_{-3.8}$ & ...\\
20  & ... & ... & ... & ...\\
23.3  & ... & ... & ... & ...\\
26.7  & ... & ... & ... & ...\\
30  & ... & ... & ... & ...
\enddata
\end{deluxetable}

\subsubsection{Detector Performance for Measuring \trise\ and \tfall}
The left panel of Figure~\ref{fig:15lwhmrwhmfuncD} shows, in the no-oscillation case,
the  95\% uncertainty in
measuring \trise\ (the rise time of the breakout burst luminosity) as
a function of distance for the detectors in our analysis. 
Table~\ref{tab:lwhmpercenterrortable} shows, for each representative
detector and as a function of distance, the mode of the PDF obtained
in each case for \trise, as well as the 
errors associated with 95\% uncertainty values, in the no-oscillation case.
The values for \trise, for the different progenitor masses employing the
\ls, are all too close to allow a measurement of \trise\ from
any of the detectors under consideration, for any of the SN
distances under consideration, to differentiate between progenitor
masses, in the no-oscillation case.  
However, Table~\ref{tab:numpeakfit_realparm} shows that there
is a larger difference (\abt 0.5 ms) in \trise\ between the \ls\ and the \shen\ for
the 15 \Msol\ model.  In the no-oscillation case, 
Super-K and JUNO would not make a measurement with this
accuracy for an SN at any distances considered here (but 
almost could at \abt 1 kpc). DUNE and Hyper-K would achieve this
accuracy for an SN at \abt
2--3 kpc.

\begin{deluxetable}{ccccc}
\tablewidth{0pc}
\tablecolumns{6}
\tablecaption{Most Likely Value and Error for Measuring
$t_{\mathrm{rise},1/2}$ for the 15 \Msol\ Model Employing the \ls,
Based on the 95\% Error Bounds, in the No-oscillation Case\label{tab:lwhmpercenterrortable}}
\tablehead{
\multicolumn{1}{c}{Distance} & \colhead{Super-K} &
\colhead{Hyper-K} & \colhead{DUNE} & \colhead{JUNO}  \\
\colhead{(kpc)} & \colhead{(ms)} & \colhead{(ms)} &
\colhead{(ms)} & \colhead{(ms)}}
\startdata
%1  & 2.4$^{+0.08}_{-0.07}$ & 2.4$^{+0.83}_{-0.7}$ & 2.4$^{+0.14}_{-0.17}$ & 2.4$^{+0.19}_{-0.19}$ & 2.4$^{+0.61}_{-0.6}$\\
%4  & 2.4$^{+0.18}_{-0.2}$ & 1.8$^{+3.3}_{-1.0}$ & 2.4$^{+0.67}_{-0.61}$ & 2.4$^{+0.88}_{-0.7}$ & 1.8$^{+2.8}_{-0.81}$\\
%7  & 2.4$^{+0.32}_{-0.3}$ & 1.9$^{+4.1}_{-1.6}$ & 2.3$^{+1.3}_{-0.7}$ & 2.0$^{+1.8}_{-0.55}$ & 1.8$^{+3.8}_{-1.3}$\\
%10  & 2.4$^{+0.46}_{-0.44}$ & ... & 1.9$^{+2.1}_{-0.5}$ & 1.9$^{+2.4}_{-0.65}$ & ...\\
%13.3  & 2.4$^{+0.65}_{-0.63}$ & ... & 1.8$^{+2.6}_{-0.64}$ & 1.8$^{+3.0}_{-0.91}$ & ...\\
%16.7  & 2.4$^{+0.84}_{-0.7}$ & ... & 1.8$^{+3.0}_{-0.83}$ & 1.8$^{+3.3}_{-1.1}$ & ...\\
%20  & 2.4$^{+1.0}_{-0.76}$ & ... & 1.8$^{+3.4}_{-0.99}$ & 1.8$^{+3.5}_{-1.3}$ & ...\\
%23.3  & 2.3$^{+1.2}_{-0.76}$ & ... & 1.7$^{+3.7}_{-1.0}$ & 1.9$^{+3.8}_{-1.4}$ & ...\\
%26.7  & 2.3$^{+1.4}_{-0.78}$ & ... & 1.8$^{+3.7}_{-1.3}$ &
%1.9$^{+3.8}_{-1.6}$ & ...\\
%30  & 2.1$^{+1.7}_{-0.67}$ & ... & 1.8$^{+4.0}_{-1.4}$ &
%1.9$^{+3.9}_{-1.7}$ & ...
1  & 2.4$^{+0.83}_{-0.7}$ & 2.4$^{+0.14}_{-0.17}$ & 2.4$^{+0.19}_{-0.19}$ & 2.4$^{+0.61}_{-0.6}$\\
4  & 1.8$^{+3.3}_{-1.0}$ & 2.4$^{+0.67}_{-0.61}$ & 2.4$^{+0.88}_{-0.7}$ & 1.8$^{+2.8}_{-0.81}$\\
7  & 1.9$^{+4.1}_{-1.6}$ & 2.3$^{+1.3}_{-0.7}$ & 2.0$^{+1.8}_{-0.55}$ & 1.8$^{+3.8}_{-1.3}$\\
10  & ... & 1.9$^{+2.1}_{-0.5}$ & 1.9$^{+2.4}_{-0.65}$ & ...\\
13.3  & ... & 1.8$^{+2.6}_{-0.64}$ & 1.8$^{+3.0}_{-0.91}$ & ...\\
16.7  & ... & 1.8$^{+3.0}_{-0.83}$ & 1.8$^{+3.3}_{-1.1}$ & ...\\
20  & ... & 1.8$^{+3.4}_{-0.99}$ & 1.8$^{+3.5}_{-1.3}$ & ...\\
23.3  & ... & 1.7$^{+3.7}_{-1.0}$ & 1.9$^{+3.8}_{-1.4}$ & ...\\
26.7  & ... & 1.8$^{+3.7}_{-1.3}$ & 1.9$^{+3.8}_{-1.6}$ & ...\\
30  & ... & 1.8$^{+4.0}_{-1.4}$ & 1.9$^{+3.9}_{-1.7}$ & ...
\enddata
\end{deluxetable}

The right panel of Figure~\ref{fig:15lwhmrwhmfuncD} 
shows, in the no-oscillation case, the 95\% uncertainty in
measuring \tfall\ (the decay time of the breakout burst luminosity) 
as a function of distance for the detectors in our analysis. 
Table~\ref{tab:rwhmpercenterrortable} shows, for each representative
detector and as a function of distance, the mode of the PDF obtained
in each case for \tfall, as well as the 
errors associated with 95\% uncertainty values, in the no-oscillation case.
The separation of the values for \tfall\ for the \ls\ and the \shen\
for the 15 \Msol\ 
model is too small for any detector or any distance
considered here to have sufficient discriminating power between these
two models, in the no-oscillation case.  
The difference between (for the \ls) the 12 and
15 \Msol\ models is \abt 0.4 ms, and the difference between (for the
\ls) the 20 and 15 \Msol\ models is \abt 0.9 ms. In the
no-oscillation case, DUNE and Hyper-K will be able to
measure \tfall\ with an accuracy of 0.4 ms for distances up to \abt 1
kpc.  JUNO and
Super-K do not achieve this accuracy for any distances in our study.

Measurements of \trise\ and \tfall\ could be used
to show that \tfall$>$\trise.  Table~\ref{tab:numpeakfit_realparm}
shows that \tfall\ is $\sim$3--4 times larger than \trise\ across all
the models.  A measurement of \tfall$>$\trise\ would be important in
verifying current models of the \nue\ breakout burst.   In the
no-oscillation case, Super-K would be able to confirm this out to \abt
2 kpc, JUNO 
would be able to out to \abt 3 kpc, 
DUNE would be able to out to \abt 10 kpc,  and Hyper-K would be able to
out to \abt 11--12 kpc.

\begin{deluxetable}{ccccc}
\tablewidth{0pc}
\tablecolumns{6}
\tablecaption{Most Likely Value and Error for Measuring $t_{\mathrm{fall},1/2}$ for the 15 \Msol\ Model Employing the \ls, Based on the 95\% Error Bounds, in the No-oscillation Case\label{tab:rwhmpercenterrortable}}
\tablehead{
\multicolumn{1}{c}{Distance} & \colhead{Super-K} &
\colhead{Hyper-K} & \colhead{DUNE} & \colhead{JUNO}  \\
\colhead{(kpc)} & \colhead{(ms)} & \colhead{(ms)} &
\colhead{(ms)} & \colhead{(ms)}}
\startdata
%1  & 7.0$^{+0.11}_{-0.11}$ & 7.0$^{+1.7}_{-1.3}$ & 7.0$^{+0.36}_{-0.29}$ & 7.1$^{+0.41}_{-0.37}$ & 7.0$^{+1.3}_{-1.0}$\\
%4  & 7.0$^{+0.44}_{-0.37}$ & 6.1$^{+8.6}_{-2.7}$ & 7.0$^{+1.4}_{-1.0}$ & 7.0$^{+1.7}_{-1.3}$ & 6.3$^{+6.6}_{-2.3}$\\
%7  & 7.0$^{+0.72}_{-0.63}$ & ... & 6.9$^{+2.6}_{-1.7}$ & 6.8$^{+3.4}_{-1.8}$ & ...\\
%10  & 7.0$^{+1.1}_{-0.86}$ & ... & 6.5$^{+4.3}_{-1.8}$ & 6.4$^{+5.2}_{-2.0}$ & ...\\
%13.3  & 7.0$^{+1.5}_{-1.1}$ & ... & 6.2$^{+5.8}_{-2.0}$ & 6.0$^{+7.1}_{-2.2}$ & ...\\
%16.7  & 6.9$^{+1.9}_{-1.3}$ & ... & 5.9$^{+7.4}_{-2.1}$ & 5.8$^{+8.6}_{-2.6}$ & ...\\
%20  & 6.9$^{+2.3}_{-1.5}$ & ... & 5.6$^{+9.1}_{-2.2}$ & ... & ...\\
%23.3  & 6.8$^{+2.8}_{-1.7}$ & ... & ... & ... & ...\\
%26.7  & 6.7$^{+3.4}_{-1.8}$ & ... & ... & ... & ...\\
%30  & 6.6$^{+3.9}_{-1.9}$ & ... & ... & ... & ...
1  & 7.0$^{+1.7}_{-1.3}$ & 7.0$^{+0.36}_{-0.29}$ & 7.1$^{+0.41}_{-0.37}$ & 7.0$^{+1.3}_{-1.0}$\\
4  & 6.1$^{+8.6}_{-2.7}$ & 7.0$^{+1.4}_{-1.0}$ & 7.0$^{+1.7}_{-1.3}$ & 6.3$^{+6.6}_{-2.3}$\\
7  & ... & 6.9$^{+2.6}_{-1.7}$ & 6.8$^{+3.4}_{-1.8}$ & ...\\
10  & ... & 6.5$^{+4.3}_{-1.8}$ & 6.4$^{+5.2}_{-2.0}$ & ...\\
13.3  & ... & 6.2$^{+5.8}_{-2.0}$ & 6.0$^{+7.1}_{-2.2}$ & ...\\
16.7  & ... & 5.9$^{+7.4}_{-2.1}$ & 5.8$^{+8.6}_{-2.6}$ & ...\\
20  & ... & 5.6$^{+9.1}_{-2.2}$ & ... & ...\\
23.3  & ... & ... & ... & ...\\
26.7  & ... & ... & ... & ...\\
30  & ... & ... & ... & ...
\enddata
\end{deluxetable}

\subsection{Results from Normal Hierarchy Neutrino Oscillations}
In the NH case, the \nue\ flux exchanges with \nuxpart.  This
means that the original \nue's no longer dominate in the
electron-scattering cross section, nor do they undergo CC
interactions with \ar40\ nuclei, which are the dominant detection
channels in the detectors under consideration here.  Because of this,
a clear detection of the \nue\ breakout burst is more difficult
in the NH case than in the no-oscillation case.

%For Gd-doped water-\cer\ and scintillation detectors, IBD interactions
%can be tagged.  Additionally, NC scatterings off of oxygen or
%carbon nuclei can be tagged as well, because of the emission of
%photons from the deexcitation of the nucleus.  
%For these proton-rich detectors, subtracting the 
%IBD interactions in the NH case will subtract much of the \background\ without
%subtracting any 
%\nue\ signal.  Subtracting NC scatterings off of oxygen or carbon
%nuclei will subtract some \nue's from the signal but, since these
%interactions have relatively large thresholds (\abt 15-20 MeV), it is
%the \nux's with higher average energy than \nue's 
%that are primarily subtracted and thus
%such a subtraction improves the detectability of the \nue\ breakout signal
%overall.  This helps beat down the \backgrounds.

For Gd-doped water-\cer\ and scintillation detectors, IBD interactions
    can be tagged with high efficiency.  Scintillation detectors can
    tag IBDs with \abt 99\% 
    efficiency, while water-Cherenkov detectors can tag IBDs with
    \abt 90\% efficiency.  The 
    remaining, untagged IBDs can be statistically subtracted using the measured rate of 
    the tagged IBDs.  Additionally, NC scatterings off of oxygen or carbon nuclei can be 
    tagged as well, because of the emission of photons from the de-excitation of the nucleus.  
    For these proton-rich detectors, subtracting the IBD interactions in the NH case will 
    subtract much of the \background\ without subtracting any \nue\ 
    signal.  Subtracting NC scatterings off of oxygen or carbon nuclei will subtract some 
    \nue's from the signal, but since these interactions have relatively large thresholds  
    (\abt 15--20 MeV), it is the \nux's with higher average energy than \nue's that are 
    primarily subtracted, and thus such a subtraction improves the detectability of the \nue\ 
    breakout signal overall.  This helps beat down the \backgrounds.  In 
    the case of the \abt 90\% IBD tagging efficiency for water-Cherenkov detectors, the 
    statistical subtraction of the \abt 10\% of IBDs that are not tagged would introduce 
    additional statistical errors.  However, these errors are modest relative to the signal 
    extracted, and so we take the simplifying assumption of having the ability to tag all of the IBD events, as well 
    as all of the NC scatterings off of oxygen and carbon.
Figure~\ref{fig:hyperk_superk_nh_backgrounds}
shows, in the NH case, the expected count rate in Hyper-K and Super-K 
for an SN at 4, 7, and
10 kpc and for all neutrinos types, with
neutrinos detected via IBDs and oxygen NC scattering events
subtracted.
Figure~\ref{fig:juno_dune_nh_backgrounds} shows the same for JUNO and
DUNE, except that DUNE has no neutrino signal subtracted and JUNO has
neutrinos detected via IBDs and carbon NC scattering events subtracted.
For Hyper-K in the NH case, the peak of the \nue\ breakout burst 
would not be detectable at 4, 7, or 10 kpc (based on the
size of the error bars relative to the difference in values in each
time bin).  Since 
the fall from the peak is dominated by the 
\backgrounds, fitting the
peak using the procedure in the previous subsection will not
provide an accurate measurement of the properties of the \nue\
breakout burst in the NH case.  The
preshock neutronization peak in the NH case is unlikely to be 
discernible owing to the expected noise.  

Based on Figures~\ref{fig:hyperk_superk_nh_backgrounds} and \ref{fig:juno_dune_nh_backgrounds}, the peak
will not be discernible in the NH case for either Super-K or JUNO at
any of the distances in that figure (4, 7, and 10 kpc).  The
preshock neutronization peak is also indiscernible in the
NH owing to the expected noise.

%Long-string detectors, unlike scintillation or Gd-doped water-\cer\
%detectors, are
%unable to make any discriminations between neutrinos.  Because of
%this, in the NH case, 
%detection of the \nue\ breakout burst peak is currently impossible for such
%detectors, even for nearby SNe.

%Figure~\ref{fig:icecube_dune_nh_backgrounds} shows the expected count rate
%in IceCube for all neutrino types and 
%for SNe at 4, 7, and 10 kpc.  Even for the small noise 
%expected due to the strong
%signal expected from such a large detector and relatively close SNe, 
%the \nue\ breakout burst peak
%cannot be made out against the \backgrounds.  The pre-shock
%neutronization peak is also unlikely to be seen, based on the noise
%levels 
%relative to the difference in the count rates between adjacent bins.

\ar40\ detectors have as their detection channels CC
absorption of \nue's and \anue's on the \ar40\ nuclei and electron
scattering.  Since, for the NH, all the original \nue\ flux becomes
\nuxpart's, the signal in \ar40\ detectors is dominated by the \nuxpart's that
have become \nue's.  The signal of the original \nue\ flux is lost to
this dominating \nuxpart\ background.  This can be seen in
Figure~\ref{fig:juno_dune_nh_backgrounds}, which shows, in the NH
case, the expected count rate
in DUNE for all neutrino types and 
for SNe at distances of 4, 7, and 10 kpc.  
Neither the \nue\ breakout burst peak nor the preshock
neutronization peak can be made out against the
\backgrounds.  

\subsection{Results from Inverse Hierarchy Neutrino Oscillations}
In the IH hierarchy case, \abt 30\% of the original \nue\ flux remains
intact.  This makes it easier to detect the \nue\ breakout burst
against the \backgrounds\ than in the NH case.  For
Gd-doped water-\cer\ and scintillation detectors (in which signals from
IBDs and oxygen/carbon NC scatterings can be subtracted), a clear peak
should be discernible in an appropriately close SN (with
``appropriately close'' depending on the size of the detector).  
Figure~\ref{fig:hyperk_superk_ih_backgrounds}
shows, in the IH case, the expected count rate in Hyper-K and Super-K 
for all neutrinos types, with
backgrounds from IBDs and oxygen NC scattering events subtracted, for
SNe at distances of 4, 7, and 10 kpc.  
Figure~\ref{fig:juno_dune_ih_backgrounds} shows the same for JUNO and
DUNE, except that DUNE has no neutrino signal subtracted and JUNO has
neutrinos detected via IBDs and carbon NC scattering events subtracted.
For all four detectors, a
cleaner peak is seen in the IH case than in the NH case (NH case shown
in Figures~\ref{fig:hyperk_superk_nh_backgrounds} and~\ref{fig:juno_dune_nh_backgrounds}).  For Hyper-K, 
a clear detection of
the \nue\ breakout burst peak in the IH case 
should be possible at 4 kpc, is
marginally possible at 7 kpc, and is unlikely at 10 kpc. The
preshock neutronization peak is not likely to be discernible in
Hyper-K at
any of these distances in the IH case.

For Super-K and JUNO,
Figures~\ref{fig:hyperk_superk_ih_backgrounds} and~\ref{fig:juno_dune_ih_backgrounds} show that
the \nue\ peak may be discernible in the IH case for an SN at 4
kpc but is not likely to be discernible at 7 or 10 kpc.  The preshock
neutronization peak is not discernible at any of these
distances in the IH case.

%Despite the partial maintenance of the \nue\ flux relative to the NH
%case, in the IH case 
%long-string detectors (which cannot subtract any \backgrounds) 
%are still unable to see a clearly defined
%peak for the \nue\ breakout burst.  This is seen in  
%Figure~\ref{fig:icecube_dune_ih_backgrounds}, which
%shows the expected count rate in IceCube for all neutrino types in
%the IH case, for
%SNe at distances of 4, 7, and 10 kpc.
%Even though, in the IH case, \nue's are able to provide a 
%stronger contribution to the
%signal than in the NH case, IBDs due to the \nuxanti's that
%have oscillated to \anue's (Equation~\ref{eq:anue_ih}) provide a
%relatively larger signal that is active through the peak in \nue\
%luminosity, starting \abt 4 ms prior to \tmax.  However, the preshock
%neutronization peak may be discernible at 4 kpc in the IH case.

It is the \ar40\ detectors that show the greatest improvement in
measuring the \nue\ signal in the IH case over the NH case.  Since
the cross section for \nue\ absorption on \ar40\ is so large relative to
the other cross sections considered in this work, the partial
maintenance of the original \nue\ flux makes a big difference in
the detectability of the \nue\ signal in these detectors.  
Figure~\ref{fig:juno_dune_ih_backgrounds}
shows, in the IH case, the expected count rate in DUNE 
for all neutrino types, for
SNe at distances of 4, 7, and 10 kpc.  In the IH case, 
the \nue\ breakout burst
peak should be discernible at 4 kpc, is marginally discernible at 7 kpc,
and is not likely to be discernible at 10 kpc.  The pre-breakout neutronization
peak is not discernible  at any of these distances.

Because the IH case allows for certain detectors to have a 
discernible peak, in principle it is also possible for the properties
of the \nue\ breakout burst peak to be measured in the IH case for
those 
SNe
distances that provide discernible peaks.  We apply the same
analysis outlined in Section~\ref{sec:method} and used in the
no-oscillation case to calculate the accuracy with which the
properties of the breakout burst can be measured by those 
detectors that have the (distance-dependent) ability to measure a
clear peak in luminosity in the IH case.  
These detectors include all the detectors
focused on in this work, minus IceCube.  In doing this, we make no
attempt to correct for the \backgrounds.  We do
take into account the partial oscillation of the \nue\ flux into \nuxpart.
Since the rising \backgrounds\ dominate the tail of the peak, we focus
our fitting routine on the peak itself and do not fit the tail past \abt 5
ms after the peak.  A fitting procedure that employs a model to fit
the \backgrounds\ expected in the IH case should provide better accuracy
in measuring the breakout burst peak than the procedure outlined
here.  The rising \backgrounds\ have a strong influence on the
luminosity decay from \nue\ peak.  Since we are not accounting for the
\backgrounds\ in our fitting, the value of \tfall\ is significantly
modified by the \backgrounds, more so than \lmax, \tmax,
and \trise.  Because of this, we do not focus on \tfall\ (and $w$,
which depends in part on \tfall) in the IH case.

Figure~\ref{fig:hyperkphysicalparms_hist_ih} shows the PDFs of the
physical parameters derived from the fits to the simulated
observations for Hyper-K in the IH
case. Figure~\ref{fig:superkphysicalparms_hist_ih} shows the same for
Super-K, Figure~\ref{fig:junophysicalparms_hist_ih} shows the same
for JUNO, and Figure~\ref{fig:dunephysicalparms_hist_ih} shows the
same for DUNE, all for the IH case.

The left panel of Figure~\ref{fig:15funcD_IH} shows the 95\% uncertainty
in measuring \lmax\ (the maximum value of the breakout burst
luminosity) in the IH case, using the analysis outlined above.
Table~\ref{tab:maxvalerrortable_IH}  shows, in the IH case, for each
representative detector (except IceCube) and as a function of
distance, the mode of the PDF obtained in each case for \lmax, as well
as the percent errors associated with the 95\% uncertainty values.
The uncertainties are larger
at a given distance for a given detector than in the no-oscillation 
case.    This is attributable to the
smaller number of \nue's detected in the IH case, 
relative to the no-oscillation case.  
In general, though, the same hierarchy in the detectors'
ability to measure \lmax\ is seen:  for a given distance, 
Hyper-K (with its larger detection mass) and DUNE (with its larger
detection cross sections)
perform better than the smaller Super-K and
JUNO.
Table~\ref{tab:numpeakfit_realparm} shows that a measurement of \lmax\
needs to have a \abt 1--2\% accuracy to differentiate between the
different models employing the \ls.  This accuracy is not obtained in
the IH case for
any of the detectors in our study for any of the distances we
examine.  However, EOSs may be able to be
differentiated.  Table~\ref{tab:numpeakfit_realparm} shows that, for
the two 15 \Msol\ models, an accuracy of 10\% is needed to
differentiate between the \ls\ and the \shen.  
Super-K and JUNO do not
obtain this accuracy even at 1 kpc for the IH case, 
but DUNE should be able to make
the discrimination at 1 kpc and out to \abt 2 kpc, and
Hyper-K can make this discrimination out to \abt 2--3 kpc.

\begin{deluxetable}{ccccc}
\tablewidth{0pc}
\tablecolumns{5}
\tablecaption{Most Likely Value and Percent Error for Measuring $L^n_{\nu_e,\mathrm{max}}$
  for the 15 \Msol\ Model Employing the \ls, Based on
  95\% Error Bounds, 
for the IH Oscillation case\label{tab:maxvalerrortable_IH}}
\tablehead{
\multicolumn{1}{c}{Distance} &  \colhead{Super-K} &
\colhead{Hyper-K} & \colhead{DUNE} & \colhead{JUNO}\\
\colhead{(kpc)}  & 
\colhead{($10^{58}$ s$^{-1}$)} & \colhead{($10^{58}$ s$^{-1}$)} 
& \colhead{($10^{58}$ s$^{-1}$)} & \colhead{($10^{58}$ s$^{-1}$)}}
\startdata
%1  & 2.1$^{+16\%}_{-11\%}$ & 2.0$^{+3.6\%}_{-2.4\%}$ & 2.1$^{+5.5\%}_{-3.6\%}$ & 1.9$^{+17\%}_{-11\%}$\\
%4  & 2.2$^{+95\%}_{-34\%}$ & 2.1$^{+13\%}_{-8.5\%}$ & 2.2$^{+20\%}_{-13\%}$ & ...\\
%7  & ... & 2.1$^{+22\%}_{-14\%}$ & 2.2$^{+39\%}_{-20\%}$ & ...\\
%10  & ... & 2.1$^{+32\%}_{-18\%}$ & 2.3$^{+81\%}_{-26\%}$ & ...\\
%13.3  & ... & 2.2$^{+41\%}_{-24\%}$ & ... & ...\\
%16.7  & ... & 2.2$^{+57\%}_{-30\%}$ & ... & ...\\
%20  & ... & 2.2$^{+96\%}_{-32\%}$ & ... & ...\\
%23.3  & ... & ... & ... & ...\\
%26.7  & ... & ... & ... & ...\\
%30  & ... & ... & ... & ...
1  & 2.1$^{+19\%}_{-14\%}$ & 2.1$^{+3.8\%}_{-3.4\%}$ & 2.1$^{+4.8\%}_{-4.2\%}$ & 1.9$^{+15\%}_{-12\%}$\\
4  & ... & 2.1$^{+16\%}_{-11\%}$ & 2.2$^{+20\%}_{-13\%}$ & ...\\
7  & ... & 2.1$^{+26\%}_{-18\%}$ & 2.2$^{+37\%}_{-21\%}$ & ...\\
10  & ... & 2.2$^{+41\%}_{-23\%}$ & 2.3$^{+79\%}_{-27\%}$ & ...\\
13.3  & ... & 2.2$^{+73\%}_{-28\%}$ & ... & ...\\
16.7  & ... & ... & ... & ...\\
20  & ... & ... & ... & ...\\
23.3  & ... & ... & ... & ...\\
26.7  & ... & ... & ... & ...\\
30  & ... & ... & ... & ...
\enddata
\end{deluxetable}

The middle panel of Figure~\ref{fig:15funcD_IH} shows, in the IH case,  the 95\% uncertainty
in measuring \tmax\ (the time of the maximum luminosity of the 
breakout burst), using the analysis outlined above.  
Table~\ref{tab:maxlocerrortable_IH}  shows, in the IH case, for each
representative detector (except IceCube) and as a function of
distance, the mode of the PDF obtained in each case for \tmax\ as well
as the errors associated with the 95\% uncertainty values.
Similar to \lmax, 
the uncertainties are larger
at a given distance for a given detector than in the no-oscillation 
case.  In particular, 
 the 95\% uncertainties are larger (by a
factor of \abt 2--3) in the IH-oscillation case than in the no-oscillation case.

\begin{deluxetable}{ccccc}
\tablewidth{0pc}
\tablecolumns{5}
\tablecaption{Most Likely Value and Error for Measuring $t_{\mathrm{max}}$
  for the 15 \Msol\ Model Employing the \ls, Based on
  95\% Error Bounds, for the IH Oscillation case\label{tab:maxlocerrortable_IH}}
\tablehead{
\multicolumn{1}{c}{Distance} & \colhead{Super-K} &
\colhead{Hyper-K} & \colhead{DUNE} & \colhead{JUNO}  \\
\colhead{(kpc)} & \colhead{(ms)} & \colhead{(ms)} &
\colhead{(ms)} & \colhead{(ms)}}
\startdata
%1  & 0.0$^{+1.3}_{-0.73}$ & 0.04$^{+0.35}_{-0.27}$ & 0.18$^{+0.43}_{-0.48}$ & 0.0$^{+1.4}_{-1.0}$\\
%4  & ... & 0.09$^{+0.95}_{-0.7}$ & 0.2$^{+2.1}_{-1.4}$ & ...\\
%7  & ... & 0.0$^{+2.1}_{-1.0}$ & ... & ...\\
%10  & ... & -0.09$^{+3.8}_{-1.5}$ & ... & ...\\
%13.3  & ... & ... & ... & ...\\
%16.7  & ... & ... & ... & ...\\
%20  & ... & ... & ... & ...\\
%23.3  & ... & ... & ... & ...\\
%26.7  & ... & ... & ... & ...\\
%30  & ... & ... & ... & ...
1  & 0.0$^{+1.9}_{-1.0}$ & -0.05$^{+0.49}_{-0.24}$ & 0.13$^{+0.48}_{-0.43}$ & 0.0$^{+1.4}_{-1.0}$\\
4  & ... & 0.03$^{+1.4}_{-0.86}$ & 0.1$^{+2.2}_{-1.3}$ & ...\\
7  & ... & -0.05$^{+3.2}_{-1.7}$ & ... & ...\\
10  & ... & ... & ... & ...\\
13.3  & ... & ... & ... & ...\\
16.7  & ... & ... & ... & ...\\
20  & ... & ... & ... & ...\\
23.3  & ... & ... & ... & ...\\
26.7  & ... & ... & ... & ...\\
30  & ... & ... & ... & ...
\enddata
\end{deluxetable}

The right panel of Figure~\ref{fig:15funcD_IH} shows, in the IH case, the  95\% uncertainty
in measuring \trise\ (the rise time of the breakout burst 
luminosity), using the analysis outlined above. 
Table~\ref{tab:lwhmerrortable_IH}  shows, in the IH case, for each
representative detector (except IceCube) and as a function of
distance, the mode of the PDF obtained in each case for \trise, as well
as the errors associated with the 95\% uncertainty values.
 Similar to \lmax\
and \tmax, \trise\ has larger uncertainties for a given detector at a
given distance in the IH case than in the no-oscillation case.  
Table~\ref{tab:numpeakfit_realparm} shows that using \trise\ to
differentiate between different mass progenitors requires an accuracy
not realized in the IH case by any of the detectors at any of  
the distances examined
here.  For a 15 \Msol\ progenitor, an accuracy of \abt 0.5 ms would be
sufficient to distinguish between the \ls\ and the \shen.  Hyper-K, in
the IH case, can
realize this accuracy at \abt 1 kpc but not beyond, and none of the other
detectors in the IH case can realize this accuracy at any of the 
distances in this study. 

\begin{deluxetable}{ccccc}
\tablewidth{0pc}
\tablecolumns{5}
\tablecaption{Most Likely Value and Error for Measuring $t_{\mathrm{rise},1/2}$
  for the 15 \Msol\ Model Employing the \ls, Based on
  95\% Error Bounds, for the IH Oscillation case\label{tab:lwhmerrortable_IH}}
\tablehead{
\multicolumn{1}{c}{Distance} & \colhead{Super-K} &
\colhead{Hyper-K} & \colhead{DUNE} & \colhead{JUNO}  \\
\colhead{(kpc)} &  \colhead{(ms)} & \colhead{(ms)} &
\colhead{(ms)} & \colhead{(ms)}}
\startdata
%1  & 2.3$^{+1.4}_{-0.77}$ & 2.3$^{+0.51}_{-0.29}$ & 2.3$^{+0.73}_{-0.34}$ & 2.3$^{+1.6}_{-0.84}$\\
%4  & ... & 2.5$^{+0.97}_{-0.82}$ & 2.3$^{+2.8}_{-0.88}$ & ...\\
%7  & ... & 2.2$^{+2.4}_{-0.83}$ & ... & ...\\
%10  & ... & 1.9$^{+4.3}_{-0.81}$ & ... & ...\\
%13.3  & ... & ... & ... & ...\\
%16.7  & ... & ... & ... & ...\\
%20  & ... & ... & ... & ...\\
%23.3  & ... & ... & ... & ...\\
%26.7  & ... & ... & ... & ...\\
%30  & ... & ... & ... & ...
1  & 2.2$^{+2.3}_{-0.81}$ & 2.3$^{+0.65}_{-0.26}$ & 2.7$^{+0.37}_{-0.71}$ & 2.3$^{+1.7}_{-0.81}$\\
4  & ... & 2.4$^{+1.5}_{-0.83}$ & 2.2$^{+2.8}_{-0.84}$ & ...\\
7  & ... & 2.0$^{+3.8}_{-0.93}$ & ... & ...\\
10  & ... & ... & ... & ...\\
13.3  & ... & ... & ... & ...\\
16.7  & ... & ... & ... & ...\\
20  & ... & ... & ... & ...\\
23.3  & ... & ... & ... & ...\\
26.7  & ... & ... & ... & ...\\
30  & ... & ... & ... & ...
\enddata
\end{deluxetable}

% LocalWords:  cccccc 0pc 0pc ccccc 0pc

\section{Conclusion}
\label{sec:conclusion}
We have calculated the expected performance of several representative
terrestrial neutrino detectors in detecting and measuring the
properties of the \nue\ breakout burst light curve in the event of a Galactic
CCSN as a function of supernova distance. We have also examined
whether these measurements of the breakout burst peak would be
sufficiently accurate to allow discrimination between different CCSN
progenitor models and nuclear EOSs.  We have explored the case
of no neutrino oscillations and neutrino oscillations due to
both the normal and inverted neutrino-mass hierarchies.  Assuming
Gd doping in water-\cer\ detectors, in the no-oscillation case
 backgrounds to the \nue\ signal 
due to other neutrino flavors emitted by the CCSN
are sufficiently low as to be negligible, allowing for the best
detection of the \nue\ breakout burst peak and best measurement of its
properties. Neutrino oscillations serve to both reduce the
detectability of the original \nue\ flux and increase the
detection rate of the \background.  We show that, in the NH
case, the \backgrounds\ are too large relative to the detectable original
\nue\ flux to see the \nue\ breakout burst peak for any of the
detectors under consideration in this work.  In the IH case,
three of
the detector types examined (water-\cer, scintillation, and
\ar40) would have a distance-dependent ability to see the
\nue\ breakout burst peak and measure its properties, although to less
accuracy than in the no-oscillation case.  
A long-string detector like IceCube, even in the IH case, would be unable by itself to
    detect the breakout burst peak.  Additionally, the random fluctuations in
    IceCube's background rate would swamp any signal that could be extracted about the 
    \nue\ light curve for SNe at reasonable SN distances even in 
    the no-oscillation case.

The maximum luminosity of the breakout burst, \lmax, can be measured
to the following errors for a CCSN at 10 kpc in the case of no
oscillations:  \abt 25\% for Hyper-K and \abt 30\% for DUNE.  
Super-K and JUNO have very large errors ($>$100\%) at
10 kpc, but JUNO would be able to make a measurement to \abt 40\%
error and Super-K would be able to make a measurement to \abt 60\%
for an SN at 4 kpc.  
In the oscillation case due to the IH, \lmax\ can be measured to the
following errors for a CCSN at 10 kpc: \abt 30\% for Hyper-K and
\abt 60\% for DUNE.  Super-K and JUNO again have very
large errors at 10 kpc. At 1 kpc, JUNO would have an error of \abt 15\% 
and Super-K would have an error of \abt 20\%.
A \abt 2\% accuracy would be needed to
differentiate between the progenitor masses examined in this work (12,
15, 20, and 25 \Msol).  In the no-oscillation case, 
Hyper-K is close to attaining this accuracy 
for SNe out to \abt 1 kpc, but no other
detector could do so for distances ${\geq}1$ kpc.  In the
IH-oscillation case, no detector in this study could make a
sufficiently accurate determination of \lmax\ for any distances examined. 
A 10\% accuracy is
needed in a determination of \lmax\ 
to differentiate between the \ls\ and the \shen\ for the
15 \Msol\ progenitor.  In the no-oscillation case, 
this accuracy is attained by Hyper-K and DUNE out to \abt 4 kpc, 
 and by JUNO out to \abt 1 kpc.  Super-K is close to achieving this
 accuracy at 1 kpc but does not achieve this accuracy for any of the
 distances examined in this work.
In the IH-oscillation case, this accuracy is attained by DUNE out to
\abt 2 kpc and by Hyper-K out to \abt 2--3 kpc. 

The time of the maximum luminosity of the breakout burst, \tmax, can
be measured to the following accuracies for a CCSN at 10 kpc in the
case of no oscillations: \abt 1.3 ms for Hyper-K and \abt 1.5 ms for DUNE.
JUNO has an error of \abt 2 ms at 4 kpc, and Super-K has an error
of \abt 2.5 ms at the same distance.
In the case of IH oscillations, Hyper-K could measure \tmax\ with an
accuracy of \abt 3 ms at 7 kpc, DUNE could measure \tmax\ with an
accuracy of \abt 2 ms at a distance of 4 kpc, JUNO 
could measure \tmax\ to an accuracy of \abt 1.5 ms at 1 kpc, and
Super-K could achieve an accuracy of \abt 2 ms for the same distance.
\tmax\ is
not useful in discriminating between CCSN models and EOSs, 
but may be useful in triangulating the position of an 
SN. Back-of-the-envelope estimates performed by us predict that
a triangulation incorporating \tmax\ information from either Super-K
or Hyper-K would not provide a more accurate determination of the
location of the SN than the individual pointing information
available in these water \cer\ detectors.  However, an optimal
alignment relative to the baselines between detectors may provide a
triangulation measurement of comparable (though lesser) accuracy to an
individual detector pointing, and both location measurements could be
productively combined.  We reserve a more complete examination of the
triangulation abilities of measurements of \tmax\ for a future study.

The width of the breakout burst peak, $w$, can be measured to the
following accuracies for a CCSN at 10 kpc in the case of no
oscillations: \abt 4 ms for Hyper-K and \abt
5 ms for DUNE.  JUNO and Super-K do not observe sufficient numbers of
neutrinos for SNe at 10 kpc to make accurate determinations of
$w$, but at 1 kpc JUNO could make a measurement of $w$ to \abt 1.5 ms
accuracy and Super-K could achieve an accuracy of \abt 2 ms, in the case of no oscillations.  One is unable to measure
$w$ in the IH case because the rising \backgrounds\ make the fall from
peak \nue\ less clear than in the no-oscillation case and measuring
the width is difficult.

Measurements of \trise\ and \tfall\ could be used
to show that \tfall$>$\trise.    In the
no-oscillation case, 
Super-K would be able to confirm  \tfall$>$\trise\ out to \abt 2 kpc, JUNO would be able to out to \abt 3 kpc, 
DUNE would be able to out to \abt 10 kpc, and Hyper-K would be able to
out to \abt 11--12 kpc.  A
determination of \tfall$>$\trise\ is difficult to make in the NH and
IH cases because \tfall\ is difficult to measure owing to  
the increasing dominance of the \backgrounds\ at
the time of fall from the \nue\ peak in these cases.

If the \backgrounds\ could be removed while leaving the \nue\ signal intact, 
the results presented in this work with oscillations due to the NH and
IH would be improved.
 The loss of the detectability of the
original \nue\ flux after it oscillates (fully or partially) to
\nuxpart\ flux cannot be made up this way, but a statistical subtraction
of the backgrounds from other neutrino species could allow for a
measurement of the \nue\ breakout burst peak in the NH case and could improve peak detectability
and measurements in both cases.  In principle, a complete statistical
subtraction of all the \backgrounds\ is possible.
For the IH, the \anue\ flux is completely exchanged with one of 
the \nuxanti's.  For Gd-doped water-\cer\ or scintillation detectors,
which are particularly sensitive to \anue's,
this would allow for a good measurement of the original \nuxanti\ flux, 
which in turn can be translated to the \nuxpart\ flux well enough for
these to all be statistically subtracted.  This would leave only the
original \nue\ and \anue\ flux in the signal.
 However, since in the IH there is still some original \nue\ flux 
that remains intact, the signal in these detectors (especially through
the \nue\ peak, which occurs before the original \anue\ flux begins to
rise) will be dominated by \nue's.

A similar subtraction of the \backgrounds\ in the NH is a little more
complicated, but still possible. However, it would 
require data from multiple
detectors. The original \anue\ flux only partially oscillates to
\nuxanti, making a complete subtraction of the \nuxpart\ and \nuxanti\ flux
using the principle described above less straightforward.  
In this case, an \ar40\ detector will probably be the most useful to
subtract the \backgrounds.  The \nue's it measures are originally
\nuxpart's. Although one might not be able to disentangle the signals
from the various detection channels in the detector,  
the \nue\ signal (from what was originally the \nuxpart\ flux) 
will be the dominant signal.  Thus, an
\ar40\ detector should be able to measure the original \nuxpart\ flux and, 
using this measurement, should be able to statistically subtract 
all the \nuxpart\ and \nuxanti\ flux.  The 
measurement of \anue's in proton-rich detectors, with the 
\nuxanti-oscillated-to-\anue\ flux subtracted off, will provide a 
measurement of \anue's, which can then be statistically 
subtracted off as well, leaving behind only the \nue\ flux.  Both the
IH and NH cases can also benefit from a measurement of the \nux\ flux
from scintillation detectors (\citealt{lahabeacom2014}).

The success of these procedures in subtracting the \backgrounds\ 
depends on distance, since a larger distance means a lower flux and
a less precise measurement of the \background, which precision would
propagate to the extracted \nue\ signal.  We encourage the various
collaborations associated with the extant and future neutrino detectors
to examine this topic and continue to investigate methods to identify
and subtract the \backgrounds.

The improvements that have been made in neutrino detection
technologies since SN 1987A have put the scientific
community in a good position to take full advantage of the neutrino
emission from the next Galactic CCSN.  In particular, the
\nue\ breakout burst peak (if it exists) from a Galactic CCSN 
could be detectable (depending on distance) 
in current and near-future neutrino detectors in
the case of the IH, but it likely won't be detectable in the NH case
(although sufficient \background\ subtraction could allow the \nue\
peak to be detected).  
A detection or nondetection of the \nue\ breakout
burst peak by itself should be sufficient to identify the
neutrino-mass 
hierarchy ({\citealt{mirizzietal2015}), and a measurement of the
  properties of the breakout burst could constrain 
progenitor mass and the nuclear EOS.  The rapidly
maturing fields of neutrino physics and neutrino astrophysics will be
greatly served by the next Galactic CCSN.

\acknowledgments

We thank Kate Scholberg, Andr\'e Rubbia, Masayuki Nakahata, and 
Lutz K\"opke
for useful conversations.
We also thank Gabriel Martinez-Pinedo for providing us with tables
for the \ar40\ cross sections.
The authors acknowledge support provided by the NSF PetaApps 
program, under award OCI-0905046 via a subaward no. 44592 from 
Louisiana State University to Princeton University, and by the 
Max-Planck/Princeton Center (MPPC) for Plasma Physics 
(NSF PHY-1144374). The authors employed computational resources 
provided by the TIGRESS high-performance computer center at 
Princeton University, which is jointly supported by the Princeton 
Institute for Computational Science and Engineering (PICSciE) and 
the Princeton University Office of Information Technology and by 
the National Energy Research Scientific Computing Center (NERSC), 
which is supported by the Office of Science of the US Department 
of Energy under contract DE-AC03-76SF00098. This work is part of 
the ``Three Dimensional Modeling of Core-Collapse Supernovae" PRAC 
allocation support by the National Science Foundation (award number 
ACI-1440032). In addition, this research is part of the Blue Waters 
sustained-petascale computing project, which is supported by the 
National Science Foundation (awards OCI-0725070 and ACI-1238993) 
and the state of Illinois. Blue Waters is a joint effort of the 
University of Illinois at Urbana-Champaign and its National Center 
for Supercomputing Applications.

\bibliographystyle{apj}
\bibliography{bibliography}

\begin{appendix}
\setcounter{table}{0}
\renewcommand{\thetable}{A\arabic{table}}
\renewcommand{\theequation}{A\arabic{equation}}
This Appendix lists the sources used for the neutrino
interaction cross sections relevant to our calculations. 
If the analytic cross section is known, then it is presented here.
Otherwise, reference is made to the source of the tabulated values of
the cross section.  
Figure~\ref{fig:sigma} shows the neutrino-matter-interaction
cross sections for \nue\ and \anue.
In what follows, it is useful to
define the quantity $\sigma_0$ as
\begin{equation}
\sigma_0 = \frac{4G_\textrm{F}^2\cos^2{\theta_c}(m_ec^2)^2}{\pi(\hbar c)^4}
\simeq 1.705 \times 10^{-44}~\textrm{cm}^2,
\end{equation}
where $G_\textrm{F}$ the Fermi constant, 
 $\theta_c$ is the Cabbibo angle, and $m_e$ is the electron mass.

Elastic scattering off of electrons ($\nu_i + e^- \rightarrow \nu_i +
e^-$)  is the primary $\nu_e$ detection channel for all but
\ar40\ detectors. \cite{tomasetal2003} provide the
differential cross section, given by
\beq   \label{el}
 \frac{d\sigma}{dy} = 
 \frac{\sigma_0}{8\cos^2{\theta_c}}\frac{E_\nu}{m_ec^2} \left[ A+B\,(1-y)^2-C\,\frac{m_e}{E_\nu}\,y
 \right] ~,
\eeq         
where $y=E_e/E_\nu$ is the energy fraction transferred to the electron.  The
coefficients $A$, $B$ and $C$ differ for the four different reaction
channels and are given in Table~\ref{ABC} (based on a similar table in
\citealt{tomasetal2003}). The vector and axial-vector
coupling constants have the usual values $C_V
=-\frac{1}{2}+2\sin^2\Theta_W$ and $C_A=-\frac{1}{2}$, with
$\sin^2\Theta_W \approx 0.231$ (\citealt{oliveetal2014}) being the Weinberg angle.

In an electron scattering, the relationship between the energy fraction transferred to the
electron ($y$) and the scattering angle
$\theta$ is given by (\citealt{tomasetal2003})
\beq
 y = \frac{2\,(m_ec^2/E_\nu)\,\cos^2\theta}{(1+m_ec^2/E_\nu)^2-\cos^2\theta}
 ~. 
\eeq
The total cross sections for electron scattering are given by
\cite{marcianoparsa2003}.
 Ignoring corrections of order $m_ec^2/E_\nu$,  the following are the 
  total cross sections:
\begin{align}
\sigma (\nu_e + e^- \rightarrow \nu_e + e^-) &= \frac{\sigma_0}{8\cos^2{\theta_c}}\left(\frac{E_\nu}{m_ec^2}\right) [1+4\sin^2\theta_W + \frac{16}{3} \sin^4\theta_W],
\end{align}
\begin{align}
\sigma (\bar\nu_e + e^- \rightarrow \bar\nu_e + e^-)  &= \frac{\sigma_0}{8\cos^2{\theta_c}}\left(\frac{E_\nu}{m_ec^2}\right)[\frac13+\frac43\sin^2\theta_W + \frac{16}{3} \sin^4\theta_W].
\end{align}
The \nux\ total cross sections are given by (neglecting terms of
 order $m_ec^2/E_\nu$)
\begin{align}
\sigma (\nu_{\mu,\tau} + e^- \rightarrow \nu_{\mu,\tau} + e^-)  &= \frac{\sigma_0}{8\cos^2{\theta_c}}\left(\frac{E_\nu}{m_ec^2}\right) [1-4\sin^2\theta_W + \frac{16}{3} \sin^4\theta_W],
\end{align}
\begin{align}
\sigma (\bar\nu_{\mu,\tau} + e^- \rightarrow \bar\nu_{\mu,\tau} + e^-) &= \frac{\sigma_0}{8\cos^2{\theta_c}}\left(\frac{E_\nu}{m_ec^2}\right)[\frac13-\frac43\sin^2\theta_W + \frac{16}{3} \sin^4\theta_W].
\end{align}
For IBD ($\bar\nu_e + p \rightarrow n + e^-$)
we use the analytic cross section of \cite{burrowsetal2006}, given by
\begin{equation}
\label{eq:anuep}
\sigma({\bar\nu_e p \rightarrow N + e^-}) =
\sigma_0\frac{1+3g_A^2}{4}\left(\frac{E_{\bar\nu_e}-\Delta_{np}}{m_ec^2}\right)^2
\left[1-\left(\frac{m_ec^2}{E_{\bar\nu_e} - \Delta_{np}}\right)^2\right]^{1/2} W_{\overline{M}},
\end{equation}
where $g_A$ is the axial-vector coupling constant, $\Delta_{np}$ is the mass-energy difference between a
proton and a neutron $(m_n - m_p)c^2$, and
$W_{\overline{M}}$ 
is the correction for weak magnetism and
recoil, $(1-7.1 E_{\bar\nu_e}/m_nc^2)$.
For both \nue\ and \anue\ CC absorption on \ar40, 
we use the cross sections from \cite{kolbeetal2003}, their Figure 9.
The data for the oxygen cross sections (which include CC
absorption by \nue\ and \anue\ and NC scattering by all species) are taken from tables in
\cite{kolbeetal2002}.  In our
analysis, we assume that all oxygen is $^{16}$O.
The cross sections for CC absorption of \nue\ and \anue\ on carbon are taken from tables in
\cite{kolbeetal1999}.  The cross sections for NC
scattering of all neutrino flavors on carbon are taken from tables in  \cite{fukugitaetal1988}.
In our analysis, we assume that all carbon is $^{12}$C.  
Neutrino
elastic scattering off of protons ($\nu_i + p \rightarrow \nu_i + p$)
is also expected to be detectable in scintillation detectors owing to
their low detection thresholds (\citealt{beacom2002}).  
However, the flux primarily probed by
this channel will be the \nux\ flux (referring to the what is
the \nux\ flux before oscillations occur) because of its higher average
energy.  \cite{oberaueretal2005} state that the 
signal from the recoil protons can be easily separated from the other
signals.  We assume this ability in our analysis and do not include
contributions from NC scattering on protons.
\begin{deluxetable}{lccc}
\tablecolumns{4}
\tablecaption{\label{ABC} 
Coefficients used in Eq.~(\ref{el}) for the elastic scattering of
neutrinos on electrons.}
\tablehead{ Neutrino Type & \colhead{$A$} & \colhead{$B$} & \colhead{$C$} }
%   & $A$ & $B$ & $C$ \\[0.5ex] \hline  
\startdata
$\nu_e$ & $(C_V{+}C_A{+}2)^2$ & $(C_V-C_A)^2$ & $(C_V+1)^2-(C_A+1)^2$ \\
$\bar\nu_e$ & $(C_V-C_A)^2$ & $(C_V{+}C_A{+}2)^2$ & $(C_V+1)^2-(C_A+1)^2$\\
$\nu_{\mu,\tau}$ & $(C_V+C_A)^2$ & $(C_V-C_A)^2$ & $C_V^2-C_A^2$\\
$\bar\nu_{\mu,\tau}$ & $(C_V-C_A)^2$ & $(C_V+C_A)^2$ &
$C_V^2-C_A^2$ 
\enddata
\end{deluxetable}

\end{appendix}

\begin{figure}[h]
\centering
\includegraphics[width=0.7\linewidth]{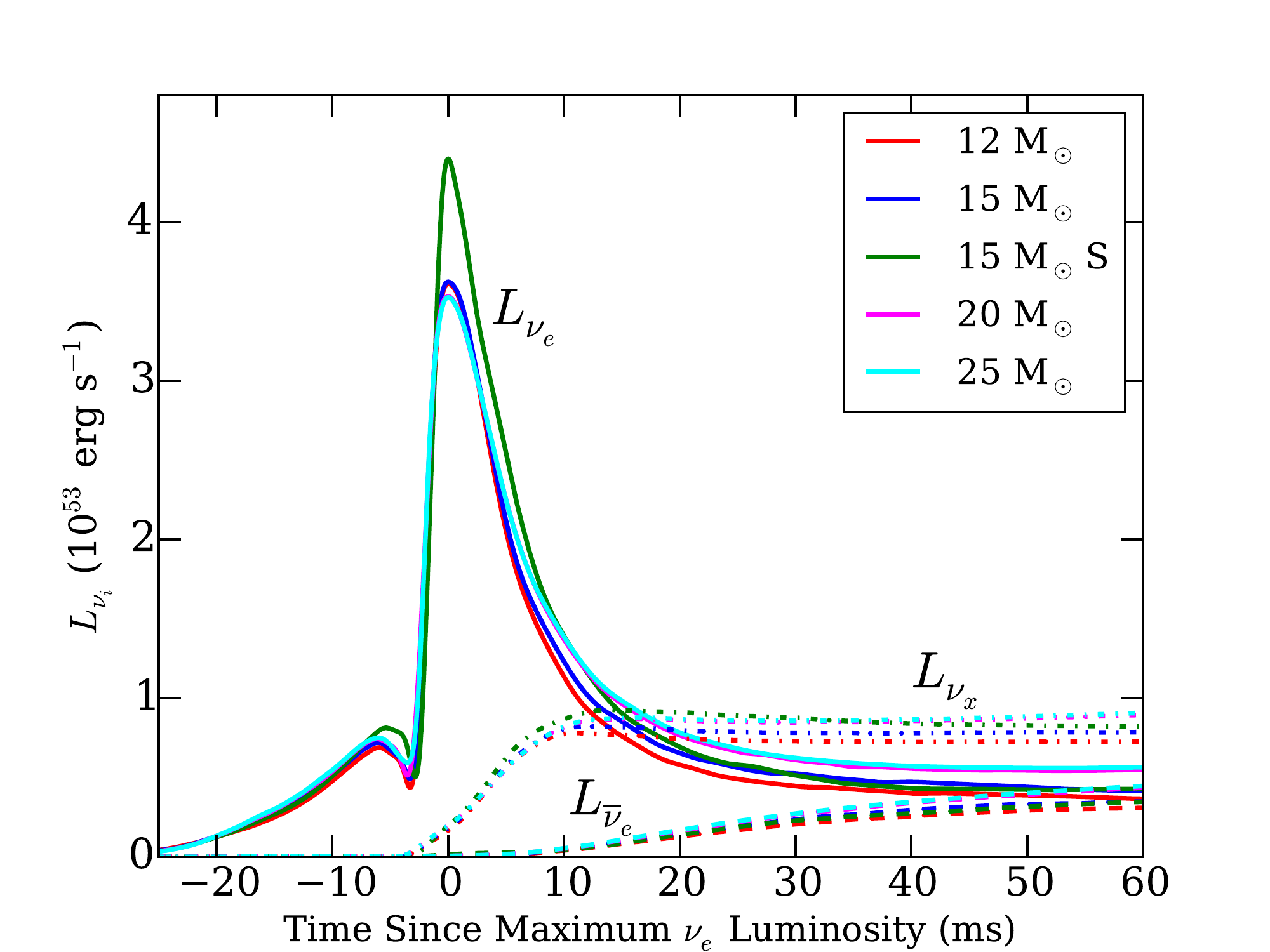}
\caption{\label{fig:lumallt}Unoscillated 
energy luminosity as a function of time for all
three main neutrino channels (\nue, \anue, and \nux) for our five
progenitor 
models.  The solid line represents the \nue\ energy luminosity, the
dashed line represents the \anue\ luminosity, and the dot-dashed line
represents the \nux\ luminosity.  ``S'' designates the \shen\ while
the rest of the models use the \ls\ with $K=220$ MeV. 
Shown is a cubic spline fit to the numerical
data. Time is calculated since the peak of
the \nue\ luminosity.  The \nux\ luminosity is shown for the four
neutrino types ($\nu_\mu,\nu_\tau,\bar\nu_\mu$, and
$\bar\nu_\tau$). The luminosity for any one of these four
neutrino types
will be one-quarter of the value shown here for \nux.}
\end{figure}

\begin{figure*}[h]
\centering
\includegraphics[width=0.75\linewidth]{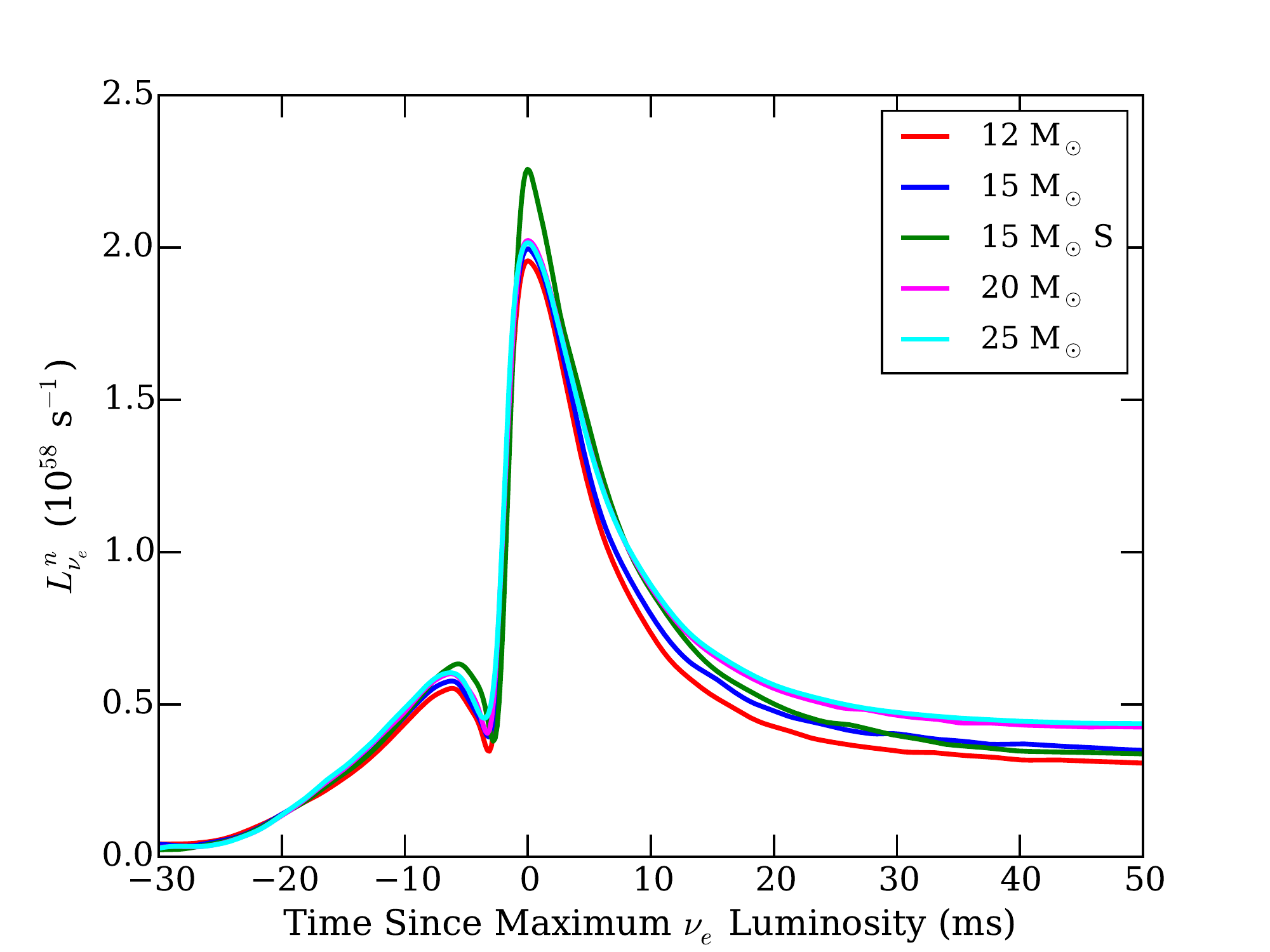}
\caption{\label{fig:nuelumt} 
Unoscillated \nue\ number luminosity as a
function of time over breakout for the various models. 
The ``S'' in the figure legend
refers to the \shen; the rest of the models use the \ls\ with $K=220$ MeV.  
The time is centered on the
time of maximum \nue\ luminosity.   The luminosity shows 
two peaks: a small peak on the
initial rise, and a large peak following a sharp rise.  The first,
smaller peak is due to neutrinos from the neutronization of
the collapsing core; the second, larger peak is from \nue's created by
electron capture on free protons liberated by the 
shock. }
\end{figure*}

\begin{figure*}[h]
\centering
\includegraphics[width=0.85\linewidth]{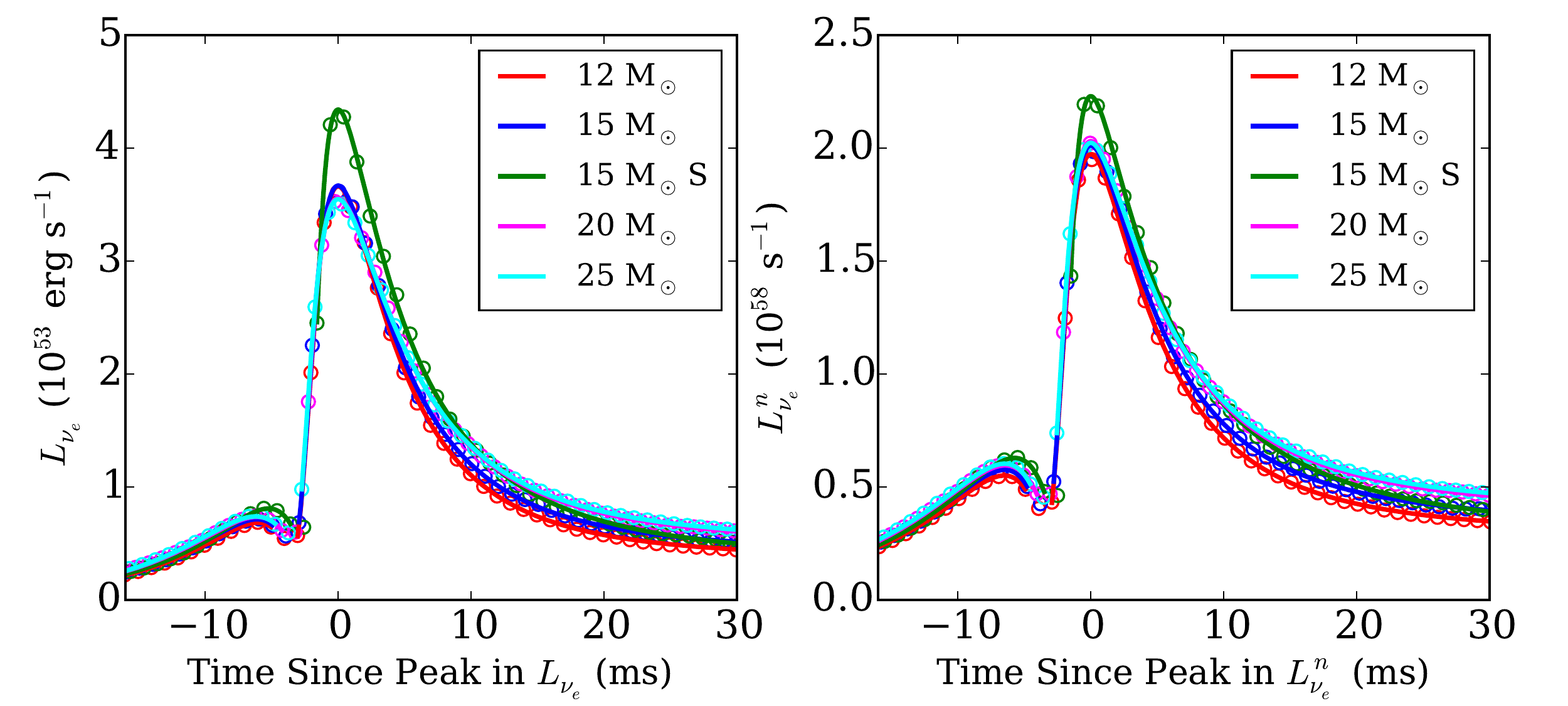}
\caption{\label{fig:fit}Fits to the models, using
  Equations~(\ref{eq:analytic}) and~(\ref{eq:smallpeak}).  
Left: fits to the unoscillated energy
  luminosity.  The parameters used in the fits are in
  Tables~\ref{tab:peakfit} and~\ref{tab:smallbumpfit}.
 Right: fits to the unoscillated number
  luminosity.  The parameters used in these fits are in
  Tables~\ref{tab:numpeakfit} and~\ref{tab:numsmallbumpfit}.
The numerical model data
  points are shown as circles, while the fits of
  Equations~(\ref{eq:analytic}) and~(\ref{eq:smallpeak})  are shown as
  lines.  For each model, the local minimum between the
  preshock neutronization peak and the breakout burst peak is not
  well fit by either of Equations~(\ref{eq:analytic}) and
  ~(\ref{eq:smallpeak}), and so no attempt is made to fit it in this figure.
}
\end{figure*}

\begin{figure*}[h]
\centering
\centerline{\includegraphics[width=0.95\linewidth]{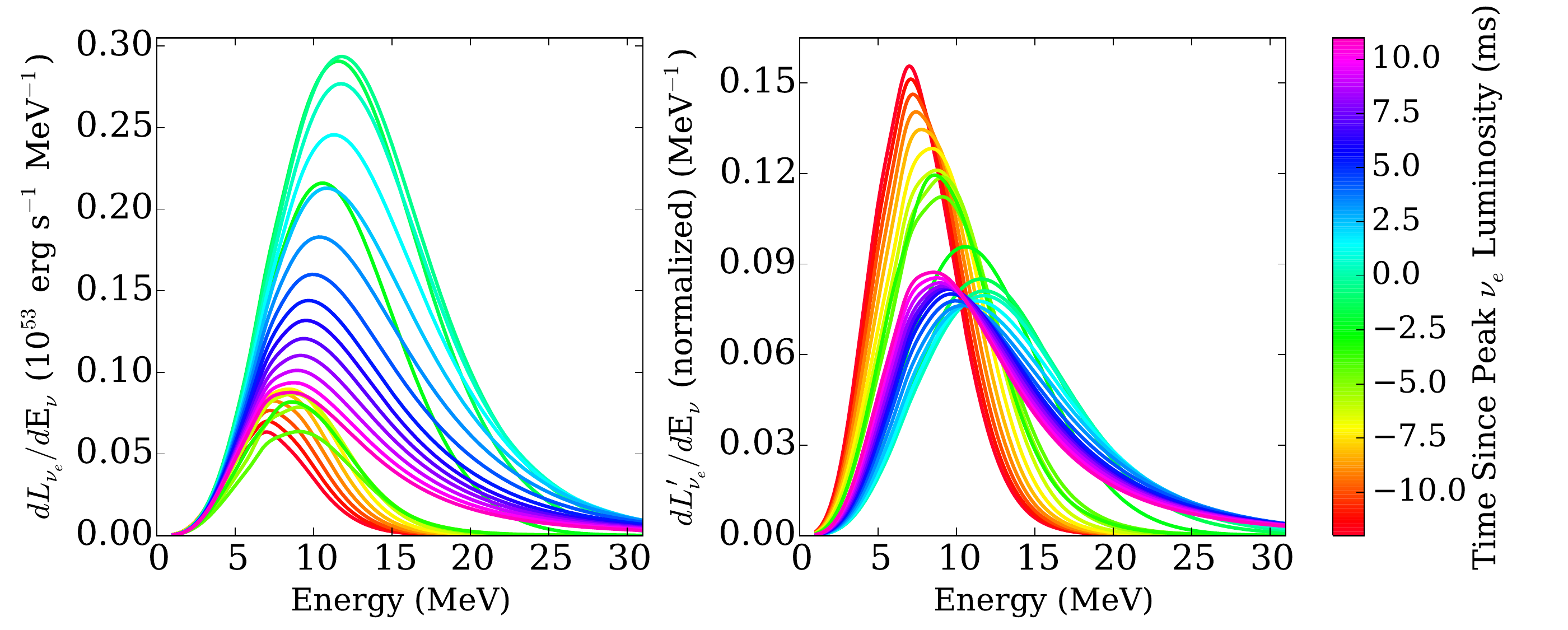}}
\caption{\label{fig:energyspectrum}Unoscillated energy spectra for the 15 \Msol\
  \ls\ model for a range of  times through the breakout burst.  
  Left: the
  full, true spectra.  Right: the normalized spectra, normalized so
  that the area under the normalized spectrum of each time (integrated
  over energy) is 1.  Both panels show spectra over the same time range.}
\end{figure*}
%\afterpage{\clearpage}
\begin{figure}[h]
\centering
\includegraphics[width=0.65\linewidth]{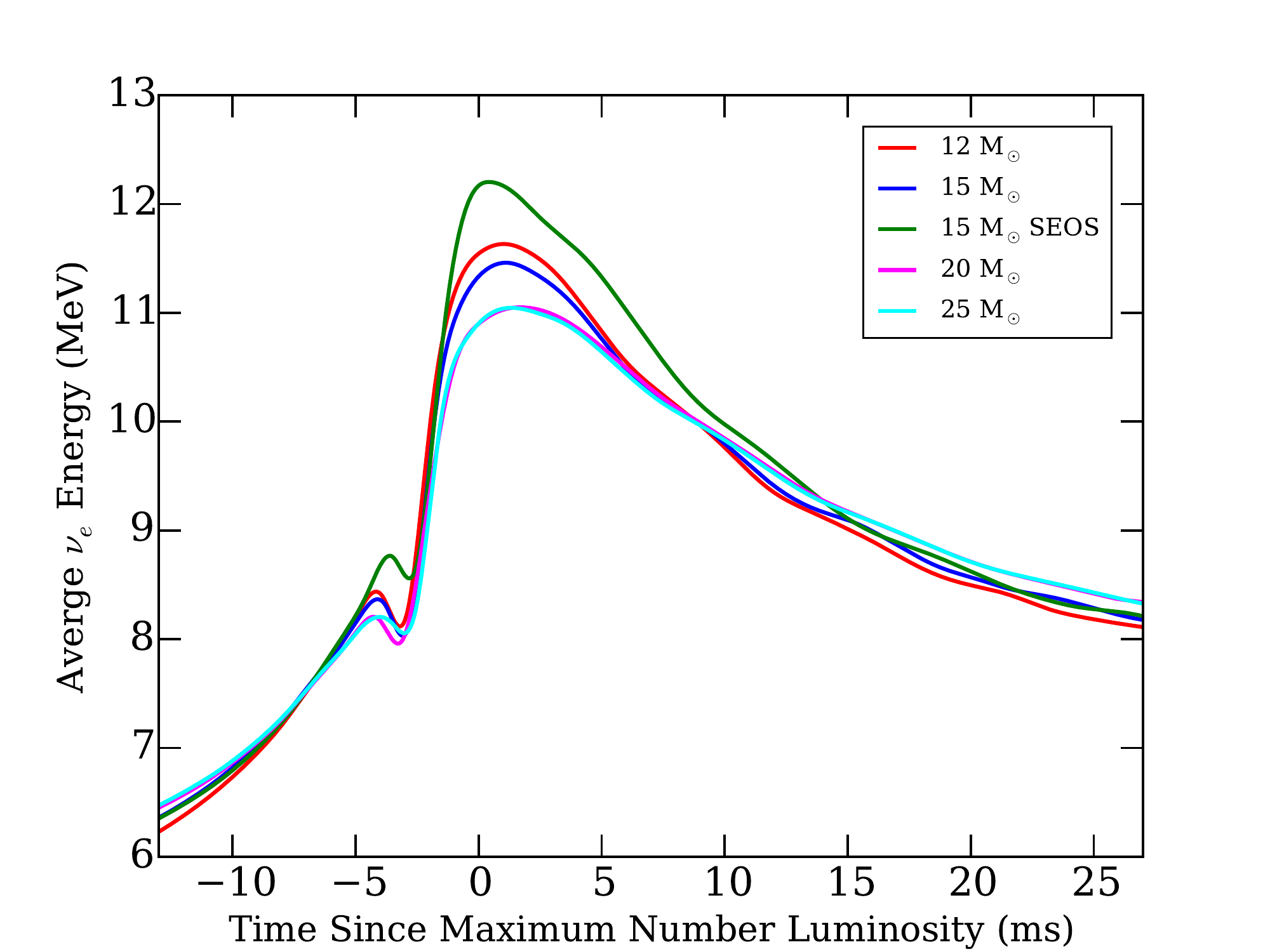}
\caption{\label{fig:avgenergy}Average \nue\ energy as a function of
time shown for all the models through the breakout burst. The average
energy is defined by Equation~(\ref{eq:averageenergy}).  Time is
calculated as time since maximum number luminosity for the fits to the
number luminosity for each model.  For the same 
EOS, the peak average energy decreases with progenitor
mass down to 20 \Msol, with the 20 and 25 \Msol\ progenitors
showing comparable values, while the average energy in the tail after 
peak increases with progenitor mass (again, with the 20  and 
25 \Msol\ progenitors
showing comparable values).
The models show a smaller peak in average energy, which is 
associated with the 
pre-breakout neutronization peak. 
The 15 \Msol\ \shen\ model has a slightly higher average
average energy during both the breakout burst peak and preshock
neutronization peak compared to its \ls\
counterpart, but comparable average energy coming into and leaving the
breakout burst.  The average energy for all the models peaks at a time
slightly after the
time of maximum number luminosity.} 
\end{figure}

%Figure 6
\begin{figure}[h]
\centering
\includegraphics[width=0.7\linewidth]{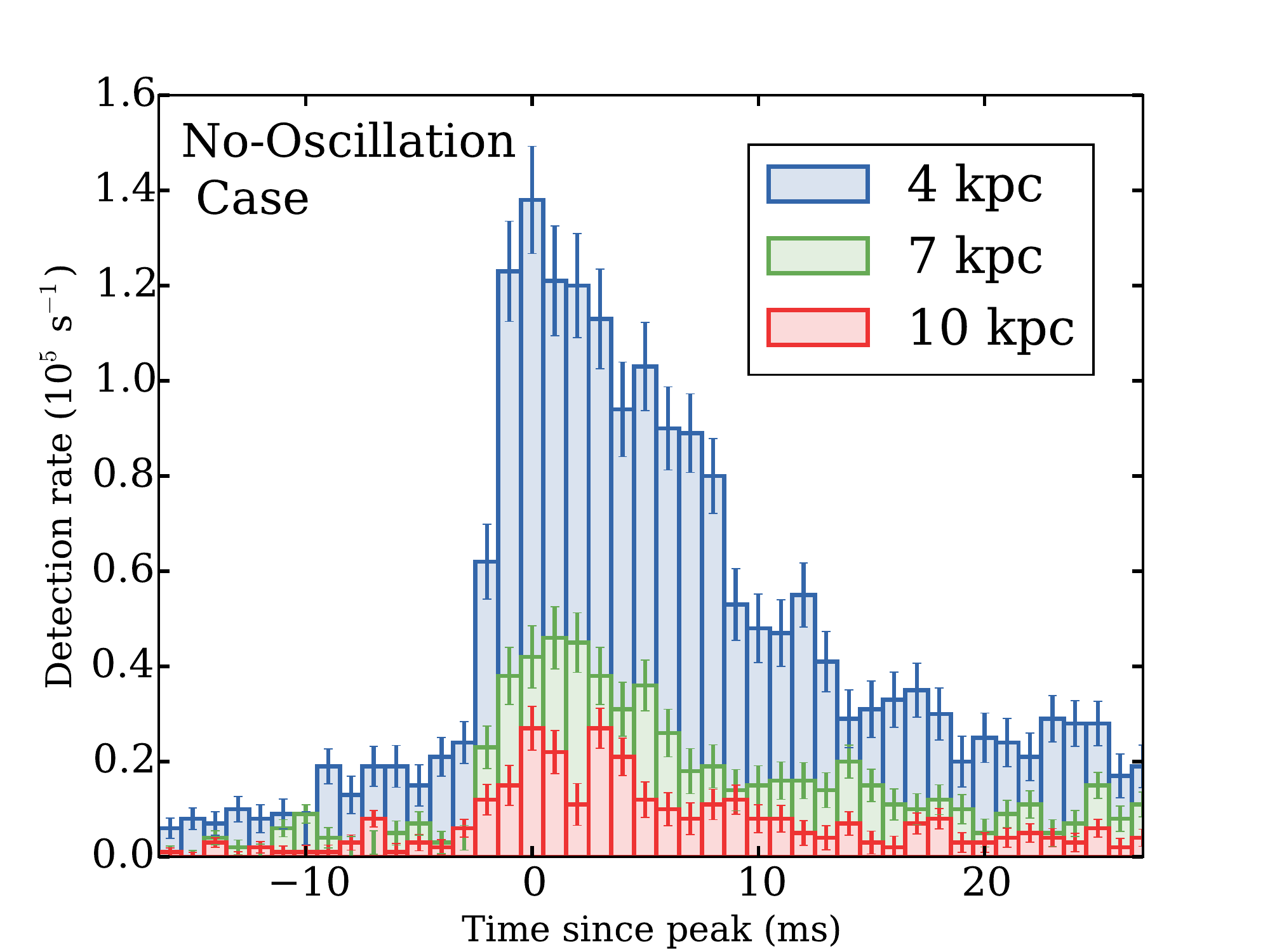}
\caption{\label{fig:hyperkhistogram}Example realization of detection
  rates in the no-oscillation case with
  1$\sigma$ error bars for CCSN neutrino detections in
  Hyper-K at distances of 4, 7, and 10 kpc, binned in 1 ms time bins.
  This figure not only shows the overall increase in signal expected
  in Hyper-K as the distance to the SN decreases but also gives a
  general sense of how the expected noise and error bars in each time bin
  depend on $D$.}
\end{figure}

\begin{figure*}[h]
\centerline{\includegraphics[width=.943\linewidth]{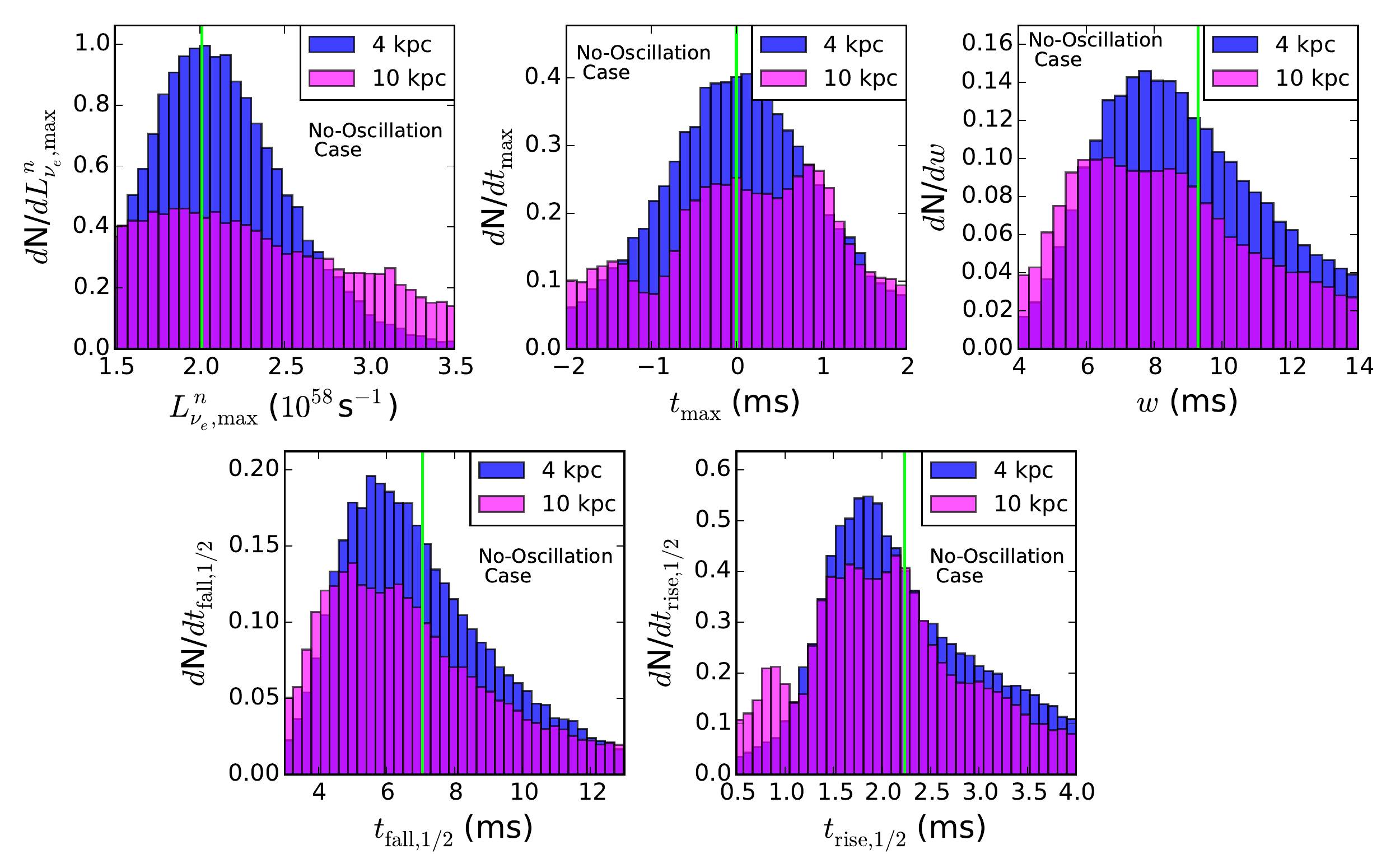}}
\caption{\label{fig:superkphysicalparms_hist} 
 PDFs  for the physical
  parameters derived from fits of Equation~(\ref{eq:analytic}) to the 
simulated observations for Super-K for the 
15 \Msol\ \ls\ model in the no-oscillation case.  
For each parameter, blue shows the 
PDF corresponding to a supernova
at a distance of 4 kpc and magenta shows the PDF
corresponding to a supernova at a distance of
10 kpc.  Overlap between the two PDFs
 is shown in purple.  The
model value is shown as a green vertical line.}
\end{figure*}

\begin{figure*}[h]
\centerline{\includegraphics[width=0.943\linewidth]{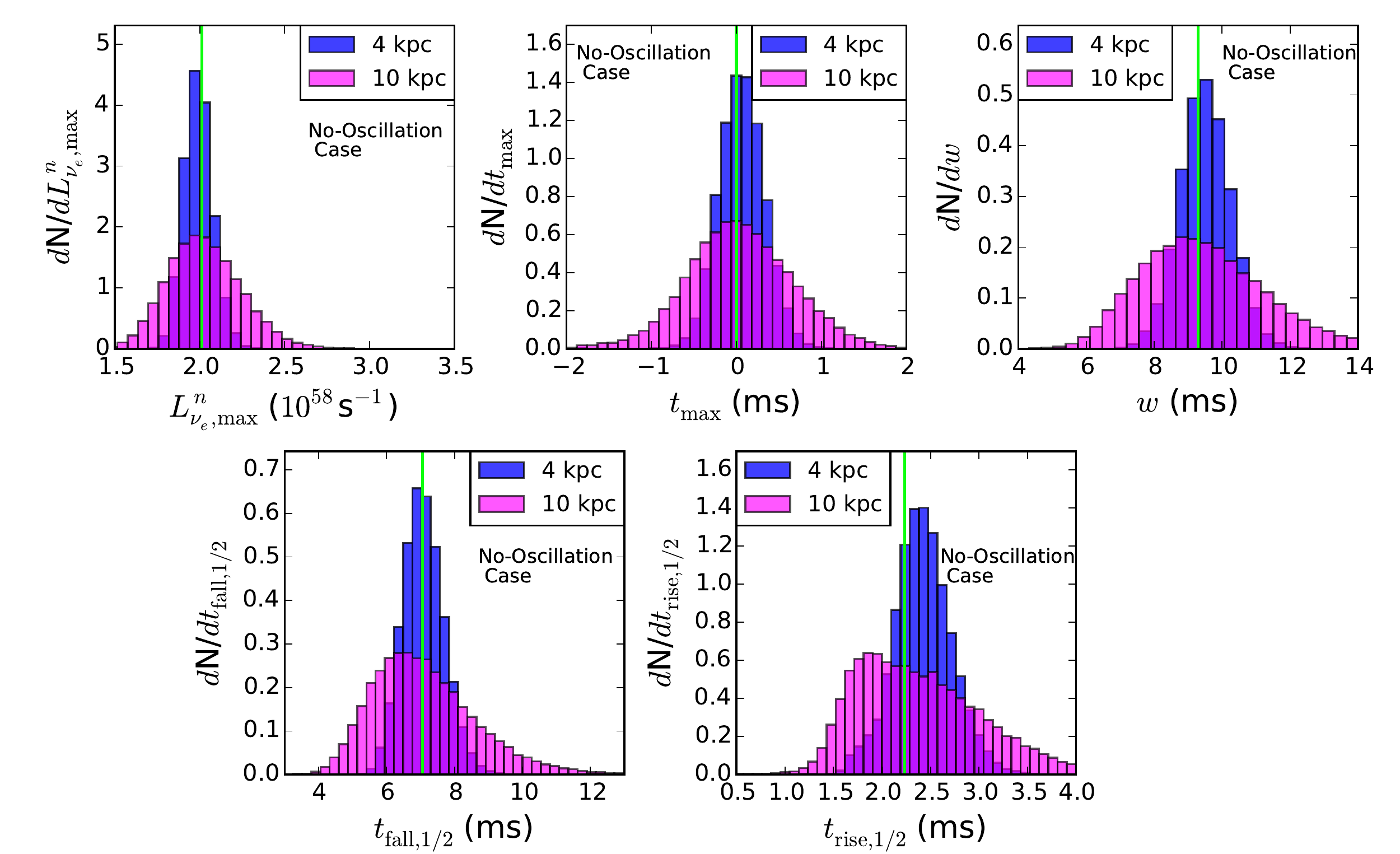}}
\caption{\label{fig:hyperkphysicalparms_hist} Same as
  Figure~\ref{fig:superkphysicalparms_hist}, but for Hyper-K in the
  no-oscillation case.}
\end{figure*}

%\afterpage{\clearpage}

%Figure 10
\begin{figure*}[h]
\centerline{\includegraphics[width=0.943\linewidth]{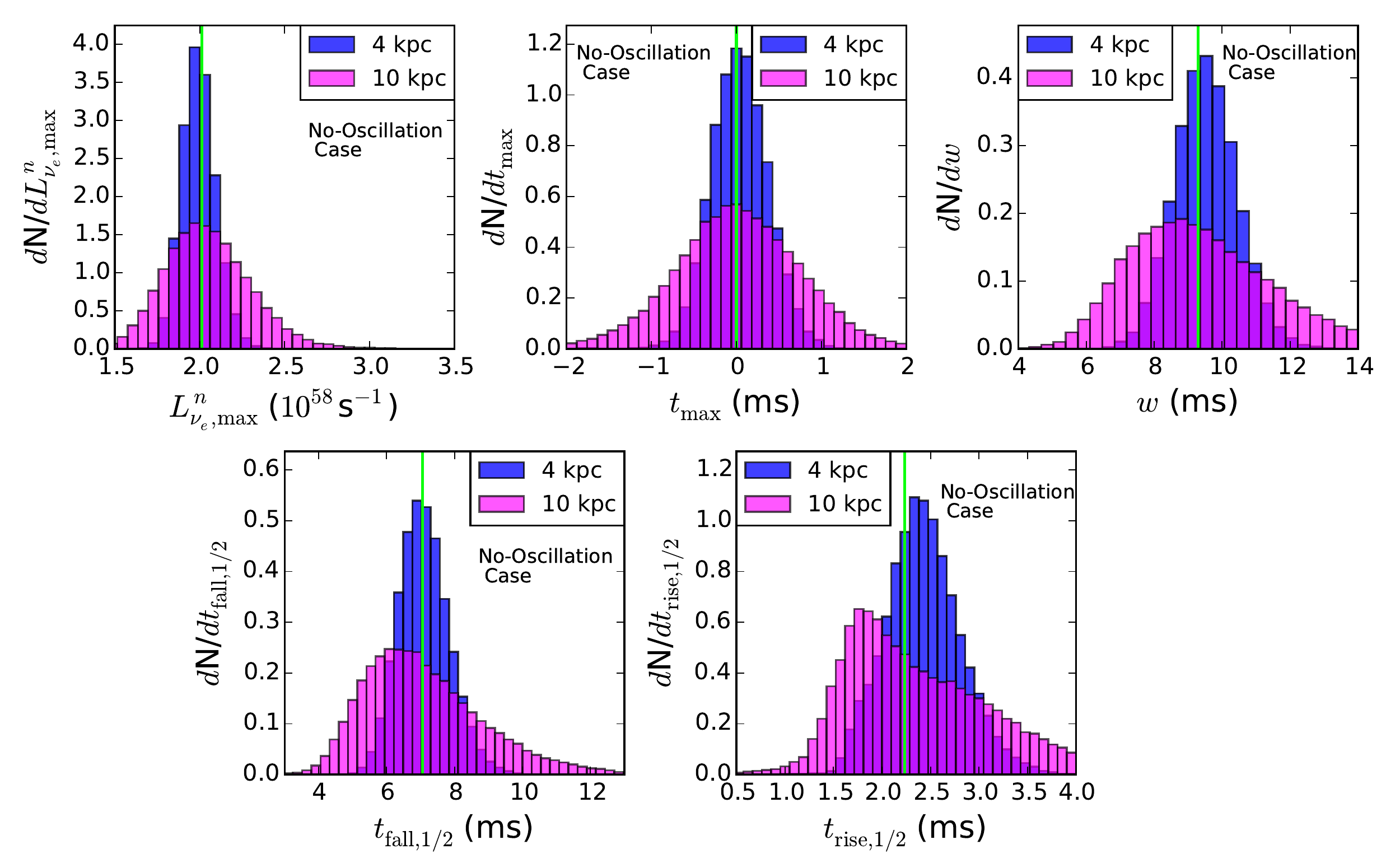}}
\caption{\label{fig:dunephysicalparms_hist} Same as
  Figure~\ref{fig:superkphysicalparms_hist}, but for DUNE in the
  no-oscillation case.}
\end{figure*}

\begin{figure*}[h]
\centerline{\includegraphics[width=0.943\linewidth]{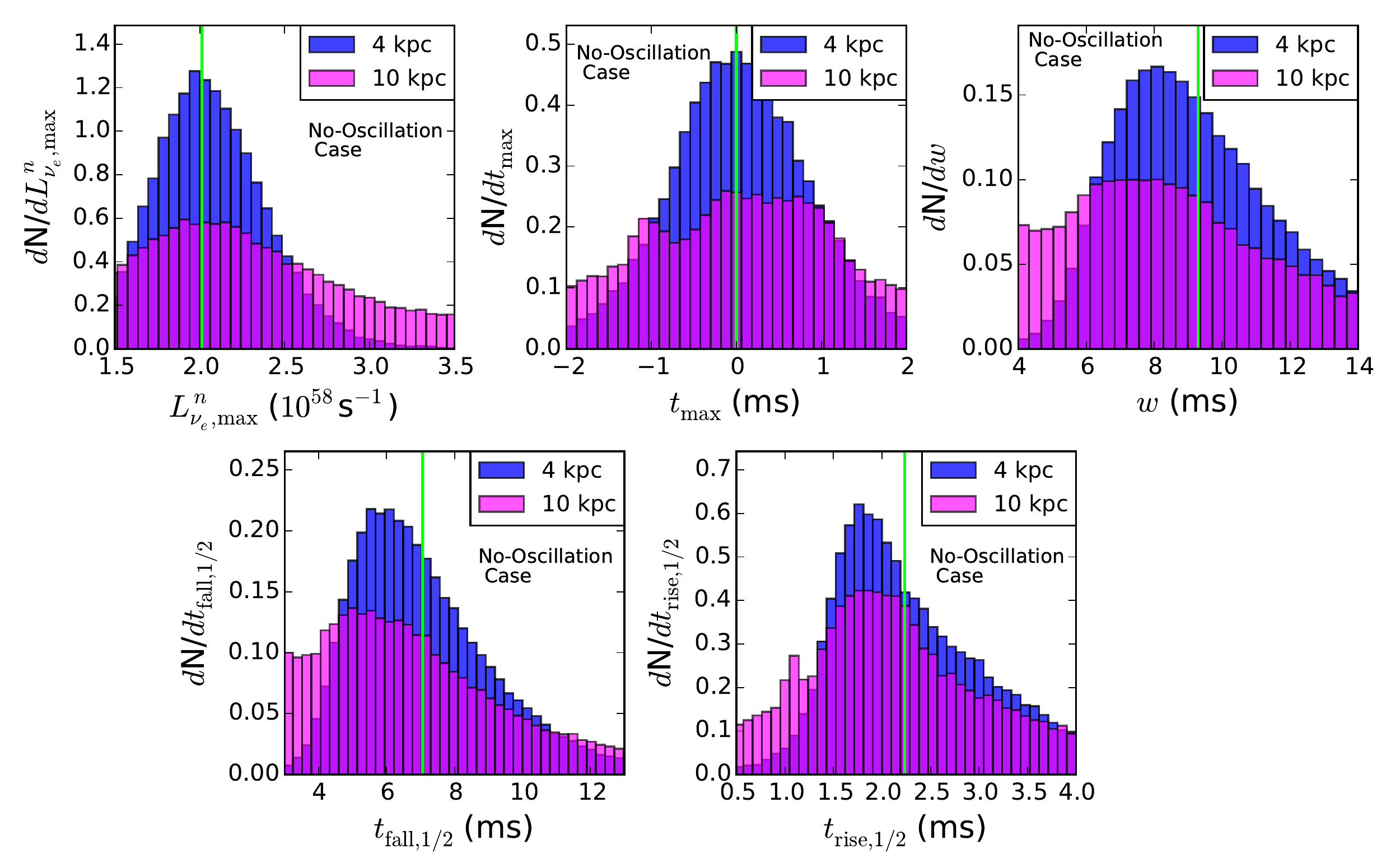}}
\caption{\label{fig:junophysicalparms_hist} Same as
  Figure~\ref{fig:superkphysicalparms_hist}, but for JUNO in the
  no-oscillation case.}
\end{figure*}

%\afterpage{\clearpage}

\begin{figure*}[h]
\centerline{\includegraphics[width=.6\linewidth]{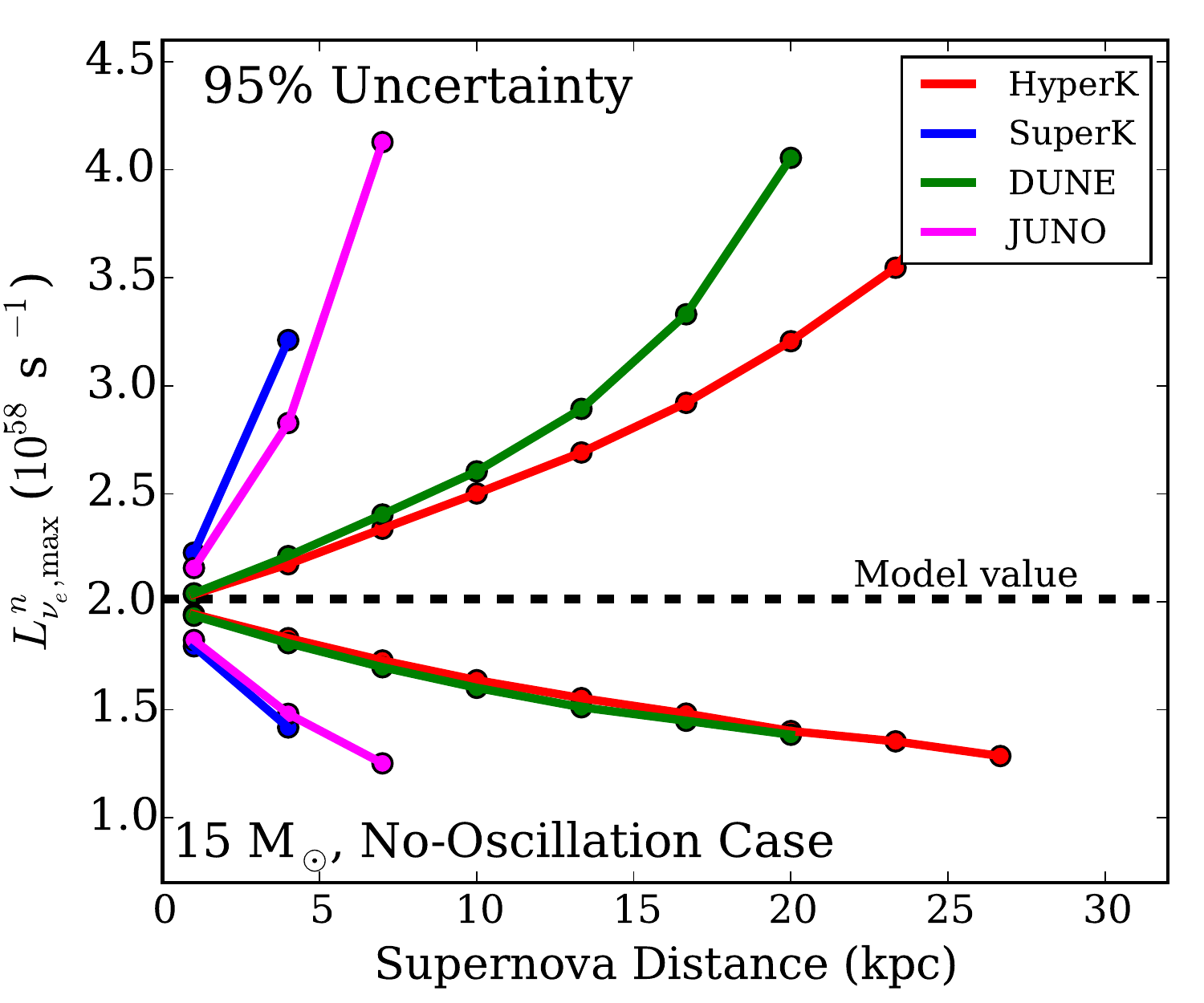}}
\caption{\label{fig:15maxvalfuncD} The 95\% uncertainty in measuring \lmax\ as a
  function of distance for various detectors for the 15
  \Msol\ \ls\ model, in the no-oscillation case. For each detector, the lines represent the span
  needed to include 95\% of the \lmax's calculated from the set of
  $5\times10^4$ 
  sampled observations. 
 When the uncertainty values for a
  specific detector get either
  too large or too small relative to the model value, we stop plotting
  the uncertainty at that distance and greater distances.  The uncertainty values for 
Super-K and JUNO were cut off
  at 7 kpc if the previous criteria were not met at 7 kpc because of
  the small number of events for an SN beyond that
  distance.
}
\end{figure*}

%\afterpage{\clearpage}

\begin{figure*}[h]
\centerline{\includegraphics[width=\linewidth]{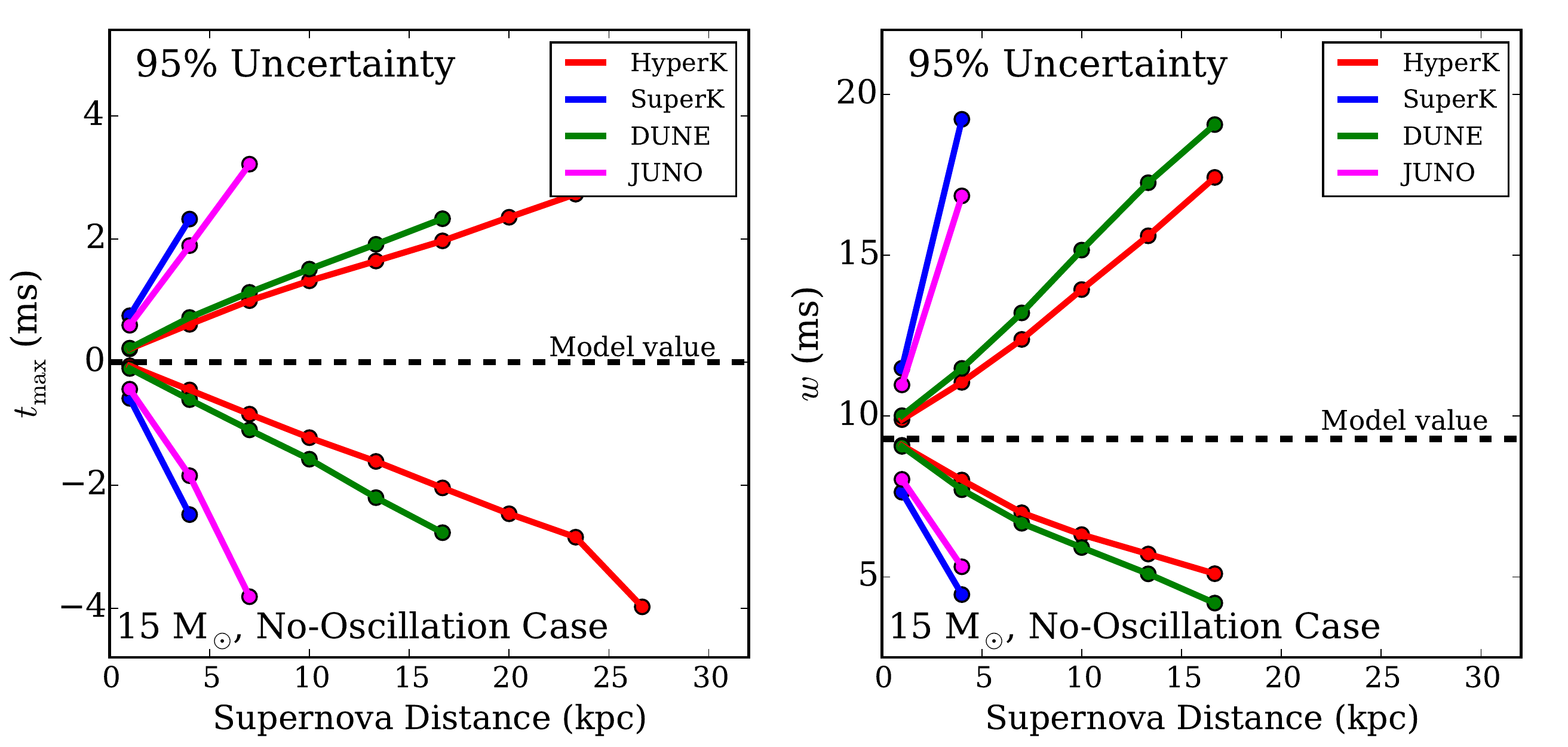}}
\caption{\label{fig:15maxlocfwhmfuncD} Same as
  Figure~\ref{fig:15maxvalfuncD}, but for \tmax\ (left) and 
     $w$ (right), in the no-oscillation case.}
\end{figure*}

\begin{figure*}[h]
\centerline{\includegraphics[width=\linewidth]{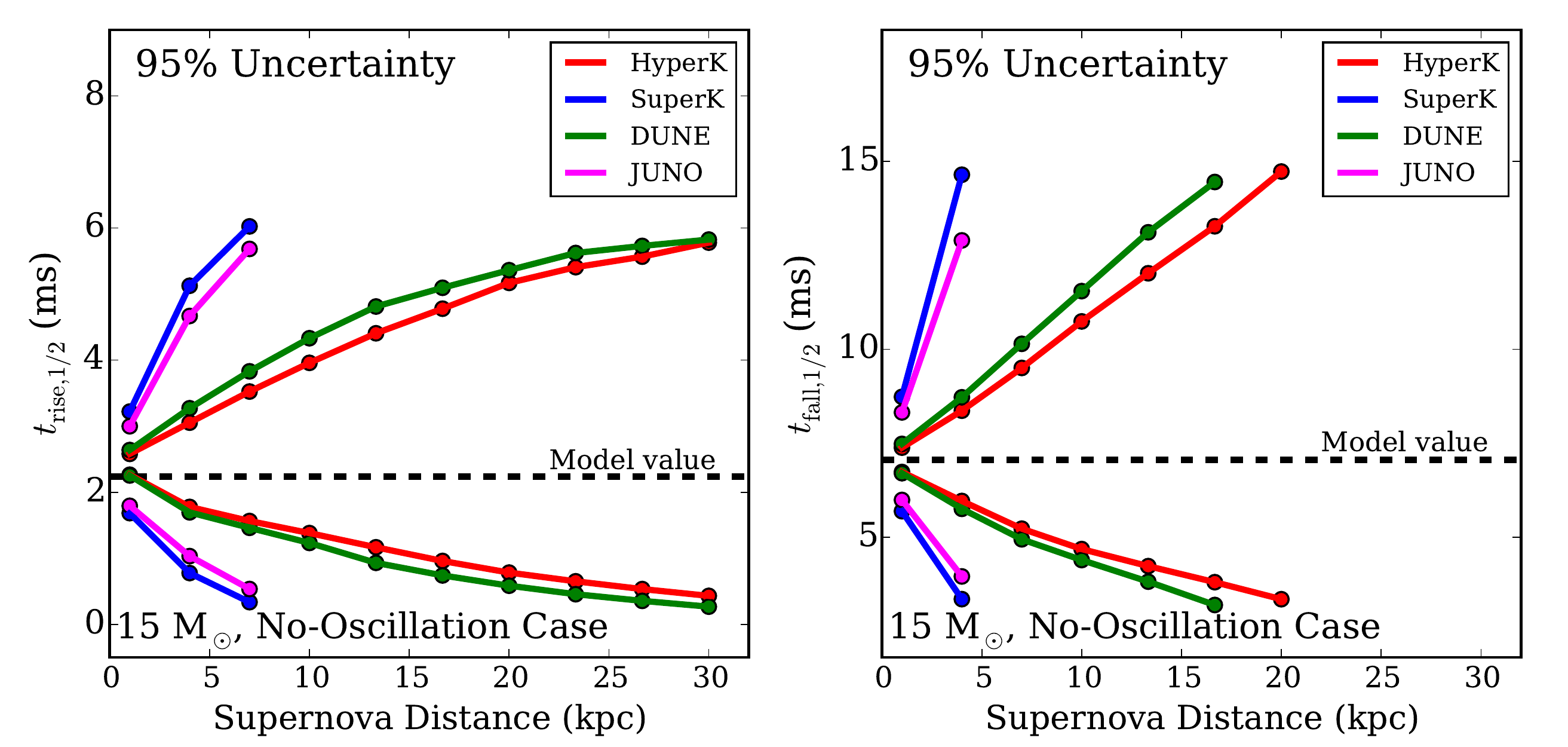}}
\caption{\label{fig:15lwhmrwhmfuncD} Same as
  Figure~\ref{fig:15maxvalfuncD}, but for \trise\ (left) and 
     \tfall\ (right), in the no-oscillation case.}
\end{figure*}

\afterpage{\clearpage}

%Figure 15
\begin{figure*}[*h]
\centerline{\includegraphics[width=\linewidth]{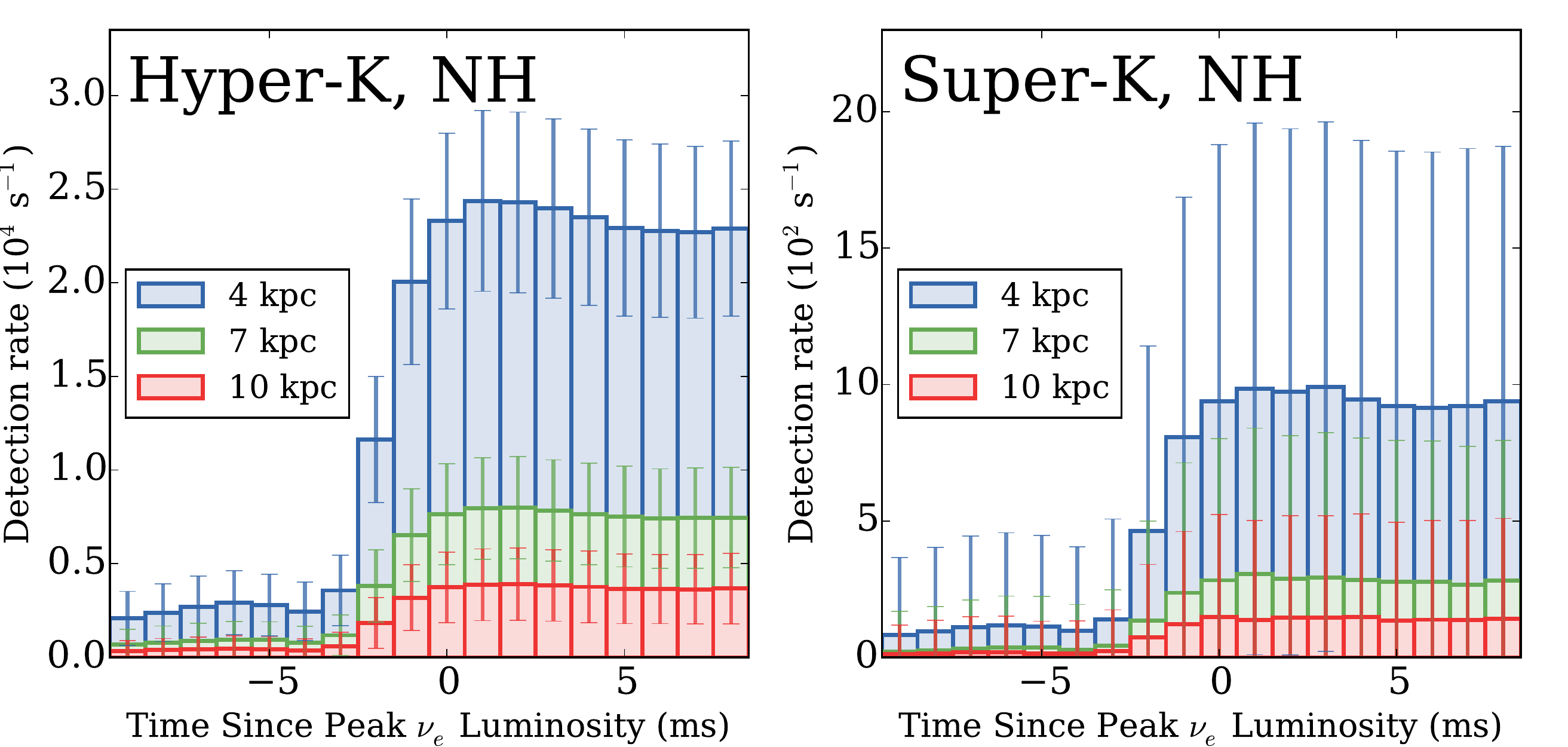}}
\caption{\label{fig:hyperk_superk_nh_backgrounds}  For
  Hyper-K (left) and
  for Super-K (right), the expected light curve
  for SNe at 4, 7, and 10 kpc, incorporating the neutrino
  oscillations expected in the case of the NH.  Detections of neutrinos of all
  flavors are taken into account, with IBDs and NC scattering off of oxygen
  subtracted, and (for JUNO) IBDs and NC scattering off of carbon
  subtracted.  Each time bin shows the mean
  count rate in that time bin over 10$^4$ realizations, and the error
  bars show the standard deviation based on the same 10$^4$
  realizations.}
\end{figure*}

%\clearpage

\begin{figure*}[h]
\centerline{\includegraphics[width=\linewidth]{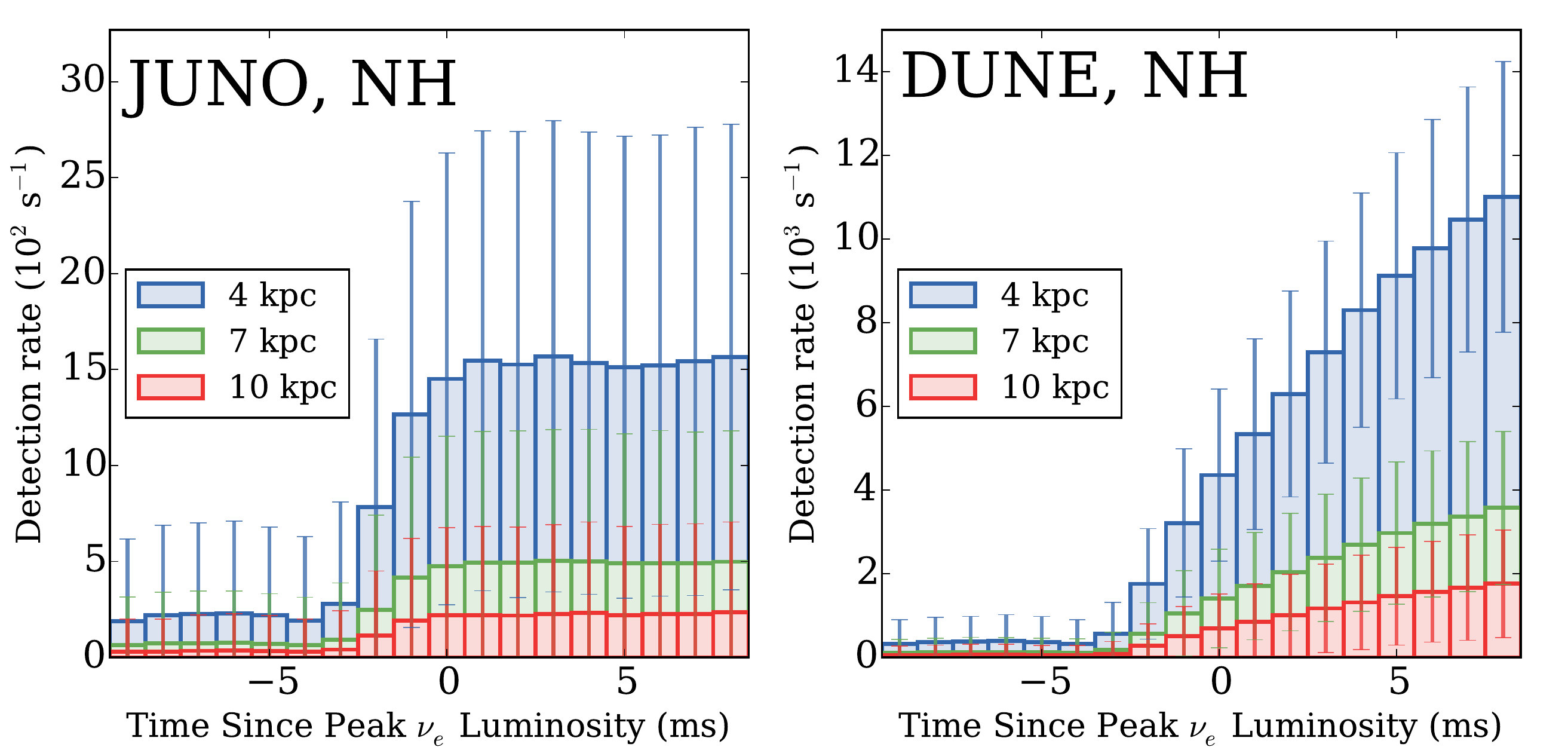}}
\caption{\label{fig:juno_dune_nh_backgrounds} Similar to
  Figure~\ref{fig:hyperk_superk_nh_backgrounds}, but for 
  JUNO (left) and DUNE (right).  For JUNO, IBDs and NC scatterings off of carbon
  have been subtracted. For DUNE, no signals from any detection channel
  have been subtracted.
}
\end{figure*}

%%IH
\begin{figure*}[h]
\centerline{\includegraphics[width=\linewidth]{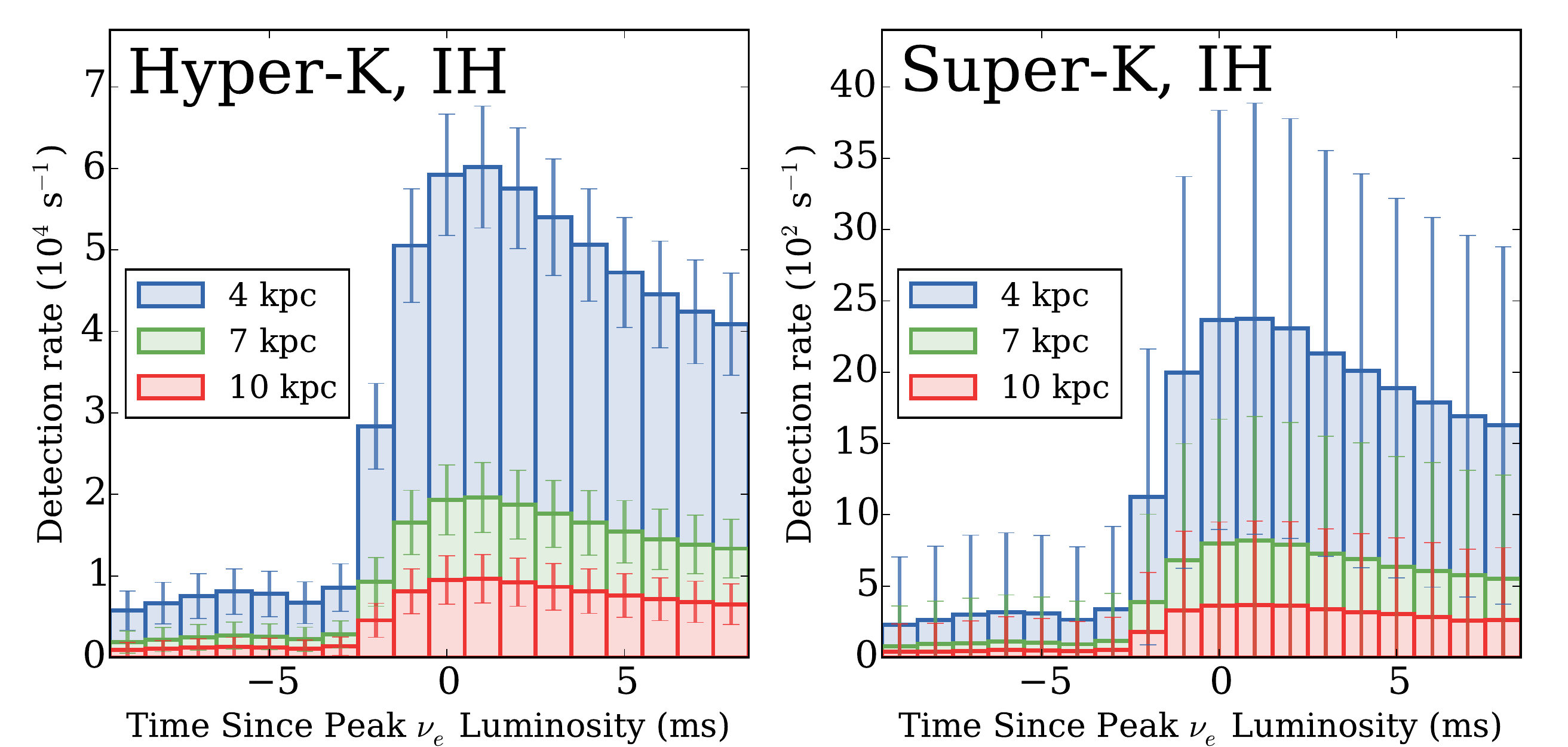}}
\caption{\label{fig:hyperk_superk_ih_backgrounds} Similar to 
  Figure~\ref{fig:hyperk_superk_nh_backgrounds}, but using the neutrino
  oscillations expected for the IH instead of the NH.
}
\end{figure*}

\clearpage

\begin{figure*}[h]
\centerline{\includegraphics[width=\linewidth]{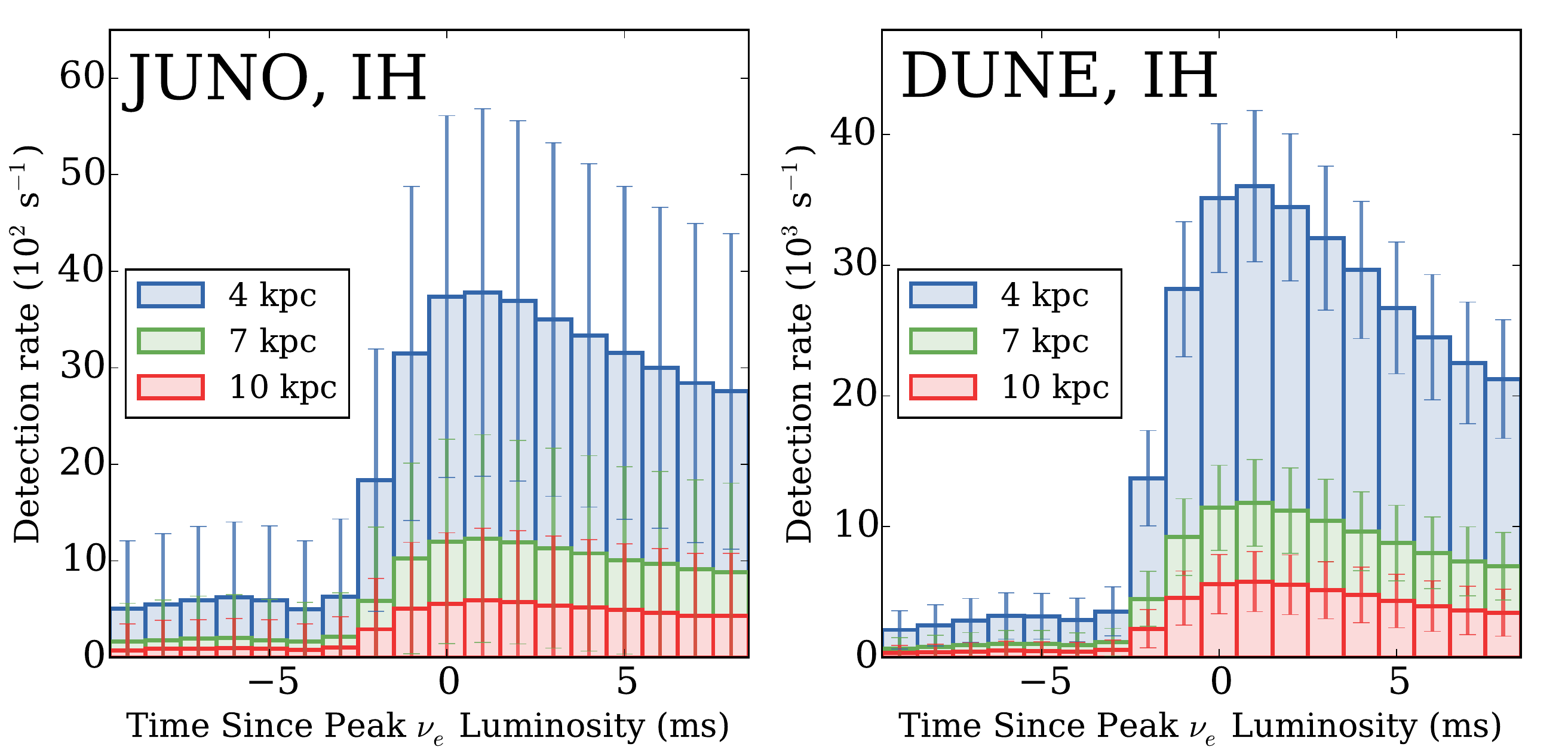}}
\caption{\label{fig:juno_dune_ih_backgrounds} Similar to
  Figure~\ref{fig:juno_dune_nh_backgrounds}, but using the neutrino
  oscillations expected for the IH instead of the NH.
}
\end{figure*}

%Figure 19
\begin{figure*}[h]
\centerline{\includegraphics[width=\linewidth]{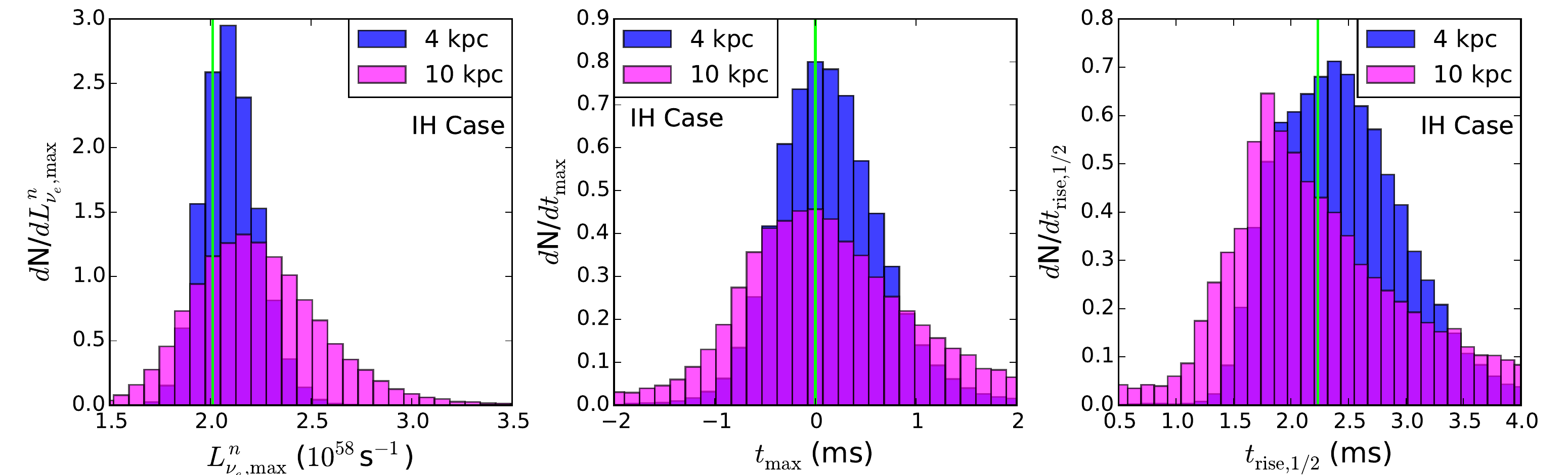}}
\caption{\label{fig:hyperkphysicalparms_hist_ih} Same as
  Figure~\ref{fig:hyperkphysicalparms_hist}, but for the IH case and
  with only \lmax, \tmax, and \trise\ shown.  For Hyper-K, the
  background signals due to IBDs and NC scatterings off of oxygen
  have been subtracted.}
\end{figure*}

\begin{figure*}[h]
\centerline{\includegraphics[width=\linewidth]{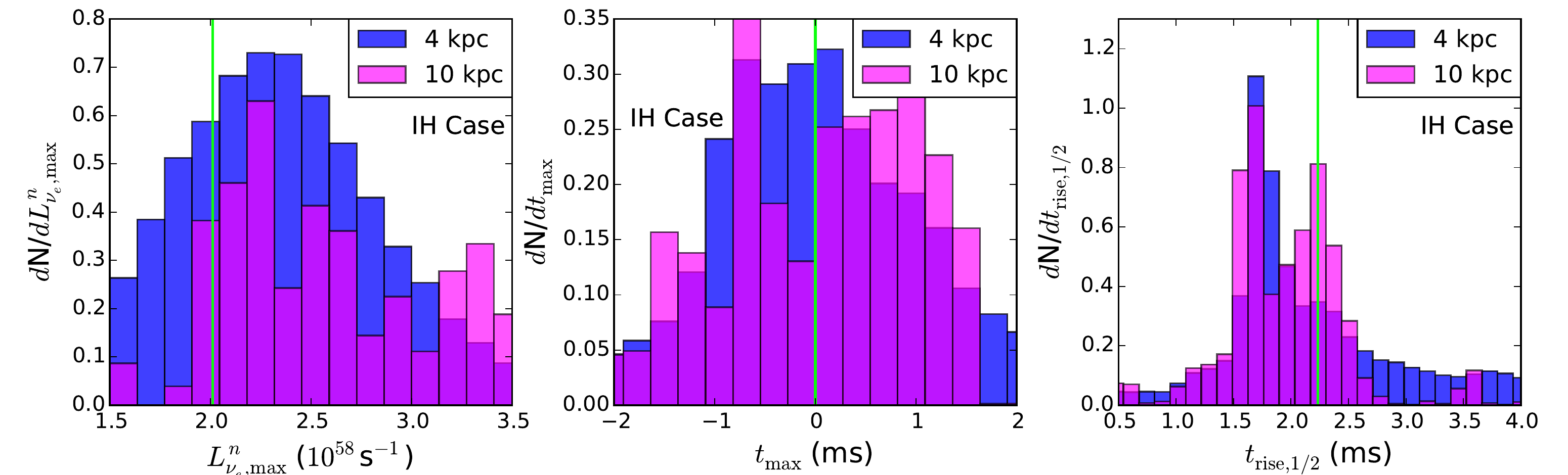}}
\caption{\label{fig:superkphysicalparms_hist_ih} Same as
  Figure~\ref{fig:hyperkphysicalparms_hist_ih}, but for Super-K in the
  IH case.  For Super-K, the
  background signals due to IBDs and NC scatterings off of oxygen
  have been subtracted.}
\end{figure*}
\afterpage{\clearpage}

\begin{figure*}[h]
\centerline{\includegraphics[width=\linewidth]{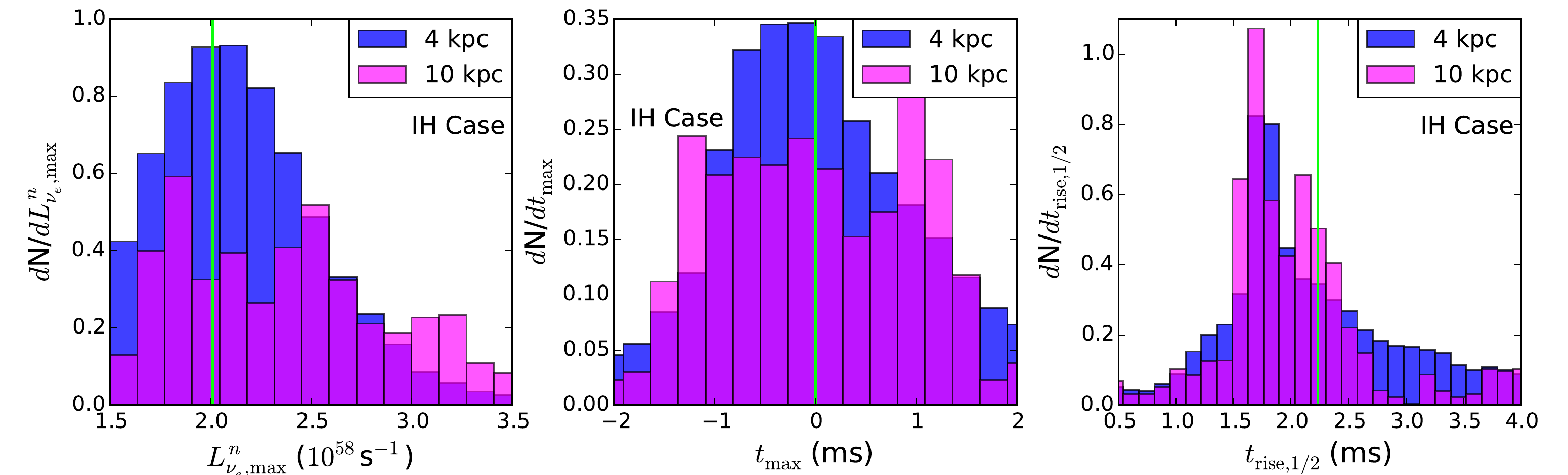}}
\caption{\label{fig:junophysicalparms_hist_ih} Same as
  Figure~\ref{fig:hyperkphysicalparms_hist_ih}, but for JUNO in the
  IH case.  For JUNO, the
  background signals due to IBDs and NC scatterings off of carbon
  have been subtracted.}
\end{figure*}

\begin{figure*}[h]
\centerline{\includegraphics[width=\linewidth]{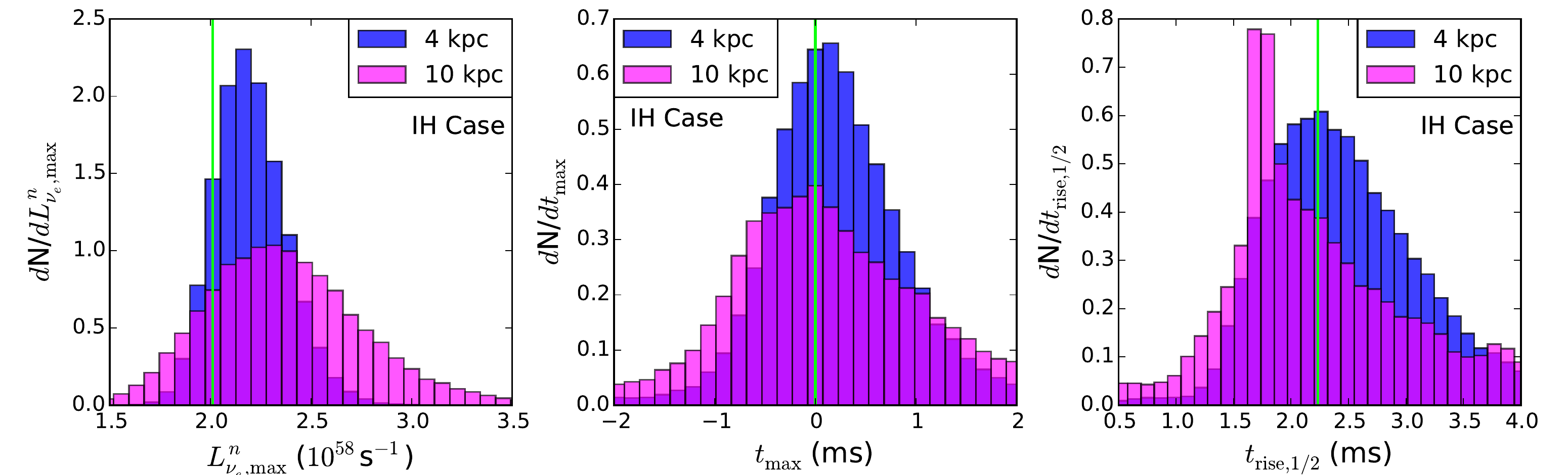}}
\caption{\label{fig:dunephysicalparms_hist_ih} Same as
  Figure~\ref{fig:hyperkphysicalparms_hist_ih}, but for DUNE in the
  IH case.  For DUNE, signals have been subtracted for any detection channel.}
\end{figure*}

%\clearpage

\begin{figure*}[h]
\centerline{\includegraphics[width=\linewidth]{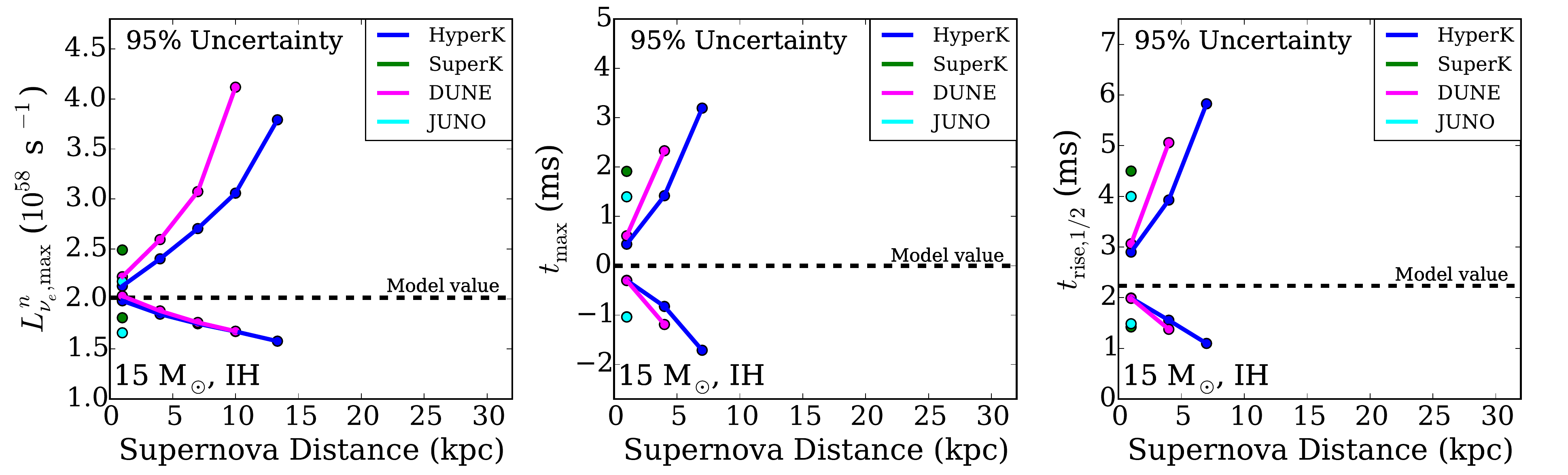}}
\caption{\label{fig:15funcD_IH} Similar to
  Figure~\ref{fig:15maxvalfuncD}, but for \lmax\ (left), 
     \tmax\ (middle), and  \trise\ (right), in the no-oscillation case.
  In all cases, the data for JUNO and/or Super-K are not
  plotted beyond 1 kpc, and the data are shown
  as a single point rather than a line connecting multiple points.
}
\end{figure*}

\setcounter{figure}{0}
\renewcommand{\thefigure}{A\arabic{figure}}

\begin{figure}[h]
\includegraphics[width=4.5in]{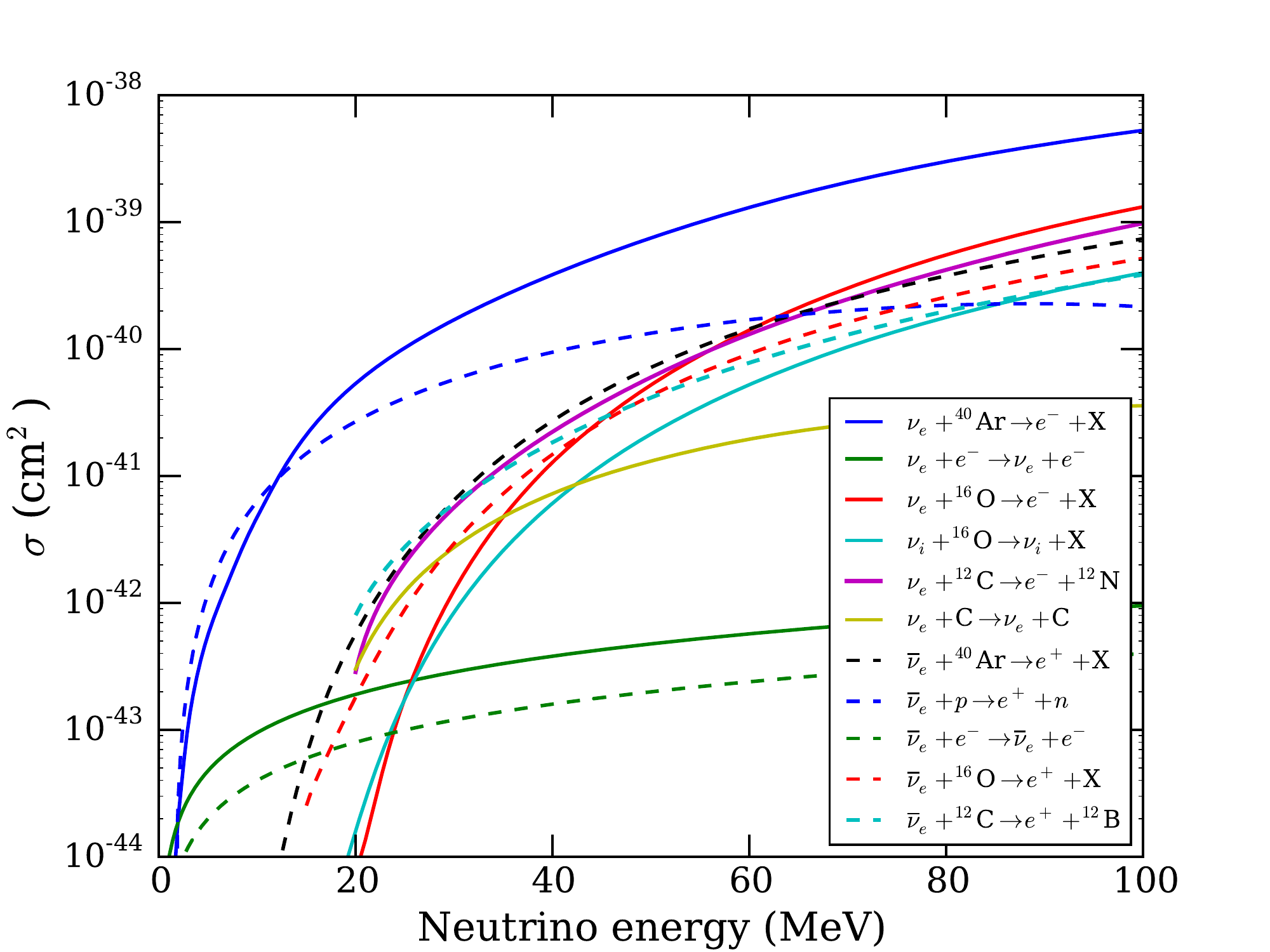}
\centering
\caption{\label{fig:sigma}The \nue\ and \anue\ matter-interaction 
cross sections used in  our study over the domain of neutrino 
energies relevant to our study.  The $^{16}$O$(\bar\nu_e,e^+)$X
cross section and both the $^{12}$C cross sections are not plotted to
zero at low energies owing to a lack of tabulated data at these energy
values from the sources used.  The cross sections are assumed to be
zero below the lowest extent of the tabulated data.}
\end{figure}

%  LocalWords:  func lumallt anue nux shen nuelumt fittingdemo.pdf

\end{document}